\documentclass[11pt,journal,draftcls,onecolumn,peerreviewca]{IEEEtran}
\usepackage{amsfonts}
\usepackage{amsmath}
\usepackage{amssymb}
\usepackage{bm}
\usepackage{xcolor,graphicx,float}
\usepackage{color, soul}
\usepackage{stfloats}
\usepackage[numbers,sort&compress]{natbib}
\usepackage[amsmath,thmmarks]{ntheorem}
\usepackage{theorem}
\usepackage{algorithm}
\usepackage{algorithmic}
\usepackage{graphicx}
\usepackage{subfigure}
\usepackage{pict2e}
\newtheorem{theorem}{Theorem}

\newtheorem{property}{Property}

\theoremheaderfont{\sc}\theorembodyfont{\upshape}
\theoremstyle{nonumberplain}
\theoremseparator{}
\theoremsymbol{\rule{1ex}{1ex}}
\newtheorem{proof}{Proof}

\begin{document}
\title{Grid-less Variational Bayesian Inference of Line Spectral from Quantized Samples}

\author{Jiang~Zhu, Qi Zhang and Xiangming Meng \thanks{Jiang Zhu and Qi Zhang are with the Key Laboratory
of Ocean Observation-imaging Testbed of Zhejiang Province, Ocean College,
Zhejiang University, No.1 Zheda Road, Zhoushan, 316021, China. Xiangming Meng is with Huawei Technologies, Co. Ltd., Shanghai, 201206, China. }}
%The corresponding author of this work is Xiangming Meng.
%\thanks{Jiang Zhu and Qi Zhang are with the key laboratory
%of ocean observation-imaging testbed of Zhejiang Province, Ocean College,
%Zhejiang University, No.1 Zheda Road, Zhoushan, 316021, China. Xiangming Meng is with Huawei Technologies, Co. Ltd., Shanghai, 201206, China. }
%\thanks{Jiang Zhu and Qi Zhang are with Ocean College, Zhejiang University. Xiangming Meng is with Huawei company. Jun Fang is with the University of Electronic Science and Technology of China. }
%\date{}
%{\center\today}
%In addition, the multi snapshot VALSE-EP (MVALSE-EP) is developed to solve the LSE with the multiple measurement vectors (MMVs).
\maketitle
\begin{abstract}
Efficient estimation of line spectral from quantized samples is of significant importance in information theory and signal processing, e.g., channel estimation in energy efficient massive MIMO systems and direction of arrival estimation. The goal of this paper is to recover the line spectral as well as its corresponding parameters including the model order, frequencies and amplitudes from heavily quantized samples. To this end, we propose an efficient grid-less Bayesian algorithm named VALSE-EP, which is a combination of the variational line spectral estimation (VALSE) and expectation propagation (EP). The basic idea of VALSE-EP is to iteratively approximate the challenging quantized model of line spectral estimation as a sequence of simple pseudo unquantized models so that the VALSE can be applied. Note that the noise in the pseudo linear model is heteroscedastic, i.e., different components having different variances, and a variant of the VALSE is re-derived to obtain the final VALSE-EP. Moreover, to obtain a benchmark performance of the proposed algorithm, the Cram\'{e}r Rao bound (CRB) is derived. Finally, numerical experiments on both synthetic and real data are performed, demonstrating the near CRB performance of the proposed VALSE-EP for line spectral estimation from quantized samples.
\end{abstract}
%\begin{keywords}
\section{Introduction}
Line spectral estimation (LSE) is a fundamental problem in information theory and statistical signal processing which has widespread applications, e.g., channel estimation \cite{Bajwa}, direction of arrival (DOA) estimation \cite{Ottersten}. To address this problem, on one hand, many classical methods have been proposed, such as the fast Fourier transform (FFT) based periodogram \cite{Stoica}, subspace based MUSIC \cite{Schmidt} and ESPRIT \cite{Roy}. On the other hand, to exploit the frequency sparsity of the line spectral signal, sparse representation and compressed sensing (CS) based methods have been proposed to estimate frequencies for multiple sinusoids.

Depending on the model adopted, CS based methods for LSE can be classified into three categories, namely, on-grid, off-grid and grid-less, which also correspond to the chronological order in which they have been developed \cite{yangzaibook}. At first, on-grid methods where the continuous frequency is discretized into a finite set of grid points are proposed \cite{Malioutov}. It is shown that grid based methods will incur basis mismatch when the true frequencies do not lie exactly on the grid \cite{MMis}. Then, off-grid compressed sensing methods have been proposed. In \cite{Mamandipoor}, a Newtonalized orthogonal matching pursuit (NOMP) method is proposed, where a Newton step and feedback are utilized to refine the frequency estimates. Compared to the incremental step in updating the frequencies in NOMP, the iterative reweighted approach (IRA) \cite{Fang} estimates the frequencies in parallel, which improves the estimation accuracy at the cost of increasing complexity. In \cite{Hansen}, superfast LSE method is proposed based on fast Toeplitz matrix inversion algorithm. In \cite{Hu1, Hu2}, a sparse Bayesian learning method is proposed, where the grid bias and the grid are jointly estimated \cite{Hu2}, or the Newton method is applied to refine the frequency estimates \cite{Hu1}. To completely overcome the grid mismatch problem, grid-less based methods have been proposed  \cite{Yang2, Tang, Chi, Yang1}. The atomic norm-based methods involve solving a semidefinite programming (SDP) problem \cite{boyd}, whose computation complexity is prohibitively high for large problem size. In \cite{Badiu}, a grid-less variational line spectral estimation (VALSE) algorithm is proposed, where posterior probability density function (PDF) of the frequency is provided. In \cite{Qi}, the multisnapshot VALSE (MVALSE) is developed for the MMVs setting, which also shows the relationship between the VALSE and the MVALSE.

In practice, the measurements might be obtained in a nonlinear way, either preferably or inevitably. For example, in the mmWave multiple input multiple output (MIMO) system, since the mmWave accompanies large bandwidths, the cost and power consumption are huge due to high precision (e.g., 10-12 bits) analog-to-digital converters (ADCs) \cite{PC}. Consequently, low precision ADCs (often $1-3$ bits) are adopted to alleviate the ADC bottleneck. Another motivation is wideband spectrum sensing in bandwidth-constrained wireless networks \cite{FS, SL}. In order to reduce the communication overhead, the sensors quantize their measurements into a single bit, and the spectrum is estimated from the heavily quantized measurements at the fusion center (FC). There are also  various scenarios where measurements are inevitably obtained nonlinearly such as phase retrieval \cite{GAMPPR, MIMOPR}. As a result, it is of great significance in designing efficient nonlinear LSE algorithms. This paper will consider in particular LSE from low precision quantized observations \cite{Fangsign, FangAda} but extension to general nonlinear scenarios could easily fit into our proposed framework without much difficulty.
\subsection{Related Work}
Many classical methods have been extended to solve the LSE from quantized samples. In \cite{Franceschetti}, the spectrum of the one-bit data is analyzed, which consists of plentiful harmonics. It shows that under low signal to noise ratio (SNR) scenario, the amplitudes of the higher order harmonics are much smaller than that of the fundamental frequency, thus the classical FFT based method still works well for the SAR imaging experiment. However, the FFT based method can overestimate the model order (number of spectrums) in the high SNR scenario. As a consequence, the quantization effects must be taken into consideration. The CS based methods have been proposed to solve the LSE from quantized samples, which can also be classified into on-grid, off-grid and grid-less methods.
\begin{itemize}
  \item on-grid methods: The on-grid methods can be classified into $l_1$ minimization based approach \cite{Gianelli, Lijian, Yu} and generalized sparse Bayesian learning (Gr-SBL) \cite{MengZhu} algorithm. For $l_1$ minimization approach, the regularization parameter is hard to determine the tradeoff between the fitting error and the sparsity. While the reconstruction accuracy of the Gr-SBL is high, its computation complexity is high since it involves a matrix inversion in each iteration.
  \item off-grid methods: The SVM based \cite{Gaoyu} and 1bRelax algorithm \cite{Gianelli2} are two typical approaches. For the SVM based approach, the model order needs to be known a priori, while the 1bRelax algorithm \cite{LiC} get rid of such need by using the consistent Bayesian information criterion (BIC) to determine the model order.
  \item grid-less methods: The grid-less approach can completely overcome the grid mismatch problem and the atomic norm minimization approach has been proposed \cite{Zhou, Fu, Wen}. However, its computational complexity is high as it involves solving a SDP.
\end{itemize}

From the point of view of CS, many Bayesian algorithms have been developed, such as approximate message passing (AMP) \cite{Kabashima, AMP}, vector AMP (VAMP) \cite{VAMP}. It is shown in \cite{AMPEP, Wu} that AMP  can be alternatively derived via expectation propagation (EP) \cite{Minka} , an effective approximate Bayesian inference method.  To deal with nonlinear observations, i.e., generalized linear models (GLM), AMP and VAMP are extended to GAMP \cite{GAMP} and GVAMP \cite{GVAMP} respectively  using different methods. The authors in  \cite{meng1}  propose a unified Bayesian inference framework for the GLM inference which shows that GLM could be iteratively approximated as a standard linear model (SLM) using EP \footnote{The extrinsic message in \cite{meng1} could be equivalently obtained through EP.}. This unified framework provides new insights into some existing algorithms, as elucidated by a concise derivation of GAMP in \cite{meng1},  and motivates the design of new algorithms such as the generalized SBL (Gr-SBL) algorithm \cite{meng1, MengZhu}. This paper extends the idea further and utilize EP to solve LSE from quantized samples.
\subsection{Main Contributions}
This work studies the LSE problem from quantized measurements. Utilizing the EP \cite{Minka}, the generalized linear model can be iteratively decomposed as two modules (a standard linear model \footnote{In fact, it is a nonlinear model instead of a standard linear model, which is different from \cite{meng1} since the frequencies are unknown.} and a componentwise minimum mean squared error (MMSE) module) \cite{meng1}. Thus the VALSE algorithm is run in the standard linear module where the frequency estimate is iteratively refined. For the MMSE module, it refines the pseudo observations of the linear model \footnote{Iteratively approximating the generalized linear model as a standard linear model is very beneficial, as many well developed methods such as the information-theoretically optimal successive interference cancellation (SIC) is developed in the SLM.}. By iterating between the two modules, the estimates of the frequency are improved gradually. The main contributions of this work are summarized as follows:
\begin{itemize}
  \item A VALSE-EP method is proposed to deal with the LSE from quantized samples. The quantized model is iteratively approximated as a sequence of pseudo unquantized models with heteroscedastic noise (different components having different variance), a variant of the VALSE is re-derived.
  \item The VALSE-EP is a completely grid-less approach. Besides, the model order estimation is coupled within the iteration and the computational complexity is low, compared to the atomic norm minimization approach.
  \item The relationship between VALSE and VALSE-EP is revealed under the unquantized case. It is shown that the major difference lies in the noise variance estimation step. For VALSE-EP, it is iteratively solved by exchanging extrinsic information between the pseudo unquantized module (module A) and the MMSE module (module B). For VALSE, the noise variance estimate is  equivalently derived through the expectation maximization (EM) step in module A, while VALSE-EP utilizes the EM step to estimate the noise variance in module B, which demonstrate that VALSE and VALSE-EP are not exactly equivalent.
  \item Utilizing the framework from \cite{meng1}, VALSE-EP is proposed which combines the VALSE algorithm with EP. The two different criterions are combined together, and numerical experiments on both synthetic and real data demonstrate the excellent performance of VALSE-EP.
  \item Although this paper focuses on the case of quantized measurements, it is believed that VALSE-EP can be easily extended to other nonlinear measurement scenarios such as phase retrieval without overcoming much difficulty.
\end{itemize}

\subsection{Paper Organization and Notation}
The rest of this paper is organized as below. Section \ref{setup} describes the system model and introduces the probabilistic formulation. Section \ref{bound} derives the Cram\'{e}r Rao bound (CRB). Section \ref{VALSEHN} develops the VALSE for heterogenous noise. The VALSE-EP algorithm and the details of the updating expressions are presented in Section \ref{Algorithm}. The relationship between VALSE and VALSE-EP in the unquantized setting is revealed in Section \ref{RelVALSEEP}. Substantial numerical experiments are provided in Section \ref{NS} and Section \ref{con} concludes the paper.

For a complex vector ${\mathbf x}\in {\mathbb C}^M$, let ${\Re}\{{\mathbf x}\}$ and $\Im\{\mathbf x\}$ denote the real and imaginary part of $\mathbf x$, respectively, let $|\mathbf x|$ and $\angle{\mathbf x}$ denote the componentwise amplitude and phase of $\mathbf x$, respectively. For the square matrix $\mathbf A$, let ${\rm diag}(\mathbf A)$ return a vector with elements being the diagonal of $\mathbf A$. While for a vector ${\mathbf a}$, let ${\rm diag}(\mathbf a)$ return a diagonal matrix with the diagonal being $\mathbf a$, and thus ${\rm diag}({\rm diag}(\mathbf A))$ returns a diagonal matrix. Let ${\rm j}$ denote the imaginary number. Let ${\mathcal S}\subset \{1,\cdots,N\}$ be a subset of indices and $|{\mathcal S}|$ denote its cardinality. For the matrix $\mathbf J\in\mathbb C^{N\times N}$, let $\mathbf J_{\mathcal S}$ denote the submatrix by choosing both the rows and columns of $\mathbf J$ indexed by $\mathcal S$. Similarly, let $\mathbf h_{\mathcal S}$ denote the subvector by choosing the elements of $\mathbf h$ indexed by $\mathcal S$. Let ${(\cdot)}^{*}_{\mathcal S}$, ${(\cdot)}^{\rm T}_{\mathcal S}$ and ${(\cdot)}^{\rm H}_{\mathcal S}$ be the conjugate, transpose and Hermitian transpose operator of ${(\cdot)}_{\mathcal S}$, respectively. For the matrix $\mathbf A$, let $|\mathbf A|$ denote the elementwise absolute value of $\mathbf A$. Let $\mathbf I_L$ denote the identity matrix of dimension $L$. $``\sim i"$ denotes the indices ${\mathcal S}$ excluding $i$. Let ${\mathcal {CN}}({\mathbf x};{\boldsymbol \mu},{\boldsymbol \Sigma})$ denote the complex normal distribution of ${\mathbf x}$ with mean ${\boldsymbol \mu}$ and covariance ${\boldsymbol \Sigma}$. Let $\phi(x)=\exp(-x^2/2)/{\sqrt{2\pi}}$ and $\Phi(x)=\int_{-\infty}^x\phi(t){\rm d}t$ denote the standard normal probability density function (PDF) and cumulative distribution function (CDF), respectively. Let ${\mathcal W}(\cdot)$ wrap frequency in radians to the interval $[-\pi,\pi]$.
\section{Problem Setup}\label{setup}
Let ${\mathbf z}\in {\mathbb C}^N$ be a line spectra consisting of $K$ complex sinusoids
\begin{align}\label{signal-generate}
{\mathbf z}=\sum\limits_{k=1}^K  {\mathbf a}(\tilde{\theta}_k)\tilde{w}_k,
\end{align}
where $\tilde{w}_k$ is the complex amplitude of the $k$th frequency, $\tilde{\theta}_k\in [-\pi,\pi)$ is the $k$th frequency, and
\begin{align}
{\mathbf a}(\theta)=[1,{\rm e}^{{\rm j}\theta},\cdots,{\rm e}^{{\rm j}(N-1)\theta}]^{\rm T}.
\end{align}
The noisy measurements of ${\mathbf z}$ are observed and quantized into a finite number of bits \footnote{Extending to the incomplete measurement scenario where only a subset of measurements ${\mathcal M}=\{m_1,\cdots,m_M\}\subseteq \{0,1,\cdots,N-1\}$ is observed is straightforward. For notation simplicity, we study the full measurement scenario. But the code that we have made available \cite{VALSE_EP} does
provide the required flexibility.}, i.e.,
\begin{align}\label{quantmodel}
{\mathbf y}={\mathcal Q}(\Re\{{\mathbf z}+{\boldsymbol \epsilon}\})+{\rm j}{\mathcal Q}(\Im\{{\mathbf z}+{\boldsymbol \epsilon}\}),
\end{align}
where ${\boldsymbol \epsilon}\sim {{\mathcal {CN}}({\boldsymbol \epsilon};{\mathbf 0},\sigma^2{\mathbf I}_N)}$, $\sigma^2$ is the variance of the noise, ${\mathcal Q}(\cdot)$ is a quantizer which is applied componentwise to map the continuous values into discrete numbers. Specifically, let the quantization intervals be $\{(t_l,t_{l+1})\}_{l=0}^{|{\mathcal D}|-1}$, where $t_0=-\infty$, $t_{{\mathcal D}}=\infty$, $\bigcup_{l=0}^{{\mathcal D}-1}[t_l,t_{l+1})={\mathbb R}$. Given a real number $a\in [t_l,t_{l+1})$, the representation is
\begin{align}
{\mathcal Q}(a)=\omega_l, \quad {\rm if}\quad a\in [t_l,t_{l+1}).
\end{align}
Note that for a quantizer with bit-depth $B$, the cardinality of the output of the quantizer is $|{\mathcal D}|=2^B$.

%Suppose that only $M\leq N$ noisy measurements of those components of ${\mathbf z}_N$ are observed and are quantized into a finite number of bits, i.e.,
%\begin{align}\label{quantmodel}
%{\mathbf y}={\mathcal Q}(\Re\{{\mathbf z}_{\mathcal M}+{\mathbf n}\})+{\rm j}{\mathcal Q}(\Im\{{\mathbf z}_{\mathcal M}+{\mathbf n}\}),
%\end{align}
%where ${\mathbf z}_{\mathcal M}=\sum\limits_{k=1}^K  {\mathbf a}_{\mathcal M}(\tilde{\theta}_k)\tilde{w}_k$,
%\begin{align}
%{\mathbf a}_{\mathcal M}(\theta)=[{\rm e}^{{\rm j}m_1\theta},{\rm e}^{{\rm j}m_2\theta},\cdots,{\rm e}^{{\rm j}m_{M}\theta}]^{\rm T},
%\end{align}
%${\mathcal M}=\{m_1,\cdots,m_M\}\subseteq \{0,1,\cdots,N-1\}$, ${\mathbf n}\sim {{\mathcal {CN}}({\mathbf n};{\mathbf 0},\sigma^2{\mathbf I}_M)}$, $\sigma^2$ is the variance of the noise. To simplify the notation, in the following text, ${\mathbf a}(\theta)$ is used instead of ${\mathbf a}_{\mathcal M}(\theta)$.

The goal of LSE is to jointly recover the number of spectrums $\hat{K}$ (also named model order), the set of frequencies $\hat{\boldsymbol \theta}=\{\hat{\theta}_k\}_{k=1}^{\hat{K}}$, the corresponding coefficients $\{\hat{w}_k\}_{k=1}^{\hat{K}}$ and the LSE $\hat{{\mathbf z}}=\sum\limits_{k=1}^{\hat{K}} \hat{\mathbf a}({\theta}_k)\hat{w}_k$ from quantized measurements ${\mathbf y}$.

Since the sparsity level $K$ is usually unknown, the line spectral consisting of $N$ complex sinusoids is assumed \cite{Badiu}
\begin{align}\label{signal-model}
{\mathbf z}=\sum\limits_{i=1}^N {w}_i {\mathbf a}({\theta}_i)\triangleq {\mathbf A}({\boldsymbol \theta}){\mathbf w},
\end{align}
where ${\mathbf A}({\boldsymbol \theta})=[{\mathbf a}({\theta}_1),\cdots,{\mathbf a}({\theta}_N)]$ and $N$ satisfies $N> K$. Since the number of frequencies is $K$, the binary hidden variables ${\mathbf s}=[s_1,...,s_N]^{\rm T}$ are introduced, where $s_i=1$ means that the $i$th frequency is active, otherwise deactive ($w_i=0$). The probability mass function (PMF) of $s_i$ is
\begin{align}\label{sprob}
p(s_i;\rho) = \rho^{s_i}(1-\rho)^{(1-s_i)},\quad s_i\in\{0,1\}.
\end{align}
Given that $s_i=1$, we assume that $w_i\sim {\mathcal {CN}}({w_i};0,\tau)$. Thus $(s_i,{w_i})$ follows a Bernoulli-Gaussian distribution, that is
\begin{align}
p({w_i}|s_i;\tau) = (1 - s_i){\delta}({w_i}) + s_i{\mathcal {CN}}({w_i};0,\tau).\label{pdfw}
\end{align}

According to (\ref{sprob}) and (\ref{pdfw}), the parameter $\rho$ denotes the probability of the $i$th component being active and $\tau$ is a variance parameter. The variable ${\boldsymbol \theta} = [\theta_1,...,\theta_N]^{\rm T}$ has the prior PDF $p({\boldsymbol \theta}) = \begin{matrix} \prod_{i=1}^N p(\theta_i) \end{matrix}$.
Without any knowledge of the frequency $\theta_i$, the uninformative prior distribution $p(\theta_i) = {1}/({2\pi})$ is used \cite{Badiu}. For encoding the prior distribution, please refer to \cite{Badiu, Qi} for further details.

Given $\mathbf z$, the PDF $p({\mathbf y}|{\mathbf z};\sigma^2)$ of $\mathbf y$ can be easily calculated through (\ref{quantmodel}). Let
\begin{align}
&{\boldsymbol \Omega}=(\theta_1,\dots,\theta_N,({\mathbf w},{\mathbf s})),\\
&{\boldsymbol \beta} =\{{\boldsymbol \beta}_w,{\beta}_z\},
\end{align}
be the set of all random variables and the model parameters, respectively, where ${\boldsymbol \beta}_w=\{\rho,~\tau\}$ and ${\beta}_z=\{\sigma^2\}$. According to the Bayes rule, the joint PDF $p({\mathbf y },{\mathbf z},{\boldsymbol \Omega};{\boldsymbol \beta})$ is
\begin{align}\label{jointpdf}
p({\mathbf y },{\mathbf z},{\boldsymbol \Omega};{\boldsymbol \beta})=p({\mathbf y}|{\mathbf z})\delta({\mathbf z}-{\mathbf A}({\boldsymbol \theta}){\mathbf w})\prod\limits_{i=1}^N p(\theta_i)p(w_i|s_i)p(s_i).
\end{align}
Given the above joint PDF (\ref{jointpdf}), the type II maximum likelihood (ML) estimation of the model parameters $\hat{\boldsymbol\beta}_{\rm ML}$ is
\begin{align}\label{MLbeta}
\hat{\boldsymbol\beta}_{\rm ML}=\underset{\boldsymbol \beta}{\operatorname{argmax}}~ p({\mathbf y };{\boldsymbol \beta})=\underset{\boldsymbol \beta}{\operatorname{argmax}}~\int p({\mathbf y },{\mathbf z},{\boldsymbol \Omega};{\boldsymbol \beta}){\rm d}{\mathbf z}{\rm d}{{\boldsymbol \Omega}}.
\end{align}
Then the minimum mean squared error (MMSE) estimates of the parameters $({\mathbf z},{\boldsymbol \Omega})$ are
\begin{align}\label{MMSE}
(\hat{\mathbf z},\hat{{\boldsymbol \Omega}})={\rm E}[({\mathbf z},{\boldsymbol \Omega})|{\mathbf y};\hat{\boldsymbol \beta}_{\rm ML}],
\end{align}
where the expectation is taken with respect to
\begin{align}
p({\mathbf z},{\boldsymbol \Omega}|{\mathbf y };\hat{\boldsymbol \beta}_{\rm ML})=\frac{p({\mathbf z},{\boldsymbol \Omega},{\mathbf y };\hat{\boldsymbol \beta}_{\rm ML})}{p({\mathbf y };\hat{\boldsymbol \beta}_{\rm ML})}.
\end{align}
Directly solving the ML estimate of $\boldsymbol \beta$ (\ref{MLbeta}) or the MMSE estimate of $({\mathbf z},{\boldsymbol \Omega})$ (\ref{MMSE}) are both intractable. As a result, an iterative algorithm is designed in Section \ref{Algorithm}.
\section{Cram\'{e}r Rao bound}\label{bound}
Before designing the recovery algorithm, the performance bounds of unbiased estimators are derived, i.e., the Cram\'{e}r Rao bound (CRB). Although the Bayesian algorithm is designed, the CRB can be acted as the performance benchmark of the algorithm. To derive the CRB, $K$ is assumed to be known, the frequencies $\tilde{\boldsymbol \theta}\in{\mathbb R}^K$ and weights $\tilde{\mathbf w}\in{C}^K$ are treated as deterministic unknown parameters, and the Fisher information matrix (FIM) ${\mathbf F}(\boldsymbol \kappa)$ is calculated first. Let $\boldsymbol \kappa$ denote the set of parameters, i.e., ${\boldsymbol \kappa}=[\tilde{\boldsymbol \theta}^{\rm T},\tilde{\mathbf g}^{\rm T},\tilde{\boldsymbol \phi}^{\rm T}]^{\rm T}\in {\mathbb R}^{3K}$, where $\tilde{\mathbf g}=|\tilde{\mathbf w}|$ and $\tilde{\boldsymbol \phi}=\angle{\tilde{\mathbf w}}$. The PMF of the measurements $p({\mathbf y}|{\boldsymbol \kappa})$ is
\begin{align}
p({\mathbf y}|{\boldsymbol \kappa})=\prod\limits_{n=1}^Np(y_n|{\boldsymbol \kappa})=\prod\limits_{n=1}^Np(\Re\{y_n\}|{\boldsymbol \kappa})p(\Im\{y_n\}|{\boldsymbol \kappa}).
\end{align}
Moreover, the PMFs of $\Re\{y_n\}$ and $\Im\{y_n\}$ are
\begin{align}
p(\Re\{y_n\}|{\boldsymbol \kappa})=\prod\limits_{\omega_l\in {\mathcal D}}p_{\Re\{y_n\}}(\omega_l|{\boldsymbol \kappa})^{{\mathbb I}_{\Re\{y_n\}=\omega_l}},\\
p(\Im\{y_n\}|{\boldsymbol \kappa})=\prod\limits_{\omega_l\in {\mathcal D}}p_{\Im\{y_n\}}(\omega_l|{\boldsymbol \kappa})^{{\mathbb I}_{\Im\{y_n\}=\omega_l}},
\end{align}
where ${\mathbb I}_{(\cdot)}$ is the indicator function,
\begin{subequations}\label{py_z}
\begin{align}
&p_{\Re\{y_n\}}(\omega_l|{\boldsymbol \kappa})={\rm P}\left(\Re\{z_n+\varepsilon_n\}\in[t_l,t_{l+1})\right)\\
=&{\Phi(\frac{t_{l+1}-\Re\{z_n\}}{\sigma/\sqrt{2}})
-\Phi(\frac{t_{l}-\Re\{z_n\}}{\sigma/\sqrt{2}})},\\
&p_{\Im\{y_n\}}(\omega_l|{\boldsymbol \kappa})={\rm P}\left(\Im\{z_n+\varepsilon_n\}\in[t_l,t_{l+1})\right)\\
=&{\Phi(\frac{t_{l+1}-\Im\{z_n\}}{\sigma/\sqrt{2}})-\Phi(\frac{t_{l}-\Im\{z_n\}}{\sigma/\sqrt{2}})}.
\end{align}
\end{subequations}
The CRB is equal to the inverse of the FIM ${\mathbf F}({\boldsymbol \kappa})\in{\mathbb R}^{3K\times 3K}$
\begin{align}
{\mathbf F}({\boldsymbol \kappa})={\rm E}\left[\left(\frac{\partial \log p({\mathbf y}|{\boldsymbol \kappa})}{\partial {\boldsymbol \kappa}}\right)\left(\frac{\partial \log p({\mathbf y}|{\boldsymbol \kappa})}{\partial {\boldsymbol \kappa}}\right)^{\rm T}\right].
\end{align}
To calculate the FIM, the following Theorem \cite{Fu} is utilized.
\begin{theorem}\label{FIMlemma}
\cite{Fu} The FIM ${\mathbf F}({\boldsymbol \kappa})$ for estimating the unknown parameter ${\boldsymbol \kappa}$ is
\begin{align}\label{quanFIM}
{\mathbf F}({\boldsymbol \kappa})=\sum\limits_{n=1}^N   \left({\lambda}_n\frac{\partial \Re\{z_n\}}{\partial {\boldsymbol \kappa}}\left(\frac{\partial \Re\{z_n\}}{\partial {\boldsymbol \kappa}}\right)^{\rm T}\right.\notag\\
\left.+\chi_n\frac{\partial \Im\{z_n\}}{\partial {\boldsymbol \kappa}}\left(\frac{\partial \Im\{z_n\}   }{\partial {\boldsymbol \kappa}}\right)^{\rm T}\right).
\end{align}
For a general quantizer, one has
\begin{align}
\lambda_n=\frac{2}{\sigma^2}\sum\limits_{l=0}^{|{\mathcal D}|-1}\frac{[\phi(\frac{t_{l+1}-\Re\{z_n\}}{\sigma/\sqrt{2}})-\phi(\frac{t_{l}-\Re\{z_n\}}{\sigma/\sqrt{2}})]^2}{\Phi(\frac{t_{l+1}-\Re\{z_n\}}{\sigma/\sqrt{2}})-\Phi(\frac{t_{l}-\Re\{z_n\}}{\sigma/\sqrt{2}})},
\end{align}
and
\begin{align}
\chi_n=\frac{2}{\sigma^2}\sum\limits_{l=0}^{|{\mathcal D}|-1}\frac{[\phi(\frac{t_{l+1}-\Im\{z_n\}}{\sigma/\sqrt{2}})-\phi(\frac{t_{l}-\Im\{z_n\}}{\sigma/\sqrt{2}})]^2}{\Phi(\frac{t_{l+1}-\Im\{z_n\}}{\sigma/\sqrt{2}})-\Phi(\frac{t_{l}-\Im\{z_n\}}{\sigma/\sqrt{2}})},
\end{align}
For the unquantized system, the FIM is
\begin{align}\label{unqFIM}
&{\mathbf F}_{\rm unq}({\boldsymbol \kappa})=\frac{2}{\sigma^2}\sum\limits_{n=1}^N   \left(\frac{\partial \Re\{z_n\}}{\partial {\boldsymbol \kappa}}\left(\frac{\partial \Re\{z_n\}}{\partial {\boldsymbol \kappa}}\right)^{\rm T}\right.\notag\\
&\left.+\frac{\partial \Im\{z_n\}}{\partial {\boldsymbol \kappa}}\left(\frac{\partial \Im\{z_n\}   }{\partial {\boldsymbol \kappa}}\right)^{\rm T}\right).
\end{align}
\end{theorem}

According to Theorem \ref{FIMlemma}, we need to calculate $\frac{\partial \Re\{z_n\}}{\partial {\boldsymbol \kappa}}$ and $\frac{\partial \Im\{z_n\}}{\partial {\boldsymbol \kappa}}$. Since
\begin{align}
z_n=\sum\limits_{k=1}^K\tilde{g}_k{\rm e}^{{\rm j}((n-1)\tilde{\theta}_k+\tilde{\phi}_k)},
\end{align}
we have, for $k=1,\cdots,K$,
\begin{align}\label{partialder}
&\frac{\partial \Re\{z_n\}}{\partial {\tilde{\theta}_k}}=-(n-1)\tilde g_k\sin((n-1)\tilde\theta_k+\tilde\phi_k),\notag\\
&\frac{\partial \Re\{z_n\}}{\partial \tilde g_k}=\cos((n-1)\tilde\theta_k+\tilde\phi_k),\notag\\
&\frac{\partial \Re\{z_n\}}{\partial \tilde \phi_k}=-\tilde g_k\sin((n-1)\tilde\theta_k+\tilde\phi_k),\notag\\
&\frac{\partial \Im\{z_n\}}{\partial {\tilde \theta_k}}=(n-1)\tilde g_k\cos((n-1)\tilde \theta_k+\tilde \phi_k),\notag\\
&\frac{\partial \Im\{z_n\}}{\partial \tilde g_k}=\sin((n-1)\tilde \theta_k+\tilde \phi_k),\notag\\
&\frac{\partial \Im\{z_n\}}{\partial \tilde \phi_k}=\tilde g_k\cos((n-1)\tilde \theta_k+\tilde \phi_k).\notag
\end{align}
The CRB for the quantized and unquantized settings are ${\rm CRB}({\boldsymbol \kappa})={\mathbf F}^{-1}({\boldsymbol \kappa})$ and ${\rm CRB}_{\rm unq}({\boldsymbol \kappa})={\mathbf F}_{\rm unq}^{-1}({\boldsymbol \kappa})$, respectively. The CRB of the frequencies are $[{\rm CRB}({\boldsymbol \kappa})]_{1:K,1:K}$, which will be used as the performance metrics.
\section{VALSE under Known Heteroscedastic Noise}\label{VALSEHN}
As shown in \cite{meng1}, according to EP, the quantized (or nonlinear) measurement model can be iteratively approximated as a sequence of pseudo linear measurement model, so that linear inference algorithms could be applied. Since diagonal EP performs better than scalar EP \footnote{The code that we have made available also provides the scalar EP.}, the noise in the pseudo linear measurement model is modeled as heteroscedastic (independent components having different known variances), as opposed to \cite{Badiu} where the noise is homogenous. As a result, a variant of VALSE is rederived in this Section, and VALSE-EP is then developed for the nonlinear measurement model in Section \ref{Algorithm}.

%the VALSE is rederived, in contrast to \cite{Badiu} where the VALSE is derived under homogenous noise.

%As shown in Section \ref{Algorithm},  with noise being heteroscedastic. In \cite{Badiu}, the VALSE is derived where the additive Gaussian noise is homogenous while here the additive Gaussian noise is heteroscedastic (independent components having different known variance). As a result, modifications are needed to ensure that the original VALSE works.

The pseudo linear measurement model is described as
\begin{align}\label{heomodelprior}
\tilde{\mathbf y}={\mathbf A}({\boldsymbol \theta}){\mathbf w}+\tilde{\boldsymbol \epsilon},
\end{align}
where $\tilde{\boldsymbol \epsilon}\sim {\mathcal {CN}}(\tilde{\boldsymbol \epsilon};{\mathbf 0},{\rm diag}(\tilde{\boldsymbol \sigma}^2))$ and $\tilde{\boldsymbol \sigma}^2$ is known.

\begin{figure}[h!t]
\centering
\includegraphics[width=3.4in]{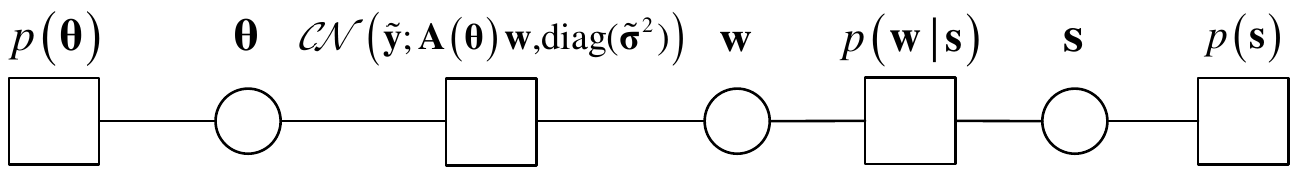}
\caption{The factor graph of (\ref{heomodelprior}) borrowed from \cite{Badiu}.}
\label{FCsub0}                                                                                                                                                                                                                                                       \end{figure}

For model (\ref{heomodelprior}), the factor graph is presented in Fig. \ref{FCsub0}. Given the pseudo measurements $\tilde{\mathbf y}$ and nuisance parameters ${\boldsymbol \beta}_w$,  the above joint PDF is
\begin{align}\label{jointpdf2}
p(\tilde{\mathbf y},{\boldsymbol \Omega};{\boldsymbol \beta}_w) \propto\left(\prod_{i=1}^Np(\theta_i)p(s_i)p({w_i}|s_i)\right)p(\tilde{\mathbf y}|\boldsymbol \theta,\mathbf w),
\end{align}
where $p(\tilde{\mathbf y}|{\boldsymbol\theta},{\mathbf w})={\mathcal {CN}}(\tilde{\mathbf y};{\mathbf A}({\boldsymbol \theta}){\mathbf w},{\boldsymbol \Sigma})$, and ${\boldsymbol \Sigma}={\rm diag}(\tilde{\boldsymbol \sigma}^2)$. Performing the type II maximum likelihood (ML) estimation of the model parameters $\hat{\boldsymbol\beta}_{w}$ are still intractable. Thus variational approach where a given structured PDF $q({\boldsymbol \Omega}|\tilde{\mathbf y})$ is used to approximate $p({\boldsymbol \Omega}|\tilde{\mathbf y})$ is adopted, where $p({\boldsymbol \Omega}|\tilde{\mathbf y})=p(\tilde{\mathbf y}, {\boldsymbol \Omega};{\boldsymbol \beta}_w)/p(\tilde{\mathbf y};{\boldsymbol \beta}_w)$ and $p(\tilde{\mathbf y};{\boldsymbol \beta}_w)=\int p(\tilde{\mathbf y}, {\boldsymbol \Omega};{\boldsymbol \beta}_w) {\rm d}{\boldsymbol \Omega}$. The variational Bayesian uses the Kullback-Leibler (KL) divergence of $p({\boldsymbol \Omega}|\tilde{\mathbf y})$ from $q({\boldsymbol \Omega}|\tilde{\mathbf y})$ to describe their dissimilarity, which is defined as \cite[p.~732]{Murphy}
\begin{align}\label{KLdef}
&{\rm{KL}}(q({\boldsymbol \Omega}|{\tilde{\mathbf y}})||p({\boldsymbol \Omega}|\tilde{\mathbf y}))=\int q({\boldsymbol \Omega}|\tilde{\mathbf y})\log \frac{q({\boldsymbol \Omega}|\tilde{\mathbf y})}{p({\boldsymbol \Omega}|\tilde{\mathbf y})}{\rm d}{\boldsymbol \Omega}.
\end{align}
In general, the posterior PDF $q({\boldsymbol \Omega}|\tilde{\mathbf y})$ is chosen from a distribution set to minimize the  KL divergence. The log model evidence $\ln p({\tilde{\mathbf y};\boldsymbol\beta}_w)$ for any assumed PDF $q({\boldsymbol \Omega}|\tilde{\mathbf y})$ is \cite[pp.~732-733]{Murphy}
\begin{align}\label{DL}
\ln p({\tilde{\mathbf y};\boldsymbol\beta}_w)= {\rm{KL}}(q({\boldsymbol \Omega}|\tilde{\mathbf y})||p({\boldsymbol \Omega}|\tilde{\mathbf y}))+ {\mathcal L}(q({\boldsymbol \Omega}|\tilde{\mathbf y})),
\end{align}
where
\begin{align}\label{DL-L}
{\mathcal L}(q({\boldsymbol \Omega}|\tilde{\mathbf y})) = {\rm E}_{q({\boldsymbol \Omega}|\tilde{\mathbf y})}\left[\ln{\tfrac{p({\mathbf y,{\boldsymbol \Omega};\boldsymbol \beta}_w)}{q({\boldsymbol \Omega}|\tilde{\mathbf y})}}\right].
\end{align}
For a given data $\tilde{\mathbf y}$, $\ln p(\tilde{\mathbf y};{\boldsymbol\beta}_w)$ is constant, thus minimizing the KL divergence is equivalent to maximizing ${\mathcal L}(q({\boldsymbol \Omega}|\tilde{\mathbf y}))$ in (\ref{DL}). Therefore we maximize ${\mathcal L}(q({\boldsymbol \Omega}|\tilde{\mathbf y}))$ in the sequel.

For the factored PDF $q({\boldsymbol \Omega}|\tilde{\mathbf y})$, the following assumptions are made:
\begin{itemize}
  \item Given $\tilde{\mathbf y}$, the frequencies $\{\theta_i\}_{i=1}^N$ are mutually independent.
  \item The posterior of the binary hidden variables $q({\mathbf s}|\tilde{\mathbf y})$ has all its mass at $\hat{\mathbf s}$, i.e., $q({\mathbf s}|\tilde{\mathbf y})=\delta({\mathbf s}-\hat{\mathbf s})$.
  \item Given $\tilde{\mathbf y}$ and $\mathbf s$, the frequencies and weights are independent.
\end{itemize}
As a result, $q({\boldsymbol \Omega}|\tilde{\mathbf y})$ can be factored as
\begin{align}
q({\boldsymbol \Omega}|\tilde{\mathbf y}) = \prod_{i=1}^Nq(\theta_i|\tilde{\mathbf y})q({\mathbf w}|{\tilde{\mathbf y}},{\mathbf s})\delta({\mathbf s}-{\hat{\mathbf s}}).\label{postpdf}
\end{align}
Due to the factorization property of (\ref{postpdf}), the frequency ${\boldsymbol \theta}$ can be estimated from $q({\boldsymbol \Omega}|\tilde{\mathbf y})$ as \cite{Badiu}
\begin{subequations}\label{ahat}
\begin{align}
&\hat{\theta}_i = {\rm arg}({\rm E}_{q(\theta_i|\tilde{\mathbf y})}[{\rm e}^{{\rm j}\theta_i}]),\label{ahat_a}\\
&\hat{\mathbf a}_i = {\rm E}_{q(\theta_i|\tilde{\mathbf y})}[{\mathbf a}_N(\theta_i)],~i\in\{1,...,N\},\label{ahat_b}
\end{align}
\end{subequations}
where ${\rm arg}(\cdot)$ returns the angle. For the given posterior PDF $q(\mathbf w|\mathbf y,\hat{\mathbf s})$, the mean and covariance estimates of the weights are calculated as
\begin{subequations}\label{w_est}
\begin{align}
&\hat{\mathbf w} = {\rm E}_{q({\mathbf w}|\tilde{\mathbf y})}[w],\\
&\hat{C}_{i,j} = {\rm E}_{q({\mathbf w}|\tilde{\mathbf y})}[{ {w}_i{ w}_j^*}] - {\hat{ w}}_i\hat{ w}_j^*,\indent i,j\in\{1,...,N\}.
\end{align}
\end{subequations}
Given that $q({\mathbf s}|\tilde{\mathbf y}) = \delta(\mathbf {s-\hat{s}})$, the posterior PDF of $\mathbf w$ is
\begin{align}\label{w_s}
q({\mathbf w}|\tilde{\mathbf y}) = \int q({\mathbf w}|\tilde{\mathbf y},{\mathbf s})\delta(\mathbf {s-\hat{s}}){\rm d}{\mathbf s} = q({\mathbf w}|\tilde{\mathbf y},\hat{\mathbf s}).
\end{align}
Let $\mathcal S$ be the set of indices of the non-zero components of $\mathbf s$, i.e.,
\begin{align}\notag
\mathcal S = \{i|1\leq i\leq N,s_i = 1\}.
\end{align}
Analogously, we define $\hat{\mathcal S}$ based on $\hat{\mathbf s}$. The model order is the cardinality of $\hat{\mathcal S}$, i.e.,
\begin{align}\notag
\hat{K} = |\hat{\mathcal S}|.
\end{align}
Finally, the line spectral ${\mathbf z}=\sum\limits_{k=1}^{K}{\mathbf a}({\widetilde\theta}_k){\widetilde{w}}_k$ is reconstructed as
\begin{align}\notag
\hat{\mathbf z} = \sum_{i\in{\hat{\mathcal S}}}\hat{\mathbf a}_i\hat{w}_i.
\end{align}

The following procedure is similar to \cite{Qi}. Maximizing ${\mathcal L}(q({\boldsymbol \Omega}|\tilde{\mathbf y}))$ with respect to all the factors is also intractable. Similar to the Gauss-Seidel method \cite{Bertsekas}, $\mathcal L$ is optimized over each factor $q({\theta_i}|\tilde{\mathbf y})$, $i=1,\dots,N$ and $q({\mathbf w},{\mathbf s}|\tilde{\mathbf y})$ separately with the others being fixed.  Maximizing ${\mathcal L}(q({\boldsymbol \Omega}|\tilde{\mathbf y});{\boldsymbol \beta}_w)$ (\ref{DL-L}) with respect to the posterior approximation $q({\boldsymbol \Omega}_d|\tilde{\mathbf y})$ of each latent variable ${\boldsymbol \Omega}_d,~d=1,\dots,N+1$ yields \cite[pp. 735, eq. (21.25)]{Murphy}
\begin{align}\label{upexpression}
\ln q({\boldsymbol \Omega}_d|\tilde{\mathbf y})={\rm E}_{q({{\boldsymbol \Omega}\setminus{\boldsymbol \Omega}_d}|\tilde{\mathbf y})}[\ln p(\tilde{\mathbf y},{\boldsymbol \Omega})]+{\rm const},
\end{align}
where the expectation is taken with respect to all the variables ${\boldsymbol \Omega}$ except ${\boldsymbol \Omega}_d$ and the constant ensures normalization of the PDF. In the ensuing three Subsections, we detail the procedures.
\subsection{Inferring the Frequencies}\label{yita}
For each $i = 1,...,N$, we maximize $\mathcal L$ with respect to the factor $q({\theta_i}|\tilde{\mathbf y})$. For $i\notin{\mathcal S}$, we have $q(\theta_i|\tilde{\mathbf y})= p(\theta_i)$. For $i\in{\mathcal S}$, according to (\ref{upexpression}), the optimal factor $q(\theta_i|\tilde{\mathbf y})$ can be calculated as
\begin{align}
\ln q(\theta_i|\tilde{\mathbf y}) = &{\rm E}_{q({\boldsymbol \Omega}\setminus{\theta_i}|\tilde{\mathbf y})}\left[
\ln p(\tilde{\mathbf y}, {\boldsymbol \Omega};{\boldsymbol\beta}_w)\right]+ \rm const,\label{q_cal}
\end{align}
Substituting (\ref{ahat}) and (\ref{w_est}) in (\ref{q_cal}), one obtains
\begin{align}\notag
&\ln q(\theta_i|\tilde{\mathbf y}) = {\rm E}_{q({\boldsymbol \Omega}\setminus{\theta_i}|\tilde{\mathbf y})}[\ln p(\tilde{\mathbf y}, {\boldsymbol \Omega};{\boldsymbol\beta}_w)] + {\rm const}\notag\\
=&{\rm E}_{q({\mathbf z}\setminus{\theta_i}|\tilde{\mathbf y})}[\ln(p(\boldsymbol\theta)p(\mathbf s)p(\mathbf {w|s})p(\tilde{\mathbf y}|\boldsymbol \theta,\mathbf w))] + {\rm const}\notag\\
%=&{\rm E}_{q({\mathbf z}\setminus{\theta_i}|{\mathbf Y})}[\sum_{j=1}^N \ln p(\theta_j)+\sum_{j=1}^N \ln p(s_j)
%+\ln p(\mathbf {w|s}) + \ln p(\tilde{\mathbf y}|\boldsymbol \theta,\mathbf w)] + {\rm const}\notag\\
=&\ln p(\theta_i)-{\rm E}_{q({\mathbf z}\setminus{\theta_i}|\tilde{\mathbf y})}[(\tilde{\mathbf y}-{{\mathbf A}_{\hat{\mathcal S}}}{\mathbf w}_{\hat{\mathcal S}})^{\rm H}\boldsymbol\Sigma^{-1}(\tilde{\mathbf y}- {\mathbf A}_{\hat{\mathcal S}}{\mathbf w}_{\hat{\mathcal S}})]+{\rm const}\notag\\
=&\ln p(\theta_i)+  \Re\{{\boldsymbol\eta}_i^{\rm H}{\mathbf a}(\theta_i)\}+{\rm const},
\end{align}
where the complex vector $\boldsymbol\eta_i$ is given by
\begin{align}\label{yita-i}
{\boldsymbol\eta}_i = 2\boldsymbol\Sigma^{-1}\left[\left(\tilde{\mathbf y}-\sum_{l\in\hat{\mathcal S} \backslash \{i\}}{\hat{\mathbf a}}_l{\hat{w}}_l\right){\hat{w}}^*_i-\sum_{l\in\hat{\mathcal S} \backslash \{i\}}{\hat{\mathbf a}}_l{\hat{C}}_{l,i}\right],
\end{align}
where $``\sim i"$ denote the indices $\hat{\mathcal S}$ excluding $i$, $\mathbf w_{\hat{\mathcal S}}$ denotes the subvector of $\mathbf w$ by choosing the $\hat{\mathcal S}$ rows of $\mathbf w$. The result is consistent with \cite[eq. (17)]{Badiu} when the diagonal covariance matrix $\boldsymbol \Sigma$ reduces to the scaled identity matrix. $q(\theta_i|\tilde{\mathbf y})$ is calculated to be
\begin{align}\label{pdf-q}
q(\theta_i|\tilde{\mathbf y})\propto p(\theta_i){\rm exp}({\Re}\{{\boldsymbol\eta}_i^{\rm H}{\mathbf a}(\theta_i)\}).
\end{align}
Since it is hard to obtain the analytical results (\ref{ahat_b}) for the PDF (\ref{pdf-q}), $q(\theta_i|\tilde{\mathbf y})$ is approximated as a von Mises distribution. For further details, please refer to \cite[Algorithm 2: Heurestic 2]{Badiu}.
\subsection{Inferring the Weights and Support}\label{W-and-C}
Next we keep $q(\theta_i|\tilde{\mathbf y}),i=1,...,N$ fixed and maximize $\mathcal L$ w.r.t. $q({\mathbf w},{\mathbf s}|\tilde{\mathbf y})$.
Define the matrices $\mathbf J$ and $\mathbf h$ as
\begin{subequations}\label{J-H}
\begin{align}
&{J}_{ij}= 	
\begin{cases}
{\rm tr}({\boldsymbol\Sigma}^{-1}),&i=j\\
{\hat{\mathbf a}}^{\rm H}_i{\boldsymbol\Sigma}^{-1}{\hat{\mathbf a}}_j,&i\neq{j}
\end{cases},\quad i,j\in\{1,2,\cdots,N\}\label{J1},\\
&{\mathbf h} = \hat{\mathbf A}^{\rm H}{\boldsymbol\Sigma}^{-1}\tilde{\mathbf y},\label{H1}
\end{align}
\end{subequations}
where $\hat{\mathbf A}=[{\hat{\mathbf a}}_1,\cdots,{\hat{\mathbf a}}_N]$.
According to (\ref{upexpression}), $q({\mathbf w},{\mathbf s}|\tilde{\mathbf y})$ can be calculated as
\begin{align}
&\ln q({\mathbf w},{\mathbf s}|\tilde{\mathbf y}) = {\rm E}_{q({\boldsymbol \Omega}\setminus{({\mathbf w},{\mathbf s})}|\tilde{\mathbf y})}\left[
\ln p(\tilde{\mathbf y},{\boldsymbol \Omega};{\boldsymbol\beta}_w)\right]+ {\rm const}\notag\\
=& {\rm E}_{q{(\boldsymbol\theta|\tilde{\mathbf y}})}[\sum_{i=1}^N \ln p(s_i) + \ln p(\mathbf {w|s})+\ln p(\tilde{\mathbf y}|\boldsymbol \theta,\mathbf w)]+{\rm const}\notag\\
=&-({\mathbf w}_{\mathcal S} - \hat{\mathbf w}_{{\mathcal S}})^{\rm H}\hat{\mathbf C}_{{\mathcal S}}^{-1}({\mathbf w}_{\mathcal S} - \hat{\mathbf w}_{{\mathcal S}}) + {\rm const},\label{Ws_pos}
\end{align}
where
\begin{subequations}\label{W-C-1}
\begin{align}
&\hat{\mathbf C}_{{\mathcal S}} = \left({{\mathbf J}_{{\mathcal S}}}+\frac{{\mathbf I}_{|{\mathcal S}|}}{\tau}\right)^{-1},\\
&\hat{\mathbf w}_{{\mathcal S}} = \hat{\mathbf C}_{{\mathcal S}}\mathbf h_{{\mathcal S}}.\label{What}
\end{align}
\end{subequations}
It is worth noting that calculating $\hat{\mathbf C}_{{\mathcal S}}$ and $\hat{\mathbf w}_{{\mathcal S}}$ involves a matrix inversion. In the Appendix \ref{findmaxz}, it is shown that $\hat{\mathbf C}_{{\mathcal S}}$ and $\hat{\mathbf w}_{{\mathcal S}}$ can be updated efficiently.

From (\ref{postpdf}), the posterior approximation $q({\mathbf w},{\mathbf s}|\tilde{\mathbf y})$  can be factored as the product of $q({\mathbf w}|\tilde{\mathbf y},{\mathbf s})$ and $\delta({\mathbf s}-{\hat{\mathbf s}})$. According to the formulation of (\ref{Ws_pos}), for a given $\hat{\mathbf s}$, $q({\mathbf w}_{\hat{\mathcal S}}|\tilde{\mathbf y})$ is a complex Gaussian distribution, and $q({\mathbf w}|\tilde{\mathbf y};\hat{\mathbf s})$ is
\begin{align}
q({\mathbf w}|\tilde{\mathbf y};\hat{\mathbf s}) = {\mathcal {CN}}({\mathbf w}_{\hat{\mathcal S}};\hat{\mathbf w}_{\hat{\mathcal S}},\hat{\mathbf C}_{\hat{\mathcal S}})\prod_{i\not\in\hat{\mathcal S}}\delta(w_{i}).\label{qwpdfiid}
\end{align}

Plugging the postulated PDF (\ref{postpdf}) in (\ref{DL-L}), one has
\begin{align}\label{L-ws}
&{\mathcal L}(q({\boldsymbol \Omega}|\tilde{\mathbf y});\hat{\mathbf s})={\rm E}_{q({\boldsymbol \Omega}|\tilde{\mathbf y})}\left[\frac{p(\tilde{\mathbf y},{\boldsymbol \Omega};\hat{\mathbf s})}{q({\boldsymbol \Omega}|\tilde{\mathbf y})}\right]\notag\\
=&{\rm E}_{q({\boldsymbol \Omega}|\tilde{\mathbf y})}[\sum_{i=1}^N \ln p(s_i) + \ln p(\mathbf {w|s})+\ln p(\tilde{\mathbf y}|{\boldsymbol \theta},{\mathbf w})-\ln q({\mathbf w}|\tilde{\mathbf y})]+{\rm const}\notag\\
=&-\ln\det({\mathbf J}_{\hat{\mathcal S}}+\frac{1}{\tau}{\mathbf I}_{|\hat{\mathcal S}|})+\mathbf h_{\hat{\mathcal S}}^{\rm H}(\mathbf J_{\hat{\mathcal S}}+\frac{1}{\tau}\mathbf I_{|\hat{\mathcal S}|})^{-1}{\mathbf h}_{\hat{\mathcal S}}+||\hat{\mathbf s}||_0\ln\frac{\rho}{1-\rho}+||\hat{\mathbf s}||_0\ln\frac{1}{\tau}+{\rm const}\notag\\
\triangleq &\ln Z(\mathbf s)|_{{\mathbf s}=\hat{\mathbf s}}
\end{align}
Then we need to find $\hat{\mathbf s}$ which maximizes $\ln Z(\mathbf s)$, i.e.,
\begin{align}\label{maxlns}
\hat{\mathbf s}=\underset{\mathbf s}{\operatorname{argmax}}\ln Z(\mathbf s).
\end{align}
The computation cost of enumerative method to find the globally optimal binary sequence $\mathbf s$ of (\ref{maxlns}) is $O(2^N)$, which is impractical for typical values of $N$. In Appendix \ref{findmaxz}, a greedy iterative search strategy similar to \cite{Badiu} is proposed. Since each step increases the objective function (which is bounded) and $\mathbf s$ can take a finite number of values (at most $2^N$), the method converges in a finite number of steps to some local optimum. In general, numerical experiments show that $O(\hat{K})$ steps is often enough to find the local optimum.

Once $\mathbf s$ is updated as ${\mathbf s}'$, the mean $\hat{\mathbf w}'_{{\mathcal S'}}$ and covariance ${\hat{\mathbf C}}'_{\mathcal S'}$ of the weights should be updated accordingly. For the active case, $\hat{\mathbf w}'_{{\mathcal S'}}$ and covariance ${\hat{\mathbf C}}'_{\mathcal S'}$ are updated according to (\ref{mean_up_act}) and (\ref{cov_up_act}), while for the deactive case, $\hat{\mathbf w}'_{{\mathcal S'}}$ and covariance ${\hat{\mathbf C}}'_{\mathcal S'}$ are updated according to (\ref{mean_up_deact}) and (\ref{cov_up_deact}).
\subsection{Estimating the Model Parameters}\label{estmodelpara}
After updating the frequencies and weights, the model parameters ${\boldsymbol\beta}_w = \{\rho,~\tau\}$ are estimated via maximizing the lower bound ${\mathcal L}(q({\boldsymbol \Omega}|\tilde{\mathbf y});{\boldsymbol \beta}_w)$ for fixed $q({\boldsymbol\Omega}|\tilde{\mathbf y})$. Straightforward calculation shows that
\begin{align}\notag
&{\mathcal L}(q({\boldsymbol \Omega}|\tilde{\mathbf y});{\boldsymbol \beta}_w)= {\rm E}_{q({\boldsymbol \Omega}|\tilde{\mathbf y})}\left[\ln\tfrac{p({\tilde{\mathbf y},{\boldsymbol \Omega};{\boldsymbol \beta}_w})}{q{({\boldsymbol \Omega}|\tilde{\mathbf y})}}\right]\notag\\
=&{\rm E}_{q({\boldsymbol \Omega}|\tilde{\mathbf y})}[\sum_{i=1}^N \ln p(s_i) + \ln p({\mathbf w}|{\mathbf s})]+{\rm const}\notag\\
=&||{\hat{\mathbf s}||_0\ln\rho} + (N - ||\hat{\mathbf s}||_0)\ln(1-\rho) + ||\hat{\mathbf s}||_0\ln\frac{1}{\pi\tau}-{\rm E}_{q({\mathbf w}|\tilde{\mathbf y})}[\frac{1}{\tau}\mathbf w^{\rm H}_{\hat{\mathcal S}}{\mathbf w}_{\hat{\mathcal S}}]  + {\rm const}.
\end{align}
Because
\begin{align}\notag
&{\rm E}_{q({\mathbf w}|\tilde{\mathbf y})}[\mathbf w^{\rm H}_{\hat{\mathcal S}}\mathbf w_{\hat{\mathcal S}})]
={\rm E}_{q({\mathbf w}|\tilde{\mathbf y})}[\sum_{i\in\hat{\mathcal S}} w^*_i w_i] =\hat{\mathbf w}^{\rm H}_{\hat{\mathcal S}}\hat{\mathbf w}_{\hat{\mathcal S}}+{\rm tr}({\hat{\mathbf C}}_{\hat{\mathcal S}})\notag,
\end{align}
we obtain\begin{align}\notag
&{\mathcal L}(q({\boldsymbol \Omega}|\tilde{\mathbf y});{\boldsymbol \beta}_w) = -\frac{1}{\tau}[(\hat{\mathbf w}^{\rm H}_{\hat{\mathcal S}}\hat{\mathbf w}_{\hat{\mathcal S}})+{\rm tr}(\hat{{\mathbf C}}_{\hat{\mathcal S}})] + ||\hat{\mathbf s}||_0(\ln\frac{\rho}{1-\rho}-{\rm ln}\tau)+N\ln(1-\rho)+{\rm const}.\notag
\end{align}
Setting $\frac{\partial\mathcal L}{\partial\rho}=0$, $\frac{\partial\mathcal L}{\partial\tau}=0$, we have
\begin{align}\label{mu-rou-tau-hat}
\hat{\rho} = &\frac{||\hat{\mathbf s}||_0}{N},\notag\\
\hat{\tau} = &\frac{\hat{\mathbf w}^{\rm H}_{\hat{\mathcal S}}\hat{\mathbf w}_{\hat{\mathcal S}}+{\rm tr}(\hat{{\mathbf C}}_{\hat{\mathcal S}})}{||\hat{\mathbf s}||_0}.
\end{align}
\section{VALSE-EP Algorithm}\label{Algorithm}
\begin{figure}[h!t]
\centering
\includegraphics[width=3.4in]{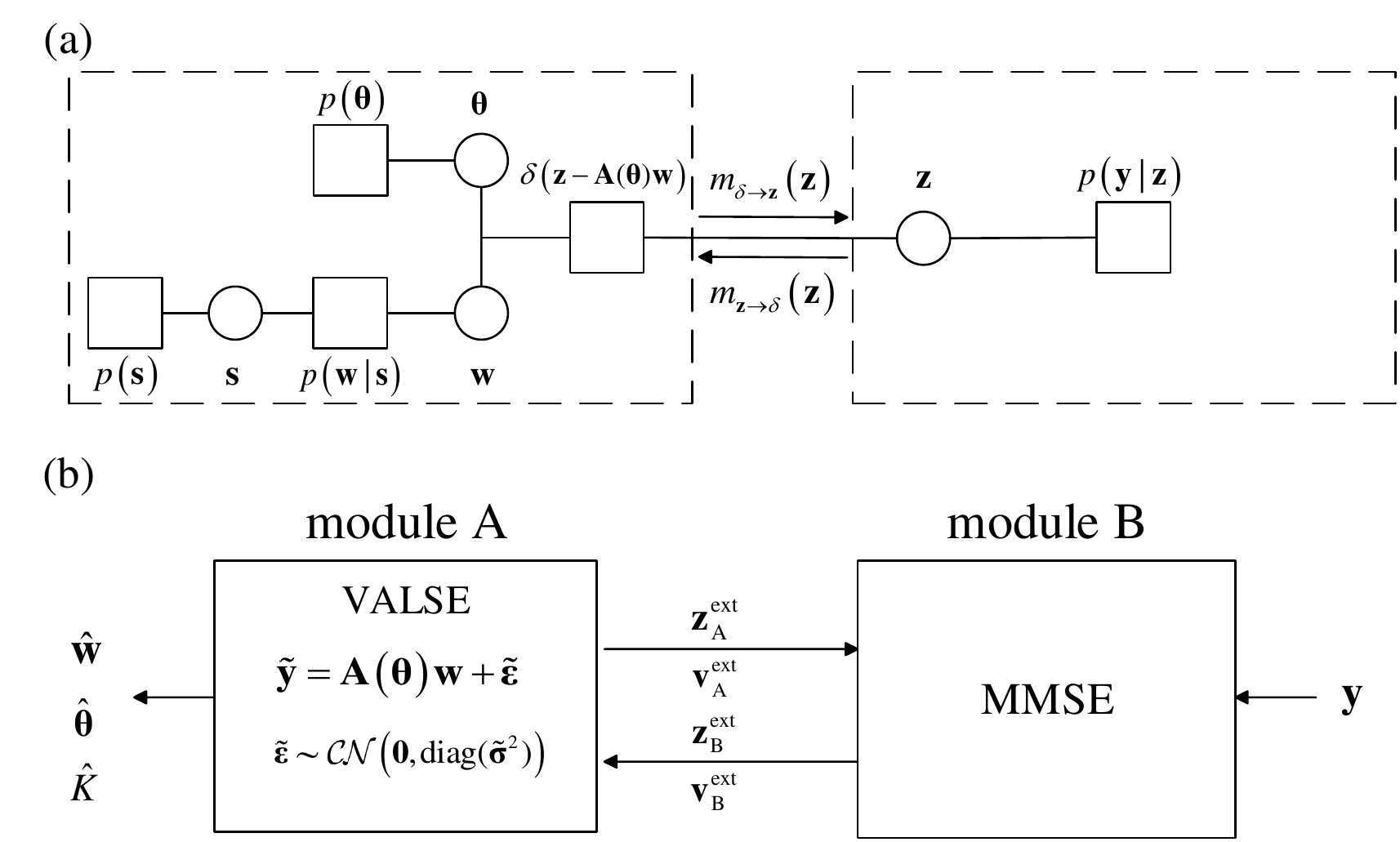}
\caption{Factor graph of the joint PDF (\ref{jointpdf}) and the module of the VALSE-EP algorithm. Here the circle denotes the variable node, and the square denotes the factor node. According to the dashed block diagram in Fig. 1 (a), the problem can be decomposed as two modules in Fig. 1 (b), where module A corresponds to the standard linear model, and module B corresponds to the componentwise MMSE estimation. Intuitively, the problem can be solved by iterating between the two modules, where module A performs a variant of VALSE algorithm, and module B performs the componentwise MMSE estimation.}
\label{FC_fig}                                                                                                                                                                                                                                                       \end{figure}

In this section, the VALSE-EP algorithm is developed based on EP \cite{Minka}. According to EP and the factor graph presented in Fig. \ref{FC_fig}, the quantization model is iteratively reduced to a sequence of pseudo unquantized models \cite{meng1}. As a result, the original quantized LSE problem is decoupled into two modules: the VALSE module named module A and the componentwise MMSE module named module B. The two modules iteratively exchange the extrinsic information and refine their estimates.

Note that the factor graph shown in Fig. \ref{FC_fig} only requires that $p({\mathbf y}|{\mathbf z})=\prod\limits_{n=1}^Np(y_n|z_n)$. As a result, it is believed that VALSE-EP is very general and can have a wider implication to a range of nonlinear identification issues in signal processing, such as phase retrieval ${\mathbf y}=|{\mathbf z}+{\boldsymbol \epsilon}|$ in noncoherent channel estimation \cite{noncoherent}, impulsive noise scenario ${\mathbf y}={\mathbf z}+{\boldsymbol \epsilon}$ where ${\boldsymbol \epsilon}$ is the impulsive noise.

The detailed VALSE-EP is presented as follows.
\subsection{Componentwise MMSE module}
The factor graph and the algorithm module are shown in Fig. \ref{FC_fig}. Specifically, in the $t$th iteration, let ${m}_{\delta\rightarrow {\mathbf z}}^t({\mathbf z})={\mathcal {\mathcal {CN}}({\mathbf z}_{\rm A}^{\rm ext}(t),{\rm diag}({\mathbf v}_{\rm A}^{\rm ext}(t)))}$ denote the message transmitted from the factor node $\delta({\mathbf z}-{\mathbf A}{\mathbf w})$ to the variable node $\mathbf z$, which can be regarded as an extrinsic information from module A. According to EP, the message $m_{{\mathbf z}\rightarrow \delta}^t({\mathbf z})$ transmitted from the variable node $\mathbf z$ to the factor node $\delta({\mathbf z}-{\mathbf A}{\mathbf w})$ can be calculated as \cite{Minka}
\begin{align}\label{extBz}
m_{{\mathbf z}\rightarrow \delta}^t({\mathbf z})\propto \frac{{\rm Proj}[{m}_{\delta\rightarrow {\mathbf z}}^t({\mathbf z})p({\mathbf y}|{\mathbf z})]}{{m}_{\delta\rightarrow {\mathbf z}}^t({\mathbf z})}\triangleq \frac{{\rm Proj}[q_{\rm B}^t({\mathbf z})]}{{m}_{\delta\rightarrow {\mathbf z}}^t({\mathbf z})},
\end{align}
where $p({\mathbf y}|{\mathbf z})=\prod_{n=1}^Np(y_n|z_n)$ and $p(y_n|z_n)$ is (\ref{py_z}), $\propto $ denotes identity up to a normalizing constant. First, the posterior means and variances of $\mathbf z$ in module B can be obtained, i.e.,
\begin{align}
&{\mathbf z}_{\rm B}^{\rm post}(t)={\rm E}[{\mathbf z}|q_{\rm B}^t({\mathbf z})],\label{comb_means}\\
&{\mathbf v}_{\rm B}^{\rm post}(t)={\rm Var}[{\mathbf z}|q_{\rm B}^t({\mathbf z})]\label{comb_vars},
\end{align}
where ${\rm E}[\cdot|q_{\rm B}^t({\mathbf z})]$ and ${\rm Var}[\cdot|q_{\rm B}^t({\mathbf z})]$ are the mean and variance operations taken componentwise with respect to the distribution $\propto q_{\rm B}^t({\mathbf z})$ and closed-form expression exists for quantized measurements \cite{AMPGRSBL}. As a result, ${\rm Proj}[q_{\rm B}^t({\mathbf z})]$ is
\begin{align}\label{postBz}
{\rm Proj}[q_{\rm B}^t({\mathbf z})]={\mathcal {CN}}({\mathbf z};{\mathbf z}_{\rm B}^{\rm post}(t),{\rm diag}({\mathbf v}_{\rm B}^{\rm post}(t))).
\end{align}
Substituting (\ref{postBz}) in (\ref{extBz}), the message $m_{{\mathbf z}\rightarrow \delta}^t({\mathbf z})$ from the variable node $\mathbf z$ to the factor node $\delta({\mathbf z}-{\mathbf A}{\mathbf x})$ is calculated as
\begin{align}\label{extBtoA}
m_{{\mathbf z}\rightarrow \delta}^t({\mathbf z})&\propto \frac{{\mathcal {CN}}({\mathbf z};{\mathbf z}_{\rm B}^{\rm post}(t),{\rm diag( {\mathbf v}_{\rm B}^{\rm post}(t))})}{{\mathcal {CN}}({\mathbf z};{\mathbf z}_{\rm A}^{\rm ext}(t),{\rm diag ({\mathbf v}_{\rm A}^{\rm ext}(t))})}\propto {\mathcal {CN}}({\mathbf z};{\mathbf z}_{\rm B}^{\rm ext}(t),{\rm diag }({\mathbf v}_{\rm B}^{\rm ext}(t))),
\end{align}
which can be viewed as the extrinsic information from module B and ${\mathbf z}_{\rm B}^{\rm ext}(t)$ and ${\mathbf v}_{\rm B}^{\rm ext}(t)$ are \cite{meng1}
\begin{subequations}
\begin{align}
&{\mathbf v}_{\rm B}^{\rm ext}(t)=\left(\frac{1}{{\mathbf v}_{\rm B}^{\rm post}(t)}-\frac{1}{{\mathbf v}_{\rm A}^{\rm ext}(t)}\right)^{-1},\label{extB_var}\\
&{\mathbf z}_{\rm B}^{\rm ext}(t)={\mathbf v}_{\rm B}^{\rm ext}(t)\odot\left(\frac{{\mathbf z}_{\rm B}^{\rm post}(t)}{{\mathbf v}_{\rm B}^{\rm post}(t)}-\frac{{\mathbf z}_{\rm A}^{\rm ext}(t)}{{\mathbf v}_{\rm A}^{\rm ext}(t)}\right),\label{extB_mean}
\end{align}
\end{subequations}
where $\odot$ denotes componentwise multiplication.

In addition, EM algorithm can be incorporated to learn the noise variance $\sigma^2$ \cite{EMSchniter}. The posterior distribution of ${\mathbf z}$ is approximated as (\ref{postBz}). We compute the expected complete log-likelihood function  $\log p({\mathbf y}|{\mathbf z};\sigma^2)+\log {m}_{\delta\rightarrow {\mathbf z}}^t({\mathbf z})$ with respect to $q({\mathbf z}|{\mathbf y};{\sigma}^2(t))$, and drop the irrelevant terms to have
\begin{align}\label{Qfun}
Q(\sigma^2;\sigma^2(t))={\rm E}_{q({\mathbf z}|{\mathbf y};\sigma^2(t))}\left[\log p({\mathbf y}|{\mathbf z};\sigma^2)\right].
\end{align}
Then $\sigma^2$ is updated as
\begin{align}\label{EMuodate}
\sigma^2(t+1)=\underset{\sigma^2}{\operatorname{argmax}}~Q(\sigma^2;\sigma^2(t)).
\end{align}
For the AWGN model
\begin{align}\label{llr}
{\mathbf y}={\mathbf z}+{\boldsymbol \epsilon},
\end{align}
where ${\boldsymbol \epsilon}\sim {\mathcal{CN}}({\boldsymbol \epsilon};{\mathbf 0},\sigma^2{\mathbf I})$, the posterior PDF $q({\mathbf z}|{\mathbf y};\sigma^2(t))$ is
\begin{align}\label{exactpostpdf}
q({\mathbf z}|{\mathbf y};\sigma^2(t))={\mathcal {CN}}({\mathbf z};{\mathbf z}_{\rm B}^{\rm post}(t),{\rm diag}({\mathbf v}_{\rm B}^{\rm post}(t))).
\end{align}
Substituting (\ref{exactpostpdf}), (\ref{llr}) and (\ref{Qfun}) in (\ref{EMuodate}), the noise variance $\sigma^2$ is estimated as
\begin{align}\label{noisevarupln}
\sigma^2(t+1)=\frac{\|{\mathbf y}-{\mathbf z}_{\rm B}^{\rm post}(t)\|^2+{\mathbf 1}^{\rm T}{{\mathbf v}_{\rm B}^{\rm post}(t)}}{M}.
\end{align}
For arbitrary $p({\mathbf y}|{\mathbf z})$ including the quantized case, we also obtain an approximate update equation. From the definition of $m_{{\mathbf z}\rightarrow \delta}({\mathbf z})$ and ${\mathbf z}={\mathbf A}({\boldsymbol \theta}){\mathbf x}$, we obtain a pseudo measurement model
\begin{align}
\tilde{\mathbf y}(t)={\mathbf z}+\tilde{\boldsymbol \epsilon}(t),
\end{align}
where $\tilde{\mathbf y}(t)={\mathbf z}_{\rm B}^{\rm ext}(t)$, $\tilde{\boldsymbol \epsilon}(t)\sim {\mathcal{CN}}({\boldsymbol \epsilon}(t);{\mathbf 0},{\rm diag}(\tilde{\boldsymbol \sigma}^2(t)))$ and $\tilde{\boldsymbol \sigma}^2(t)={\mathbf v}_{\rm B}^{\rm ext}(t)$. The noise variance $\sigma^2$ is updated as \cite{Accessmeng}
\begin{align}\label{noisevarupnln}
\sigma^2(t+1)=\frac{\|\tilde{\mathbf y}(t)-{\mathbf z}_{\rm B}^{\rm post}(t)\|^2+{\mathbf 1}^{\rm T}{{\mathbf v}_{\rm B}^{\rm post}(t)}}{M}.
\end{align}
Note that (\ref{noisevarupln}) is a special case of (\ref{noisevarupnln}), as for the AWGN model (\ref{llr}), $\tilde{\mathbf y}={\mathbf y}$ is proved later according to (\ref{zBy}) and (\ref{tildeyzB}).
\subsection{VALSE module}
\begin{figure}[h!t]
\centering
\includegraphics[width=3.4in]{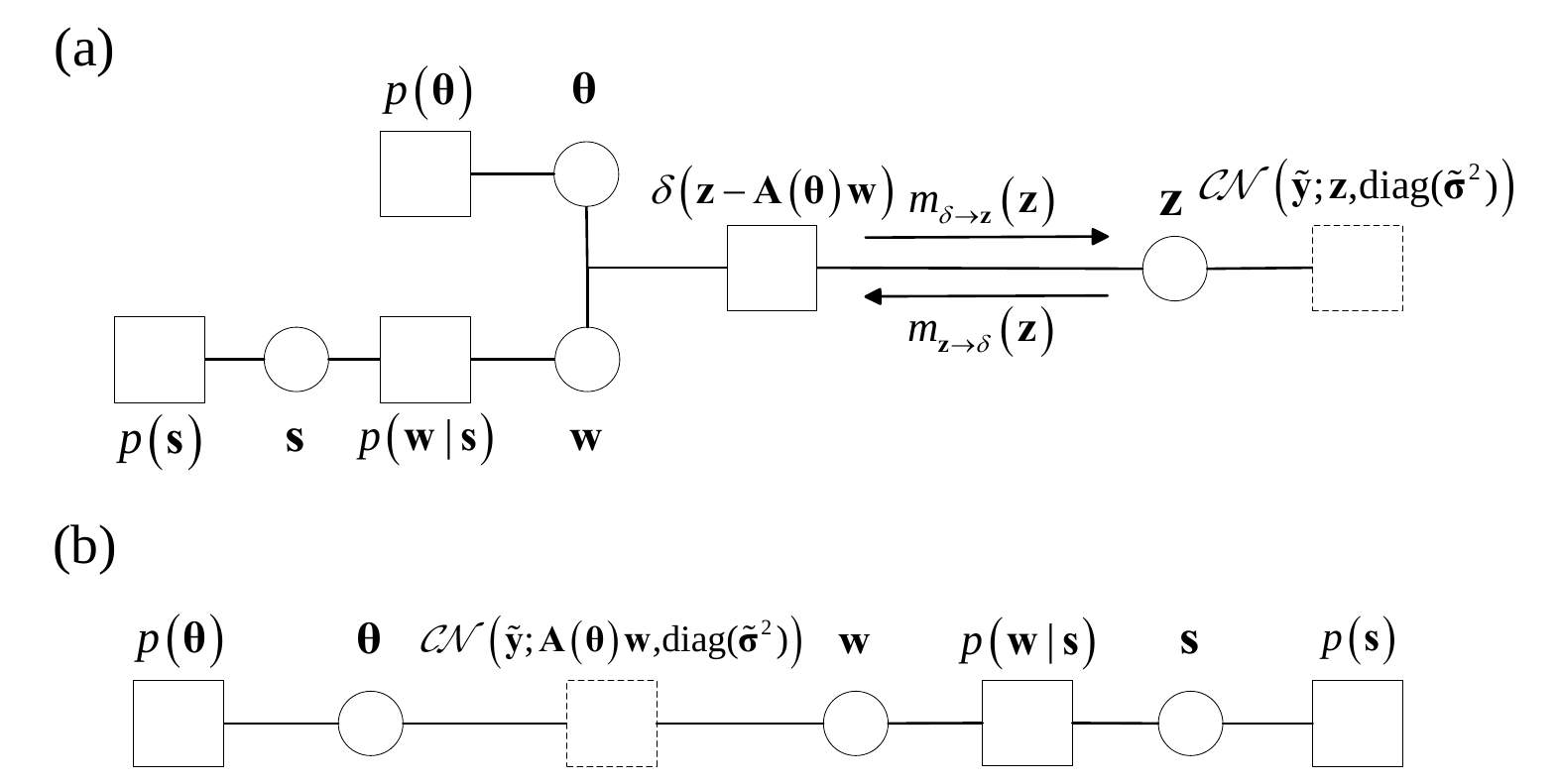}
\caption{Two equivalent factor graphs of the joint PDF (\ref{jointpdf2}). The dashed square denotes the pseudo factor node.}
\label{FC2}                                                                                                                                                                                                                                                       \end{figure}

According to (\ref{extBtoA}), the message $m_{{\mathbf z}\rightarrow \delta}^t({\mathbf z})$ (nonGaussian likelihood) transmitted from the variable node $\mathbf z$ to the factor node $\delta({\mathbf z}-{\mathbf A}{\mathbf x})$ is iteratively approximated as a Gaussian distribution (Gaussian likelihood), and the factor graph is shown in Fig. \ref{FC2} (a). Based on the definition of the factor node $\delta({\mathbf z}-{\mathbf A}{\mathbf x})$ and the message $m_{{\mathbf z}\rightarrow \delta}^t({\mathbf z})$ transmitted from the variable node $\mathbf z$ to the factor node $\delta({\mathbf z}-{\mathbf A}{\mathbf w})$, a pseudo linear measurement model
\begin{align}\label{heomodel}
\tilde{\mathbf y}(t+1)={\mathbf A}({\boldsymbol \theta}){\mathbf w}+\tilde{\boldsymbol \epsilon}(t+1),
\end{align}
is obtained, where $\tilde{\boldsymbol \epsilon}(t+1)\sim {\mathcal {CN}}(\tilde{\boldsymbol \epsilon}(t+1);{\mathbf 0},{\rm diag}(\tilde{\boldsymbol \sigma}^2(t+1)))$, and
\begin{align}
\tilde{\mathbf y}(t+1)&={\mathbf z}_{\rm B}^{\rm ext}(t),\\
\tilde{\boldsymbol \sigma}^2(t+1)&={\mathbf v}_{\rm B}^{\rm ext}(t).
\end{align}
As a result, the pseudo factor graph Fig. \ref{FC2} (b) is obtained and is equivalent to Fig. \ref{FCsub0}. In Section \ref{VALSEHN}, a variant of VALSE is rederived. Here we run the VALSE in a single iteration and output the approximated posterior PDF $q({\mathbf w}_{\hat{S}}|\tilde{\mathbf y})$ and $q({\boldsymbol \theta}|\tilde{\mathbf y})$.

\subsection{From VALSE module to MMSE module}
According to the approximated posterior PDF $q({\mathbf w}_{\hat{S}}|\tilde{\mathbf y})$ and $q({\boldsymbol \theta}|\tilde{\mathbf y})$, we calculate the message ${m}_{\delta\rightarrow {\mathbf z}}^{t+1}({\mathbf z})$ as
\begin{align}\label{extA}
&{m}_{\delta\rightarrow {\mathbf z}}^{t+1}({\mathbf z})\propto\frac{{\rm Proj}[\int q({\mathbf w}_{\hat{S}}|\tilde{\mathbf y})\delta({\mathbf z}-{\mathbf A}_{\hat{S}}({\boldsymbol \theta}){\mathbf w}_{\hat{S}})q({\boldsymbol \theta}|\tilde{\mathbf y}){\rm d}{\mathbf w}_{\hat{S}}{\rm d}{\boldsymbol \theta}]}{ m_{{\mathbf z}\rightarrow \delta}^t({\mathbf z})}\notag\\
&\triangleq \frac{{\rm Proj}[q_{\rm A}^{t+1}(\mathbf z)]}{m_{{\mathbf z}\rightarrow \delta}^t({\mathbf z})}.
\end{align}
%\begin{align}
%{m}_{\delta\rightarrow {\mathbf z}}^{t+1}({\mathbf z})\propto\frac{{\rm Proj}[\int q({\mathbf w},{\mathbf s}|\tilde{\mathbf y})\delta({\mathbf z}-{\mathbf A}({\boldsymbol \theta}){\mathbf w})q({\boldsymbol \theta}|\tilde{\mathbf y}){\rm d}{\mathbf w}{\rm d}{\mathbf s}{\rm d}{\boldsymbol \theta}]}{ m_{{\mathbf z}\rightarrow \delta}^t({\mathbf z})}
%\end{align}
According to (\ref{extA}), the posterior means and variances of $\mathbf z$ averaged over $q_{\rm A}^{t+1}(\mathbf z)$ in module A are
\begin{align}
&{\mathbf z}_{\rm A}^{\rm post}(t+1)=\hat{\mathbf A}_{\hat{S}}\hat{\mathbf w}_{\hat{S}},\label{post_means_A}\\
&{\mathbf v}_{\rm A}^{\rm post}(t+1)= {\rm diag}(\hat{\mathbf A}_{\hat{S}}\hat{\mathbf C}_{\hat{S}}\hat{\mathbf A}_{\hat{S}}^{\rm H})+\left(\hat{\mathbf w}_{\hat{S}}^{\rm H}\hat{\mathbf w}_{\hat{S}}{\mathbf 1}_{N}-|\hat{\mathbf A}_{\hat{S}}|^2|\hat{\mathbf w}_{\hat{S}}|^2\right)\notag\\
&+\left[{\rm tr}(\hat{\mathbf C}_{\hat{S}}){\mathbf 1}_N-|\hat{\mathbf A}_{\hat{S}}|^2{\rm diag}(\hat{\mathbf C}_{\hat{S}})\right],\label{post_vars_A}
\end{align}
where details of (\ref{post_vars_A}) are postponed to Appendix \ref{calpostcov}.
Thus ${\rm Proj}[q_{\rm A}^{t+1}(\mathbf z)]$ is
\begin{align}
{\rm Proj}[q_{\rm A}^{t+1}(\mathbf z)]={\mathcal {CN}}({\mathbf z};{\mathbf z}_{\rm A}^{\rm post}(t+1),{\rm diag}({\mathbf v}_{\rm A}^{\rm post}(t+1))).
\end{align}
According to (\ref{extA}), ${m}_{\delta\rightarrow {\mathbf z}}^{t+1}({\mathbf z})$ is calculated to be
\begin{align}
{m}_{\delta\rightarrow {\mathbf z}}^{t+1}({\mathbf z})={\mathcal {CN}}({\mathbf z};{\mathbf z}_{\rm A}^{\rm ext}(t+1),{\rm diag}({\mathbf v}_{\rm A}^{\rm ext}(t+1))),
\end{align}
where the extrinsic means ${\mathbf z}_{\rm A}^{\rm ext}(t+1)$ and variances ${\mathbf v}_{\rm A}^{\rm ext}(t+1)$ are given by \cite{meng1}
\begin{align}
&\frac{1}{{\mathbf v}_{\rm A}^{\rm ext}(t+1)}= \frac{1}{{\mathbf v}_{\rm A}^{\rm post}(t+1)}-\frac{1}{\tilde{\boldsymbol \sigma}_w^2(t+1)},\label{extvarA}\\
&{\mathbf z}_{\rm A}^{\rm ext}(t+1)={{\mathbf v}_{\rm A}^{\rm ext}(t+1)}\odot\left(\frac{{\mathbf z}_{\rm A}^{\rm post}(t+1)}{{\mathbf v}_{\rm A}^{\rm post}(t+1)}-\frac{\tilde{\mathbf y}(t+1)}{\tilde{\boldsymbol \sigma}_w^2(t+1)}\right),\label{extmeanA}
\end{align}
and we input them to module B. The algorithm iterates until convergence or the maximum number of iterations is reached. The VALSE-EP algorithm is summarized as Algorithm \ref{VALSE_ep}.
\begin{algorithm}[h]
\caption{VALSE-EP algorithm}\label{VALSE_ep}
\begin{algorithmic}[1]
\STATE Set the number of iterations $T$ and implement the initialization in Subsection \ref{initsubsec}.\
\FOR {$t=1,\cdots,T$ }
\STATE Update $\mathbf J$ (\ref{J1}) and $\mathbf h$ (\ref{H1}) using $\tilde{\mathbf y}(t)$ and $\tilde{\boldsymbol \sigma}^2(t)$.
\STATE Update~$\hat{\mathbf s},\hat{\mathbf w}_{\hat{\mathcal S}}~{\rm and}~\hat{\mathbf C}_{\hat{\mathcal S}}$ ({\rm Section} \ref{W-and-C}).
\STATE Update~$\hat{\rho}$, $\hat{\tau}$ (\ref{mu-rou-tau-hat}) ({\rm Section} \ref{estmodelpara}).
\STATE Update~$\boldsymbol\eta_i$~and~$\hat{\mathbf a}_i$ for all $i\in \hat{\mathcal S}$ ({\rm Section} \ref{yita}).
\STATE Calculate the posterior means ${\mathbf z}_{\rm A}^{\rm post}(t)$ (\ref{post_means_A}) and variances ${\mathbf v}_{\rm A}^{\rm post}(t)$ (\ref{post_vars_A}).
\STATE Compute the extrinsic mean and variance of $\mathbf z$ as ${\mathbf v}_{\rm A}^{\rm ext}(t)$ (\ref{extvarA}), ${\mathbf z}_{\rm A}^{\rm ext}(t)$ (\ref{extmeanA}).
\STATE Compute the post mean and variance of $\mathbf z$ as ${\mathbf z}_{\rm B}^{\rm post}(t)$ (\ref{comb_means}), ${\mathbf v}_{\rm B}^{\rm post}(t)$ (\ref{comb_vars}).
\STATE Compute the extrinsic mean and variance of $\mathbf z$ as ${\mathbf z}_{\rm B}^{\rm ext}(t)$ (\ref{extB_mean}) and ${\mathbf v}_{\rm B}^{\rm ext}(t)$ (\ref{extB_var}), and set $\tilde{\boldsymbol \sigma}^2(t+1) ={\mathbf v}_{\rm B}^{\rm ext}(t)$ and $\tilde{\mathbf y}(t+1)={\mathbf z}_{\rm B}^{\rm ext}(t)$. Implement EM to estimate the noise variance as $\sigma^2(t+1)$ (\ref{noisevarupnln}).
%run the VALSE algorithm \ref{VALSE} until the stopping criterion is satisfied. Otherwise, run the VALSE algorithm \ref{VALSE} directly with initialization provided by the previous results of the VALSE.
%\STATE Calculate the posterior means ${\mathbf z}_{\rm A}^{\rm post}(t)$ (\ref{post_means_A}) and variances ${\mathbf v}_{\rm A}^{\rm post}(t)$ (\ref{post_vars_A}).
%\STATE Compute the extrinsic mean and variance of $\mathbf z$ as ${\mathbf v}_{\rm A}^{\rm ext}(t+1)$ (\ref{extvarA}), ${\mathbf z}_{\rm A}^{\rm ext}(t+1)$ (\ref{extmeanA}).\
\ENDFOR
\STATE Return $\hat{\boldsymbol \theta}$, $\hat{\mathbf w}$, $\hat{\mathbf z}$ and $\hat{K}$.
\end{algorithmic}
\end{algorithm}

%\red{
%\begin{remark}
%It is worth noting that Step $3$ is very important for the VALSE-EP algorithm. The model order is mainly determined by both $\mathbf J$ (\ref{J1}) and $\mathbf h$ (\ref{H1}). Once $\tilde{\boldsymbol \sigma}^2$ and $\tilde{\mathbf y}$ are updated, $\mathbf J$ (\ref{J1}) and $\mathbf h$ (\ref{H1}) must be reupdated.
%\end{remark}}
%\subsection{Implementation Details}
%To implement the VALSE algorithm, several strategies are proposed to improve the robustness of the proposed algorithm:
%\begin{itemize}
%  \item The extrinsic variances $v_{\rm A}^{\rm ext}$ and $v_{\rm B}^{\rm ext}$ are restricted to the interval $[v_{\rm min},v_{\rm max}]$. Here $v_{\rm min}=10^{-8}$ and $v_{\rm max}=10^8$.
%  \item For the first iteration, the VALSE algorithm is initialized the same as \cite{Badiu}. After the first iteration, warm sart is adopted and the VALSE is initialized with the previous results. In details, ${\mathbf J}$, $\mathbf h$, $\rho$, $\tau$, $\mathbf s$, $\mathbf A$, $\mathbf C$, $\mathbf w$ are initialized with the previous results.
%\end{itemize}
\subsection{Initialization}\label{initsubsec}
The initialization of VALSE-EP is presented. First, the additive quantization noise model (AQNM) \cite{CSquant}
\begin{align}\label{AQNM}
{\mathbf y}_{\rm q}={\mathbf z}+{\boldsymbol \epsilon}+{\boldsymbol \epsilon}_{\rm q},
\end{align}
is adopted, where the output levels ${\mathbf y}_{\rm q}$ are the midpoints of the quantization interval, ${\boldsymbol \epsilon}_{\rm q}$ is the quantization error. First, the noise variance $\sigma^2$ is initialized. For $n^{'}=1,\cdots,N-1$, we calculate $\gamma_{n^{'}}=\frac{1}{N}\sum_{(k^{'},l^{'})\in \Upsilon_{n^{'}}}y_{k^{'}}y_{l^{'}}^*$ with $\Upsilon_{n^{'}}=\{(k^{'},l^{'})|1\leq k^{'},l^{'}\leq N-1,k^{'}-l^{'}=n^{'}\}$. We use ${\boldsymbol \gamma}$ to build a Toeplitz estimate of ${\rm E}[{\mathbf y}_{\rm q}{\mathbf y}_{\rm q}^{\rm H}]$. Then, we initialize $\sigma^2$ with the average of the lower quarter of the eigenvalues of the Toeplitz matrix (ignoring the quantization noise). Typically, it is found that setting ${\mathbf v}_{\rm A}^{\rm ext}$ ranging from $1\sim10^2$ works well. Here we initialize ${\mathbf z}_{\rm A}^{\rm ext}={\mathbf 0}$ and ${\mathbf v}_{\rm A}^{\rm ext}={\mathbf {10}}$. After performing the MMSE estimation in module B, $\tilde{\mathbf y}={\mathbf z}_{\rm B}^{\rm ext}$ and $\tilde{\boldsymbol \sigma}^2={\mathbf v}_{\rm B}^{\rm ext}$ are obtained. Given that ${\rm E}[\tilde{\mathbf y}{\rm diag}(\tilde{\boldsymbol \sigma}^{-2})\tilde{\mathbf y}^{\rm H}]=N+\rho N \tau{\rm tr}({\rm diag}(\tilde{\boldsymbol \sigma}^{-2}))$, we set $\hat{\rho}=0.5$ and let $\hat{\tau}=(\tilde{\mathbf y}{\rm diag}(\tilde{\boldsymbol \sigma}^{-2})\tilde{\mathbf y}^{\rm H}-N)/(\hat{\rho}N{\rm tr}({\rm diag}(\tilde{\boldsymbol \sigma}^{-2})))$. Then in step $i$, when the first $i-1$ PDFs of the frequencies are initialized, the estimates $\{\hat{w}\}_{k=1}^{i-1}$ and the residual ${\mathbf y}_{i-1}^{\rm r}=\tilde{\mathbf y}-\sum_{k=1}^{i-1}\hat{\mathbf a}_k\hat{w}_k$ are obtained. Then initialize $q(\theta_i|{\mathbf y}_{i-1}^{\rm r})\propto {\rm exp}\left(|{\mathbf y}_{i-1}^{\rm r}{\rm diag}(\tilde{\boldsymbol \sigma}^{-2}){\mathbf a}(\theta)|^2/N\right)$ and project it as a von Mises distribution and compute $\hat{\mathbf a}_i$.
\subsection{Computation Complexity}
From Algorithm \ref{VALSE_ep}, it can be seen that VALSE-EP involves the componentwise MMSE operation and VALSE. For the componentwise MMSE operation, the computation complexity is $O(N\hat{K})$. As for the VALSE, the complexity per iteration is dominated by the two steps: the maximization of $\ln Z(\mathbf s)$ and the approximations of the posterior PDF $q(\theta_i|{\mathbf y})$ by mixtures of von Mises PDFs. This work approximates the posterior PDF $q(\theta_i|{\mathbf y})$ by Heuristic $2$ in \cite[Algorithm 2]{Badiu}. Thus the complexity of the two steps are $O(N\hat{K}^3)$ and $O(N^2)$ \cite{Badiu}. Therefore, the complexity of VALSE-EP is comparable to that of VALSE.
\section{Relationship between VALSE-EP and VALSE under Unquantized Setting}\label{RelVALSEEP}
Before revealing the relation between VALSE-EP and VALSE under unquantized setting, the following properties are revealed.

\begin{property}\label{property1}
The noise variance estimate in VALSE can be equivalently derived in EM step.
\end{property}
\begin{proof}
In \cite{Badiu}, the noise variance is estimated as (reformulated as our notation)
\begin{align}
\sigma_{\rm VA}^2(t+1)=&\frac{1}{N}\|{\mathbf y}-\sum\limits_{i\in \hat{\mathcal S}}\hat{\mathbf a}_i\hat{w}_i\|_2^2+\frac{1}{N}{\rm tr}\left({\mathbf J}_{\hat{\mathcal S}}^{'}\hat{\mathbf C}_{\hat{\mathcal S}}\right)+\sum\limits_{i\in \hat{\mathcal S}}|\hat{w}_i|^2(1-\|\hat{\mathbf a}_i\|_2^2/N),
\end{align}
where ${\mathbf J}_{ii}^{'}=N$ and ${\mathbf J}_{ij}^{'}=\hat{\mathbf a}_i^{\rm H}\hat{\mathbf a}_j$. According to EM, the noise variance $\sigma^2(t+1)$ in module A should be updated as (replacing ${\mathbf z}_{\rm B}^{\rm post}(t)$ and ${\mathbf v}_{\rm B}^{\rm post}(t)$ with ${\mathbf z}_{\rm A}^{\rm post}(t)$ and ${\mathbf v}_{\rm A}^{\rm post}(t)$ in (\ref{noisevarupln}), respectively)
\begin{align}\label{noisevarupln0}
\sigma^2(t+1)=\frac{\|{\mathbf y}-{\mathbf z}_{\rm A}^{\rm post}(t)\|^2+{\mathbf 1}^{\rm T}{{\mathbf v}_{\rm A}^{\rm post}(t)}}{N}.
\end{align}
Substituting ${\mathbf z}_{\rm A}^{\rm post}(t)$ (\ref{post_means_A}) and ${\mathbf v}_{\rm A}^{\rm post}(t)$ (\ref{post_vars_A}) in (\ref{noisevarupln0}) and utilizing ${\mathbf J}^{'}=\hat{\mathbf A}^{\rm H}\hat{\mathbf A}+N{\mathbf I}_N-{\rm diag}({\rm diag}(\hat{\mathbf A}^{\rm H}\hat{\mathbf A}))$, one can show that $\sigma^2(t+1)=\sigma_{\rm VA}^2(t+1)$.
\end{proof}

Property \ref{property1} reveals that the noise variance estimates of both VALSE-EP and VALSE can be derived via the EM step. For the VALSE algorithm, the noise variance estimate of the EM step is performed in module A, while for VALSE-EP, the noise variance estimate of the EM step is performed in module B. In the following text, the relationships between the posterior means and variances of VALSE and that of VALSE-EP are derived.
\begin{property}\label{property2}
The relationships between the posterior means ${\mathbf z}_{\rm B}^{\rm post}(t)$ and variances ${\mathbf v}_{\rm B}^{\rm post}(t)$ of $\mathbf z$ in module B and the posterior means ${\mathbf z}_{\rm A}^{\rm post}(t)$ and variances ${\mathbf v}_{\rm A}^{\rm post}(t)$ of $\mathbf z$ in module A are
\begin{subequations}\label{resha}
\begin{align}
\frac{1}{{\mathbf v}_{\rm B}^{\rm post}(t+1)}&=\frac{1}{{\mathbf v}_{\rm A}^{\rm post}(t+1)}-\frac{1}{\sigma^2(t)}+\frac{1}{\sigma^2(t+1)},\\
\frac{{\mathbf z}_{\rm B}^{\rm post}(t+1)}{{\mathbf v}_{\rm B}^{\rm post}(t+1)}&=\frac{{\mathbf z}_{\rm A}^{\rm post}(t+1)}{{\mathbf v}_{\rm A}^{\rm post}(t+1)}-{\mathbf y}\left(\frac{1}{\sigma^2(t)}-\frac{1}{\sigma^2(t+1)}\right).
\end{align}
\end{subequations}
\end{property}
\begin{proof}
For the $t$th iteration, let the extrinsic means and variances of $\mathbf z$ from module A be ${\mathbf z}_{\rm A}^{\rm ext}(t)$ and ${\mathbf v}_{\rm A}^{\rm ext}(t)$ (Step 8 in Algorithm \ref{VALSE_ep}), and the noise variance be $\sigma^2(t)$. The posterior variances (\ref{comb_vars}) and means (\ref{comb_means}) of $\mathbf z$ are
\begin{subequations}\label{postplusextB}
\begin{align}
\frac{1}{{\mathbf v}_{\rm B}^{\rm post}(t)}=\frac{1}{{\mathbf v}_{\rm A}^{\rm ext}(t)}+\frac{1}{\sigma^2(t)},\\
\frac{{\mathbf z}_{\rm B}^{\rm post}(t)}{{\mathbf v}_{\rm B}^{\rm post}(t)}=\frac{{\mathbf z}_{\rm A}^{\rm ext}(t)}{{\mathbf v}_{\rm A}^{\rm ext}(t)}+\frac{\mathbf y}{\sigma^2(t)}.
\end{align}
\end{subequations}
Then the extrinsic means and variances of $\mathbf z$ from module B ${\mathbf z}_{\rm B}^{\rm ext}(t)$ and ${\mathbf v}_{\rm B}^{\rm ext}(t)$ are
\begin{subequations}\label{extplusextB}
\begin{align}
\frac{1}{{\mathbf v}_{\rm B}^{\rm ext}(t)}=\frac{1}{{\mathbf v}_{\rm B}^{\rm post}(t)}-\frac{1}{{\mathbf v}_{\rm A}^{\rm ext}(t)},\\
\frac{{\mathbf z}_{\rm B}^{\rm ext}(t)}{{\mathbf v}_{\rm B}^{\rm ext}(t)}=\frac{{\mathbf z}_{\rm B}^{\rm post}(t)}{{\mathbf v}_{\rm B}^{\rm post}(t)}-\frac{{\mathbf z}_{\rm A}^{\rm ext}(t)}{{\mathbf v}_{\rm A}^{\rm ext}(t)}.
\end{align}
\end{subequations}
According to (\ref{postplusextB}) and (\ref{extplusextB}), one has
\begin{subequations}\label{zBy2}
\begin{align}
{\mathbf z}_{\rm B}^{\rm ext}(t)&={\mathbf y},\label{zBy}\\
{\mathbf v}_{\rm B}^{\rm ext}(t)&=\sigma^2(t){\mathbf 1}.
\end{align}
\end{subequations}
In addition, EM is implemented to update $\sigma^2$ as $\sigma^2(t+1)$. By setting
\begin{subequations}\label{tildeyzBall}
\begin{align}
\tilde{\mathbf y}(t+1)&={\mathbf z}_{\rm B}^{\rm ext}(t),\label{tildeyzB}\\
\tilde{\boldsymbol \sigma}^2(t+1)&={\mathbf v}_{\rm B}^{\rm ext}(t)=\sigma^2(t){\mathbf 1},
\end{align}
\end{subequations}
we run the VALSE (noise variance aware) algorithm, and calculate ${\mathbf z}_{\rm A}^{\rm post}(t+1)$ and ${\mathbf v}_{\rm A}^{\rm post}(t+1)$. Then ${\mathbf z}_{\rm A}^{\rm ext}(t+1)$ and ${\mathbf v}_{\rm A}^{\rm ext}(t+1)$ are updated as
\begin{subequations}\label{updatevAext}
\begin{align}
\frac{1}{{\mathbf v}_{\rm A}^{\rm ext}(t+1)}=\frac{1}{{\mathbf v}_{\rm A}^{\rm post}(t+1)}-\frac{1}{\tilde{\boldsymbol \sigma}^2(t+1)},\\
\frac{{\mathbf z}_{\rm A}^{\rm ext}(t+1)}{{\mathbf v}_{\rm A}^{\rm ext}(t+1)}=\frac{{\mathbf z}_{\rm A}^{\rm post}(t+1)}{{\mathbf v}_{\rm A}^{\rm post}(t+1)}-\frac{\tilde{\mathbf y}(t+1)}{\tilde{\boldsymbol \sigma}^2(t+1)}.
\end{align}
\end{subequations}
According to (\ref{postplusextB}), (\ref{zBy2}), (\ref{tildeyzBall}) and (\ref{updatevAext}), (\ref{resha}) is obtained.
\end{proof}

According to Property \ref{property2}, ${\mathbf v}_{\rm B}^{\rm post}(t+1)={\mathbf v}_{\rm A}^{\rm post}(t+1)$ and ${\mathbf z}_{\rm B}^{\rm post}(t+1)={\mathbf z}_{\rm A}^{\rm post}(t+1)$ holds when $\sigma^2(t)=\sigma^2(t+1)$, which means that the noise variance is a constant during the iteration. Combing Property \ref{property1} and Property \ref{property2}, it is concluded that in the unquantized setting VALSE-EP is equivalent to VALSE when the noise variance is known, i.e., the noise variance estimation step is removed. In general, VALSE-EP is not exactly equivalent to VALSE even in the unquantized setting.

\section{Numerical Simulation}\label{NS}
In this section, numerical experiments are conducted to evaluate the performance of the proposed VALSE-EP algorithm. In addition, for performance comparison in the quantized setting, the AQNM is adopted and ${\mathbf y}_{\rm q}$ (\ref{AQNM}) is directly input to the VALSE to perform estimation. Furthermore, VALSE-EP is also compared with VALSE in unquantized setting.
\subsection{Simulation Setup}
The frequencies are randomly drawn such that the minimum wrap around distance is greater than $2\pi/N$. The noninformative prior, i.e., $p(\theta_i)=1/(2\pi), ~i=1,\cdots,N$, is used for both VALSE and VALSE-EP algorithm. The magnitudes of the weight coefficients are drawn i.i.d. from a Gaussian distribution ${\mathcal N}(1,0.04)$, and the phases are drawn i.i.d. from a uniform distribution between $[-\pi,\pi]$. For multi-bit quantization, a uniform quantizer is adopted and the quantization interval is restricted to $[-3\sigma_z/\sqrt{2},3\sigma_z/\sqrt{2}]$, where $\sigma_z^2$ is the variance of ${\mathbf z}$. In our setting, it can be calculated that $\sigma_z^2\approx K$.
For one-bit quantization, zero is chosen as the threshold. The SNR is defined as ${\rm SNR} = 20{\rm log}(||\mathbf A(\boldsymbol\theta){\mathbf w}||_{2}/||\boldsymbol \epsilon||_{\rm 2})$. Both VALSE and VALSE-EP stop at iteration $t$ if $\|\hat{\mathbf z}(t)-\hat{\mathbf z}(t-1)\|_2/\|\hat{\mathbf z}(t)\|_2<10^{-6}$ or the number of iterations exceeds $1000$.

The signal estimation error ${\rm NMSE}(\hat{\mathbf z})$, the frequency estimation error ${\rm MSE}(\hat{\boldsymbol \theta})$ and the correct model order estimation probability ${\rm P}(\hat{K}=K)$ are used to characterize the performance. The normalized MSE (NMSE) of signal $\hat{\mathbf z}$ (for unquantized and multi-bit quantized system) and MSE of $\hat{\boldsymbol\theta}$ are defined as ${\rm NMSE}(\hat{\mathbf z})\triangleq 10{\rm log}(||\hat{\mathbf z} - {\mathbf z}||_2^2/||{\mathbf z}||_2^2)$ and ${\rm MSE}(\hat{\boldsymbol \theta})\triangleq 10{\rm log}(||\hat{\boldsymbol\theta} - \boldsymbol\theta||_2^2)$, respectively. Please note that, due to magnitude ambiguity, it is impossible to recover the exact magnitude of $\tilde{w}_k$ from one-bit measurements in the noiseless scenario.  Thus for one-bit quantization, the debiased NMSE of the signal defined as ${\rm dNMSE}(\hat{\mathbf z})\triangleq\underset{ c}{\operatorname{min}}~10\log ({\|{\mathbf z}^*-c\hat{\mathbf z}\|_2^2}/{\|{\mathbf z}^*\|_2^2})$ is calculated. As for the frequency estimation error, we average only the trials in which all those algorithms estimate the correct model order. All the results are averaged over $500$ Monte Carlo (MC) trials unless stated otherwise. The MSE of the frequency estimation is calculated only when the model order is correctly estimated.

\begin{figure*}
  \centering
  \subfigure[]{
    \label{Rec10dB} %% label for first subfigure
    \includegraphics[width=58mm]{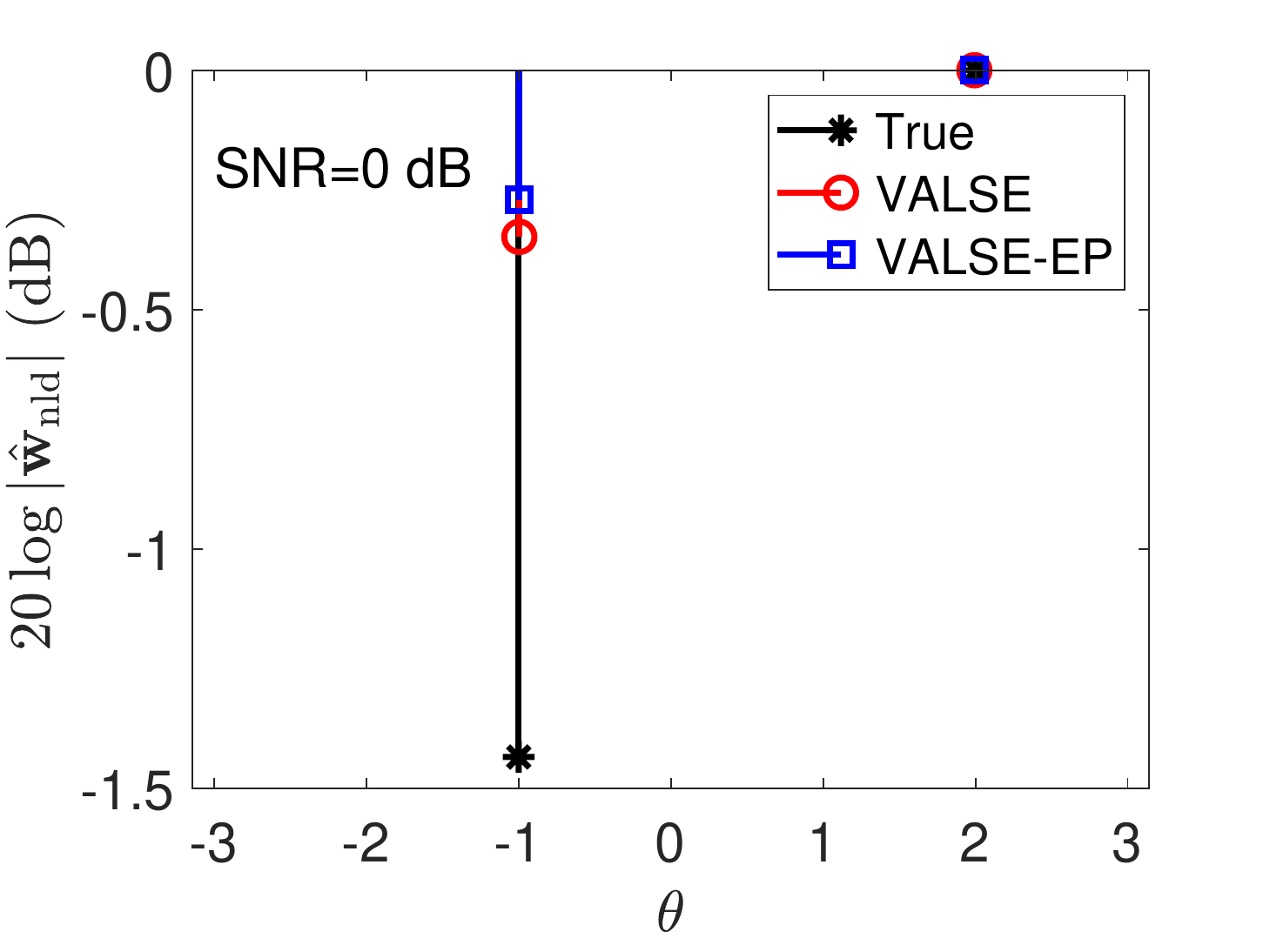}}
  \subfigure[]{
    \label{Rec20dB} %% label for first subfigure
    \includegraphics[width=58mm]{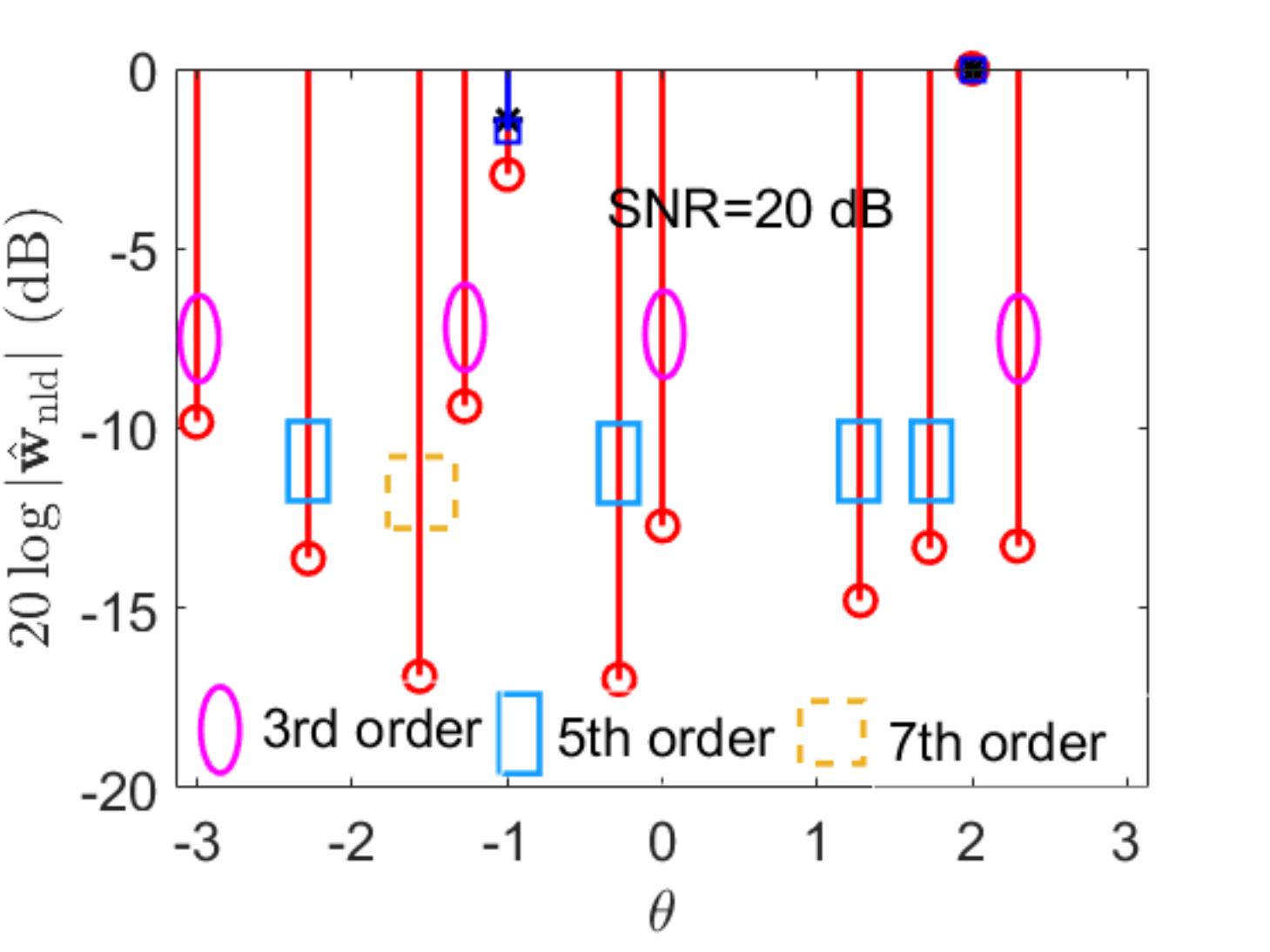}}
  \subfigure[]{
    \label{Rec40dB} %% label for second subfigure
    \includegraphics[width=58mm]{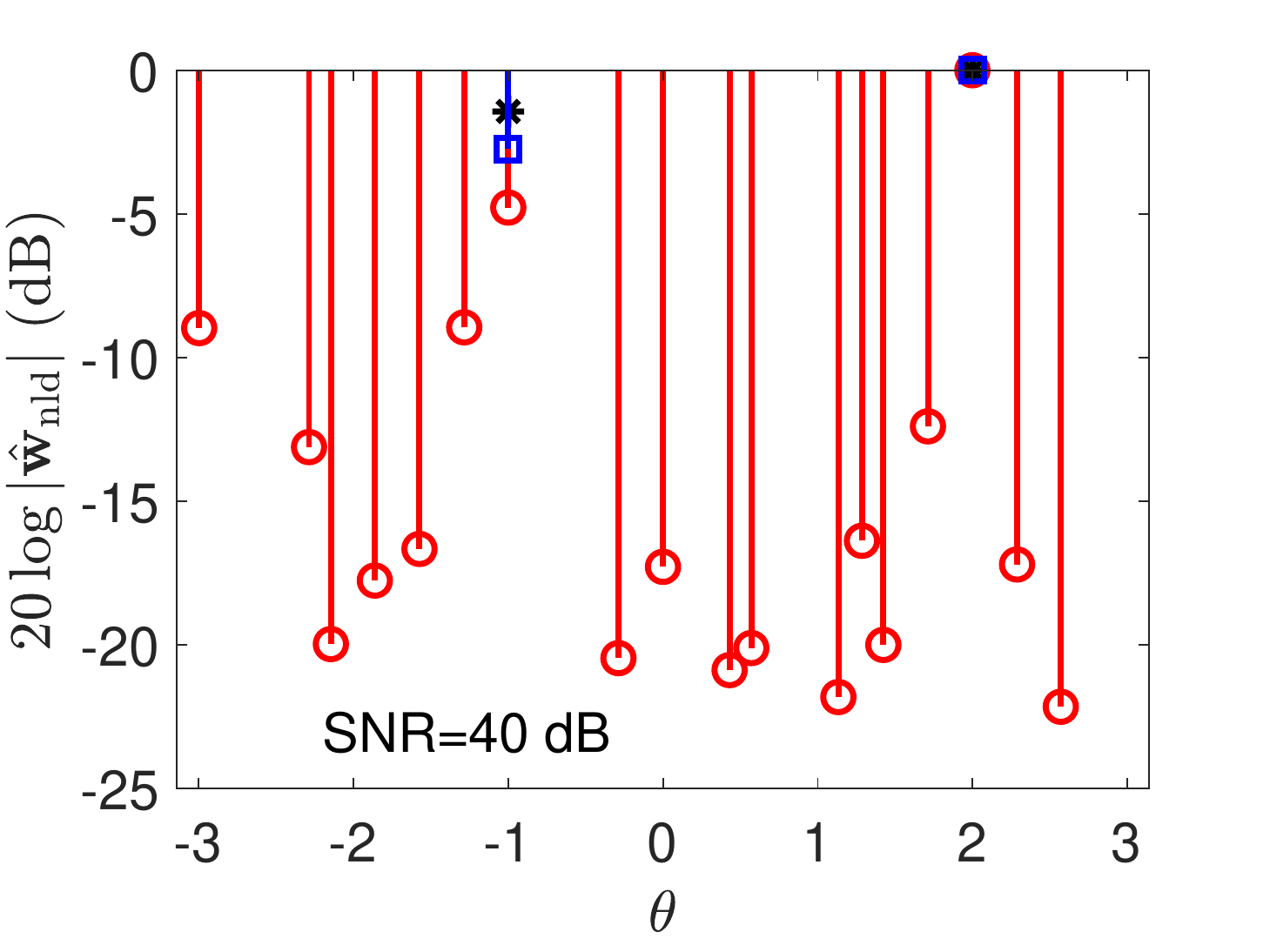}}
  \caption{A typical reconstruction results of VALSE-EP and VALSE under $1$ bit quantization at ${\rm SNR}=0$ dB, ${\rm SNR}=20$ dB and ${\rm SNR}=40$ dB. The normalized amplitudes $\hat{\mathbf w}_{\rm nld}\triangleq \hat{\mathbf w}/({\operatorname{max}_{i}} |\hat{w}_i|)$ are plotted.}
  \label{RecResults} %% label for entire figure
\end{figure*}

At first, an experiment is conducted to show that VALSE-EP is able to suppress the harmonics coming from the quantized data. The parameters are set as follows: $N=100$, $K=2$ and the true frequencies are $\tilde{\boldsymbol \theta}=[-1,2]^{\rm T}$. The results are shown in Fig. \ref{RecResults} for ${\rm SNR}=0$ dB, ${\rm SNR}=20$ dB and ${\rm SNR}=40$ dB. As stated in \cite{Jin}, the one-bit data consists of plentiful harmonics including self-generated and cross-generated harmonics. especially at high SNR. For low SNR scenario (${\rm SNR}=0$ dB), both VALSE and VALSE-EP estimate the model order successfully. For medium (${\rm SNR}=20$ dB) and high SNR (${\rm SNR}=40$ dB) scenario, VALSE overestimates the model order, outputs the fundamental (true) frequency and the self-generated and cross-generated harmonics. For example, the $3$rd order harmonics corresponding to ${\mathcal W}(-\theta_1-2\theta_2)\approx -3$, ${\mathcal W}(-\theta_1+2\theta_2)\approx -1.28$,  ${\mathcal W}(2\theta_1-\theta_2)\approx 2.29$, ${\mathcal W}(-2\theta_1-\theta_2)=0$, the $5$th order harmonics corresponding to ${\mathcal W}(-4\theta_1+\theta_2)\approx -0.28$, ${\mathcal W}(5\theta_1)\approx 1.28$,  ${\mathcal W}(-2\theta_1+3\theta_2)\approx 1.72$, ${\mathcal W}(2\theta_1+3\theta_2)\approx -2.28$, the $7$th order harmonic ${\mathcal W}(-3\theta_1+4\theta_2)\approx -1.57$, are estimated for ${\rm SNR}=20$ dB. While, VALSE-EP estimates the model order correctly for both ${\rm SNR}=20$ dB and ${\rm SNR}=40$ dB, demonstrating the effectiveness of suppressing the harmonics.

For the ensuing subsections, four numerical experiments with synthetic data and one with real data are conducted to demonstrate the excellent performance of VALSE-EP under quantizied setting, compared to VALSE.
\subsection{NMSE of the Line Spectral versus Iteration}\label{NMSEvsItersec}
\begin{figure*}
  \centering
  \subfigure[]{
    \label{NMSEvsIter0dB} %% label for first subfigure
    \includegraphics[width=58mm]{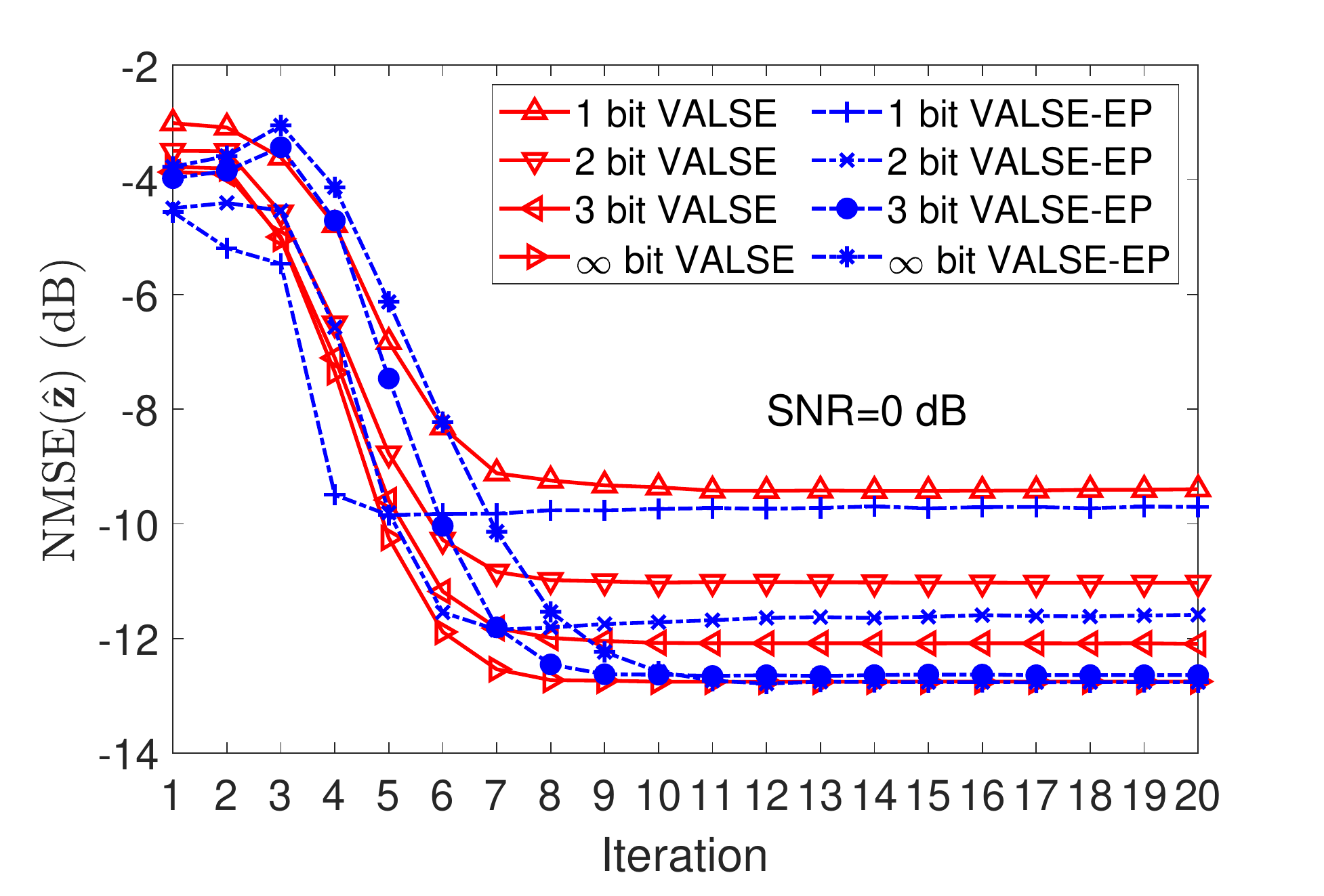}}
  \subfigure[]{
    \label{NMSEvsIter20dB} %% label for first subfigure
    \includegraphics[width=58mm]{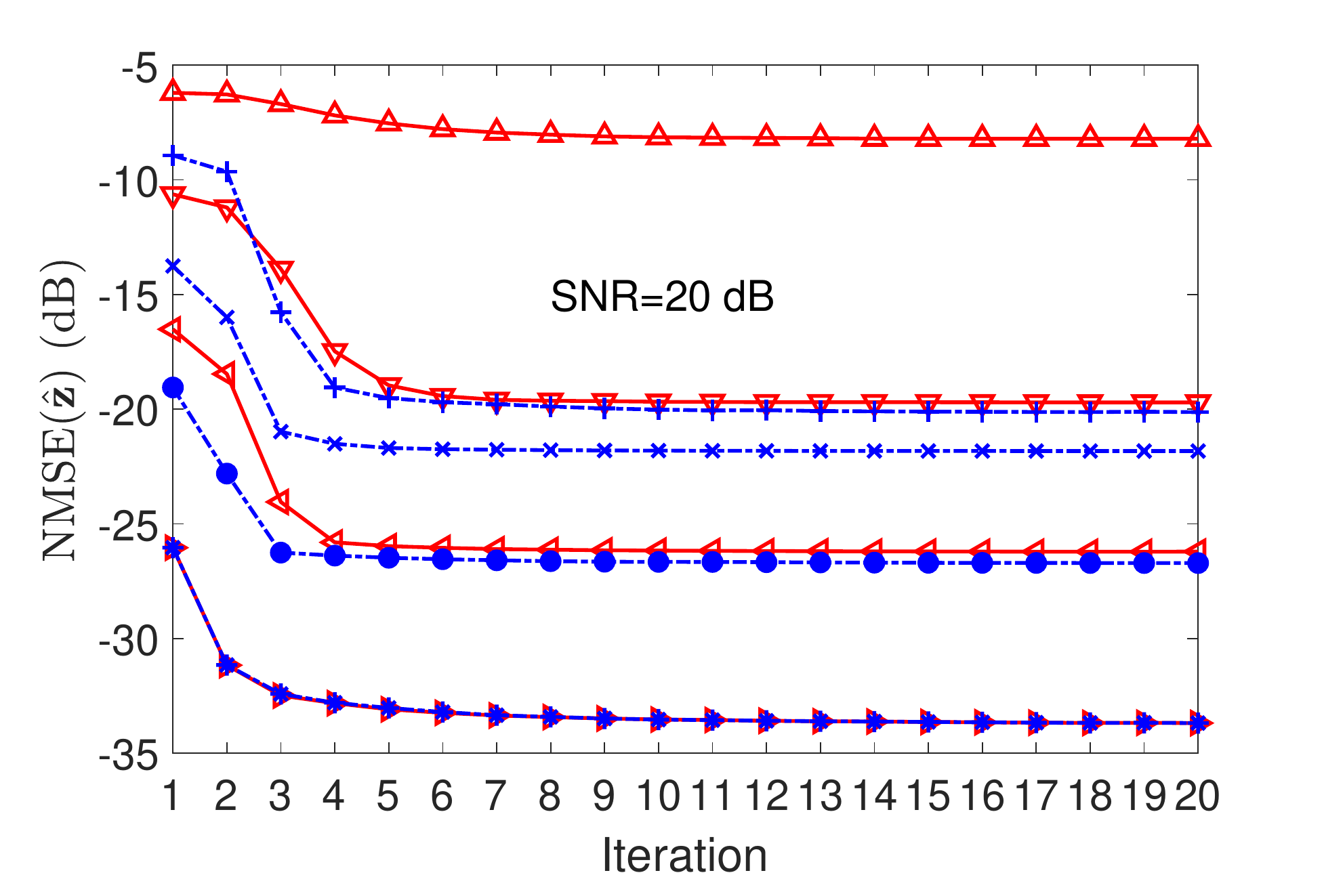}}
  \subfigure[]{
    \label{NMSEvsIter40dB} %% label for second subfigure
    \includegraphics[width=58mm]{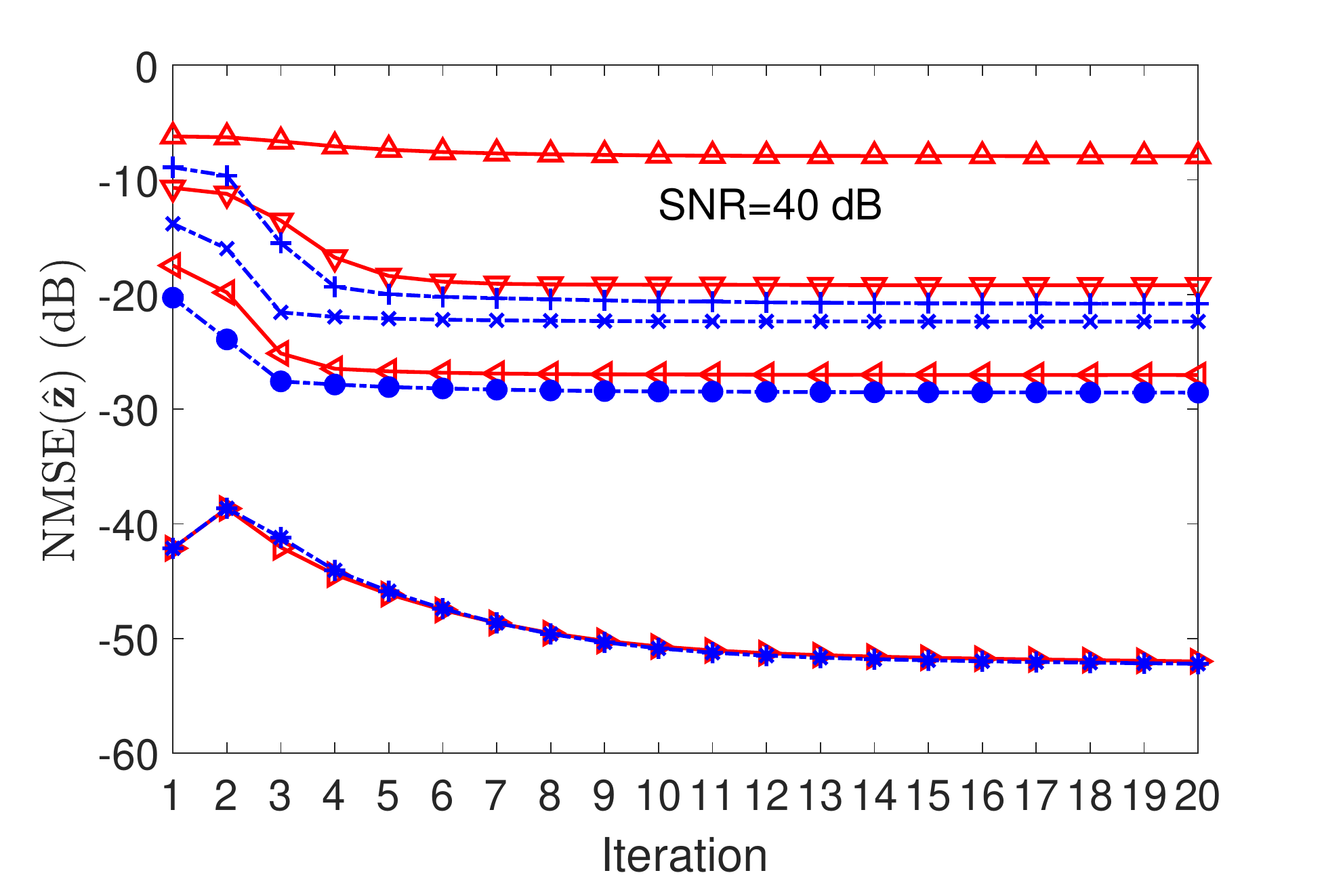}}
  \caption{The NMSE of the LSE of the VALSE-EP versus the number of iterations under ${\rm SNR}=0$ dB, ${\rm SNR}=20$ dB, ${\rm SNR}=40$ dB, respectively. Here $N=100$ and $K=3$.}
  \label{NMSEvsIter} %% label for entire figure
\end{figure*}

The NMSEs of the reconstructed line spectral versus the iteration are investigated and results are shown in Fig. \ref{NMSEvsIter}. Note that both VALSE-EP and VALSE converge in tens of iterations. For low SNR scenario (${\rm SNR}=0$ dB), VALSE-EP and VALSE are comparable. As SNR increases, the performance gap between VALSE-EP and VALSE increases under quantized settings. Besides, the performance of the VALSE-EP improves as ${\rm SNR}$ increases, especially for higher bit-depth. In contrast, for $1$ bit quantization, VALSE works well under low SNR scenario, and degrades as SNR increases. As the bit-depth increases, the performances of the VALSE-EP and VALSE improve and approach to the unquantized setting.
\subsection{Estimation versus SNR}
\begin{figure*}
  \centering
  \subfigure[]{
    \label{NMSEvsSNR} %% label for first subfigure
    \includegraphics[width=58mm]{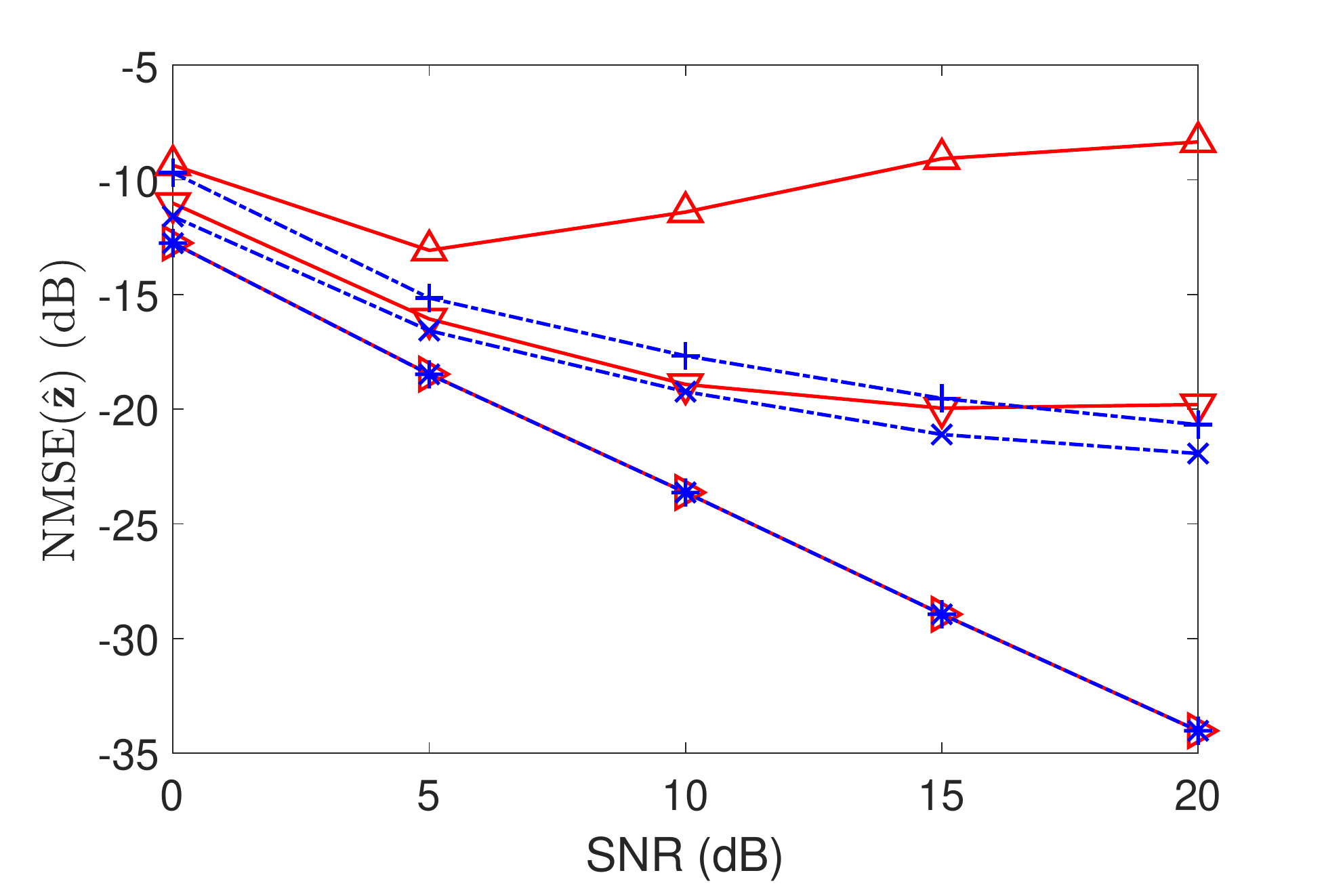}}
  \subfigure[]{
    \label{ProbvsSNR} %% label for second subfigure
    \includegraphics[width=58mm]{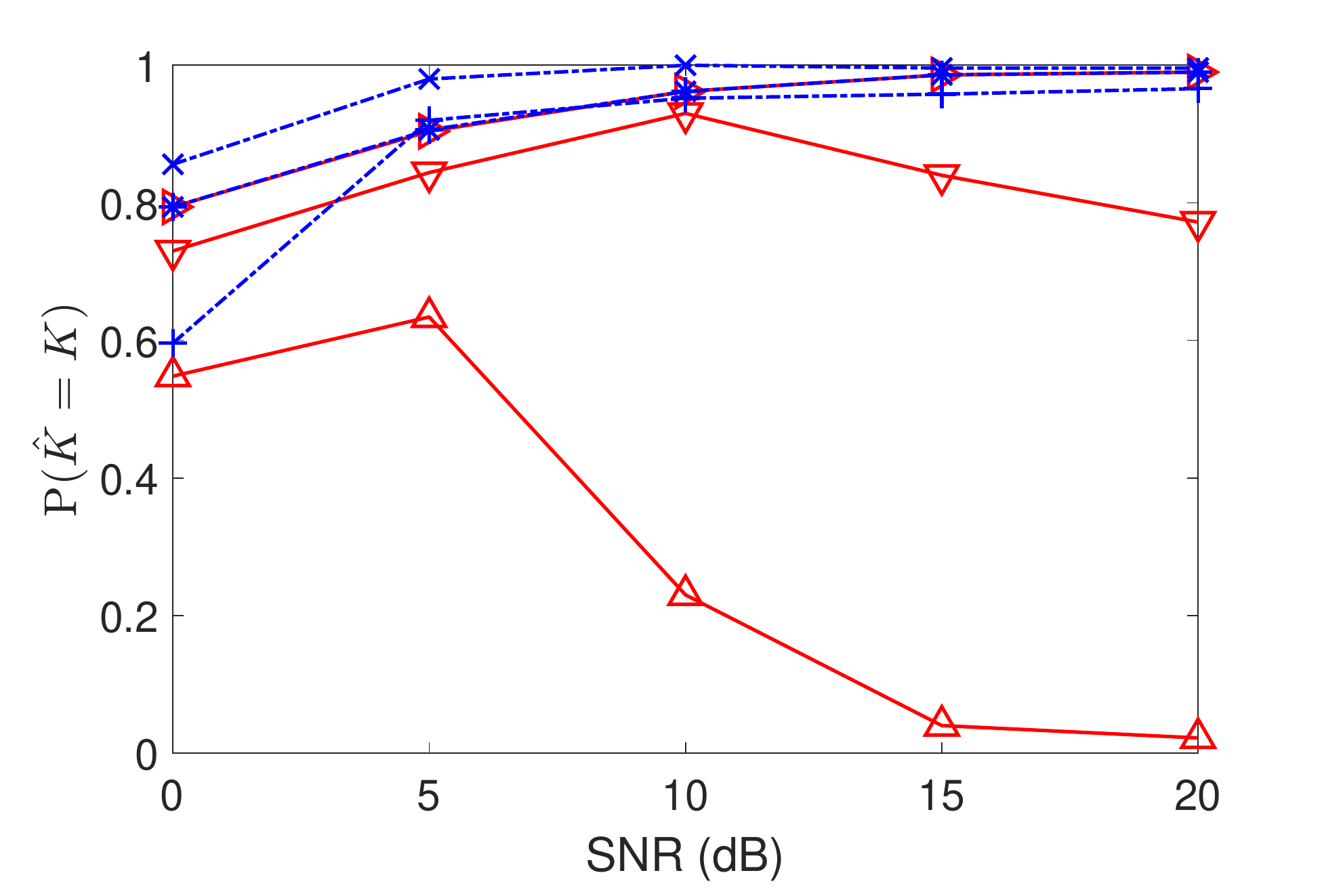}}
  \subfigure[]{
    \label{MSEvsSNR} %% label for first subfigure
    \includegraphics[width=58mm]{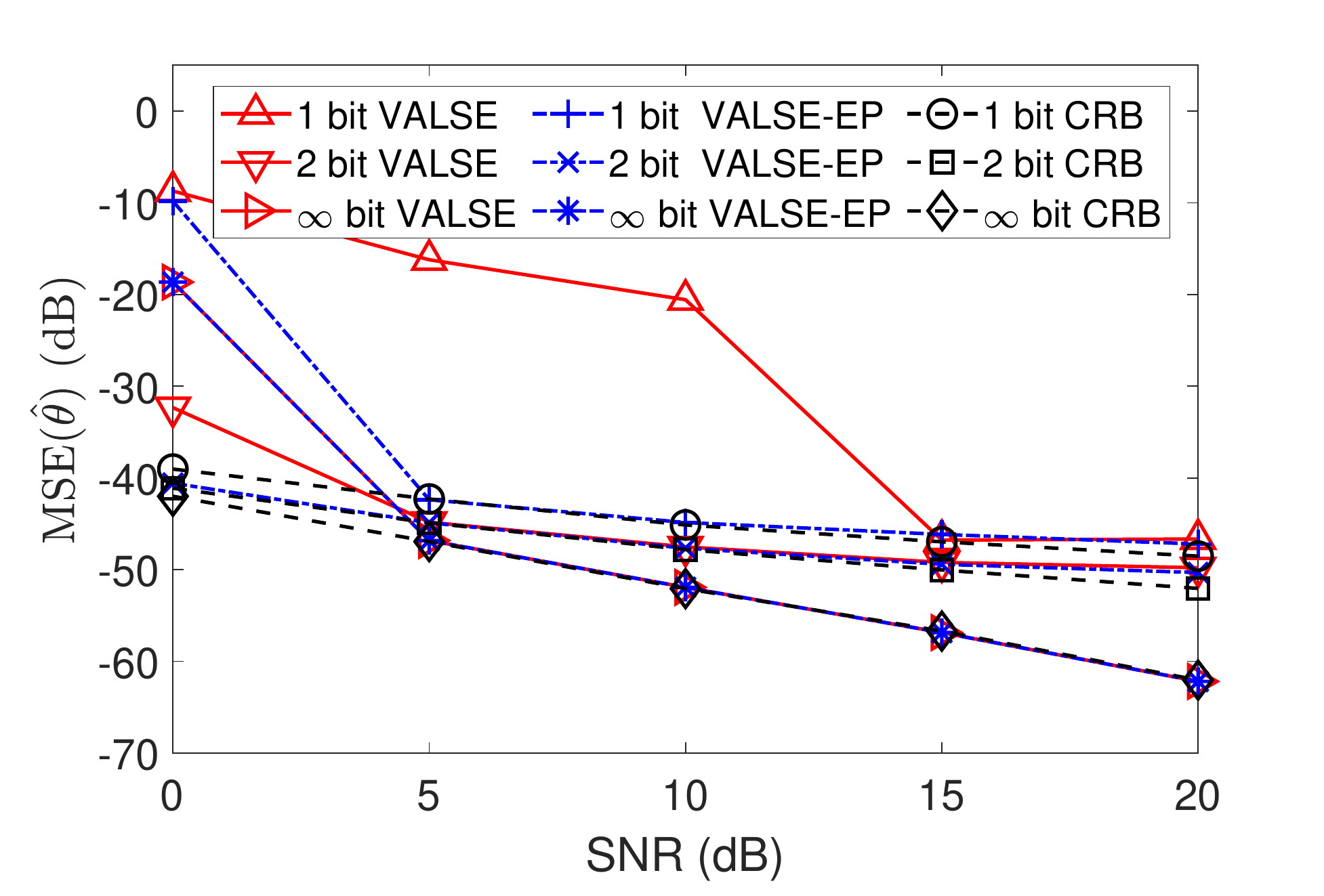}}
  \caption{Performance versus ${\rm SNR}$ for $N=100$ and $K=3$: (a) NMSE of the reconstructed signal; (b) success rate of model order estimation; (3) MSE of frequency estimation.}
  \label{PerbvsSNR} %% label for entire figure
\end{figure*}

The performance of the VALSE-EP versus SNR is investigated and the results are plotted in Fig. \ref{PerbvsSNR}. For the signal reconstruction and model order estimation probability, Fig. \ref{NMSEvsSNR} and \ref{ProbvsSNR} show that for 1 bit and 2 bit quantization, as SNR increases, the performances of VALSE first improve, and then degrade, while the performances of VALSE-EP always improve and are better than VALSE. A surprising phenomenon is that VALSE-EP under 2 bit quantization achieves the highest success rate of model order estimation and approaches to the CRB firstly. As the SNR continuous to increase, VALSE-EP deviates a little away from CRB under 1 bit and 2 bit quantization, while in the unquantized setting, both VALSE and VALSE-EP approach to the CRB asymptotically.
\subsection{Estimation versus Number of Measurements $N$}
\begin{figure*}
  \centering
  \subfigure[]{
    \label{ZNMSEvsN} %% label for first subfigure
    \includegraphics[width=58mm]{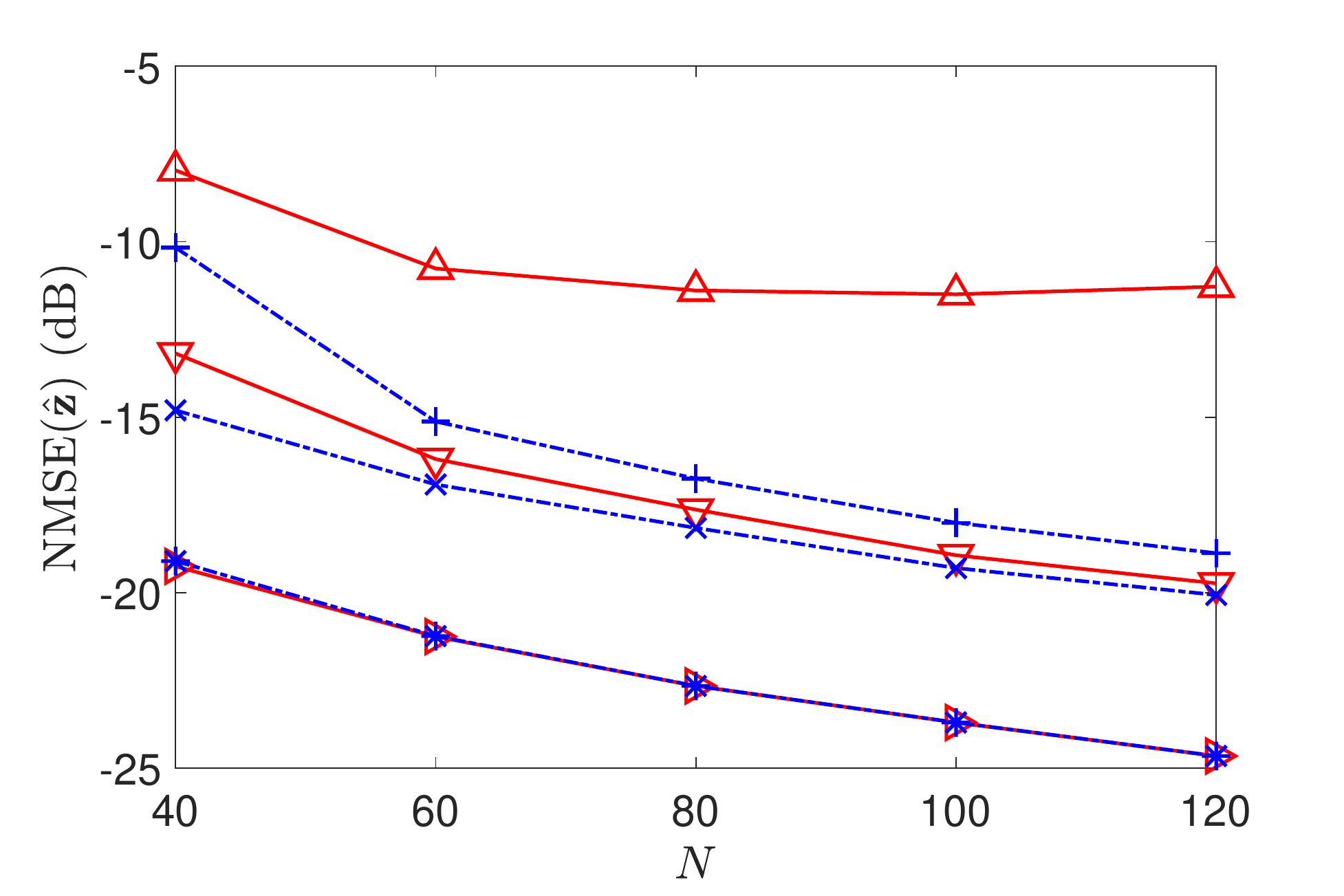}}
  \subfigure[]{
    \label{ThetaMSEvsN} %% label for first subfigure
    \includegraphics[width=58mm]{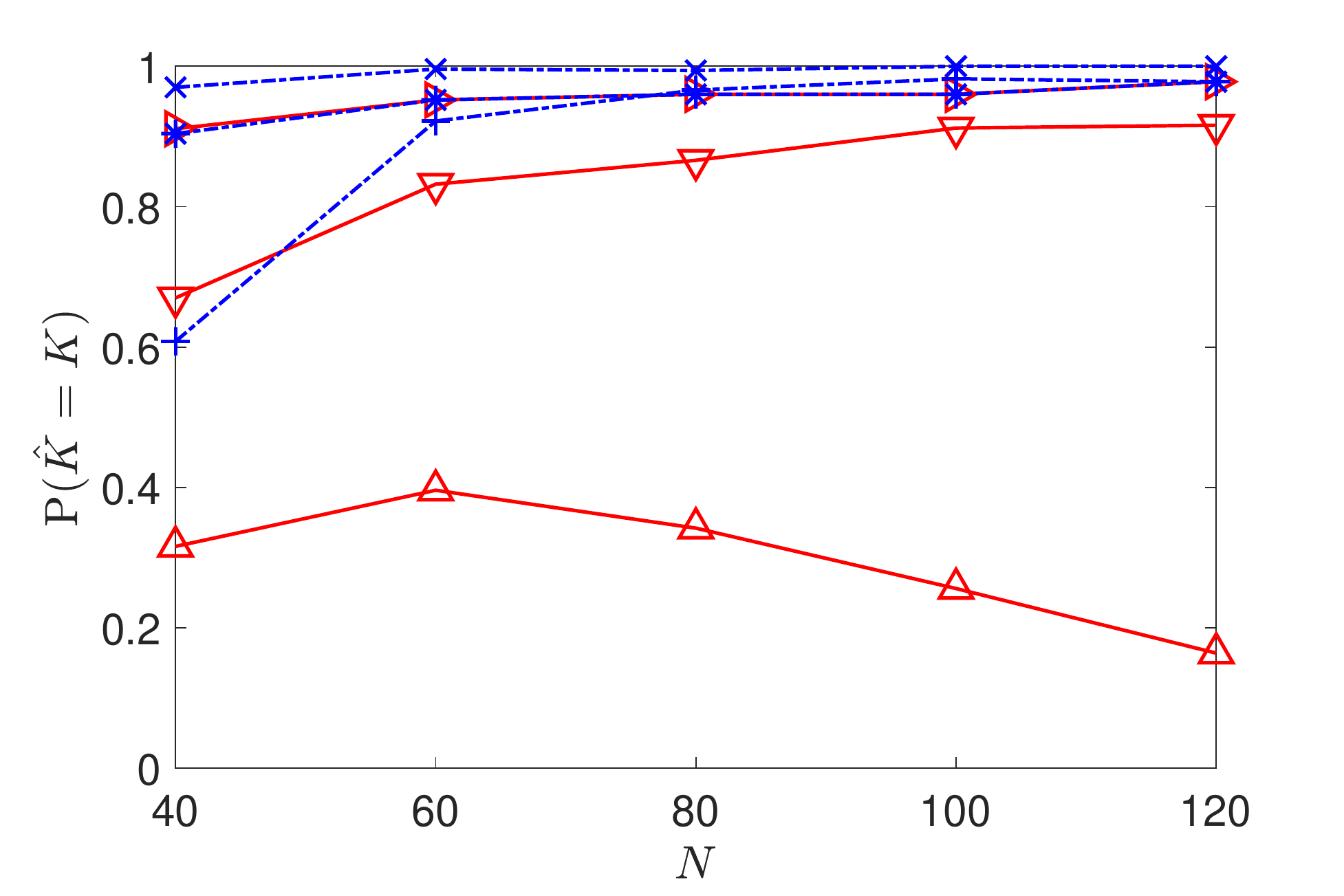}}
  \subfigure[]{
    \label{ProbvsN} %% label for second subfigure
    \includegraphics[width=58mm]{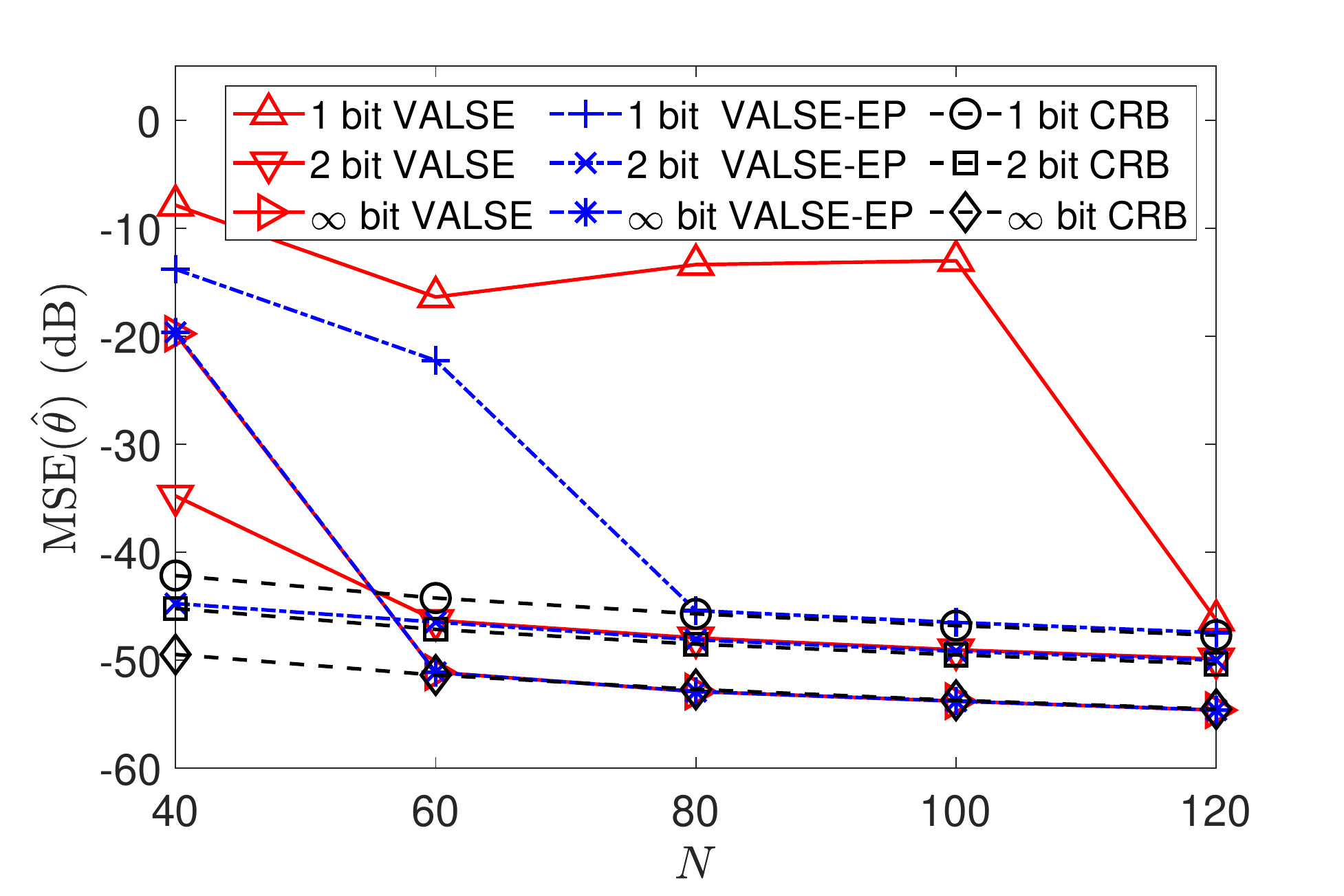}}
  \caption{The performance of VALSE-EP versus the number of measurements $N$ for ${\rm SNR}=10$ dB and $K=3$: (a) NMSE of the reconstructed signal; (b) success rate of model order estimation; (3) MSE of frequency estimation.}
  \label{PerbvsN} %% label for entire figure
\end{figure*}

The performances of both VALSE-EP and VALSE versus the number of measurements $N$ are examined and results are shown in Fig. \ref{PerbvsN}. For both algorithms, the signal reconstruction error decreases as the number of  measurements $N$ increases. It can be seen that VALSE-EP works better than VALSE under $1$ bit and $2$ bit quantization. As $N$ increases, the success rate of model order estimation of VALSE first increases and then decreases under $1$ bit quantization. In contrast, the success rate of model order estimation of VALSE-EP increases and even exceed the VALSE and VALSE-EP under unquantized setting for $N=100$. As for the frequency estimation error, VALSE-EP approaches to the CRB quickly than VALSE under quantized setting.
\subsection{Estimation versus Number of Spectral $K$}
\begin{figure*}
  \centering
  \subfigure[]{
    \label{NMSEvsK20dB} %% label for first subfigure
    \includegraphics[width=58mm]{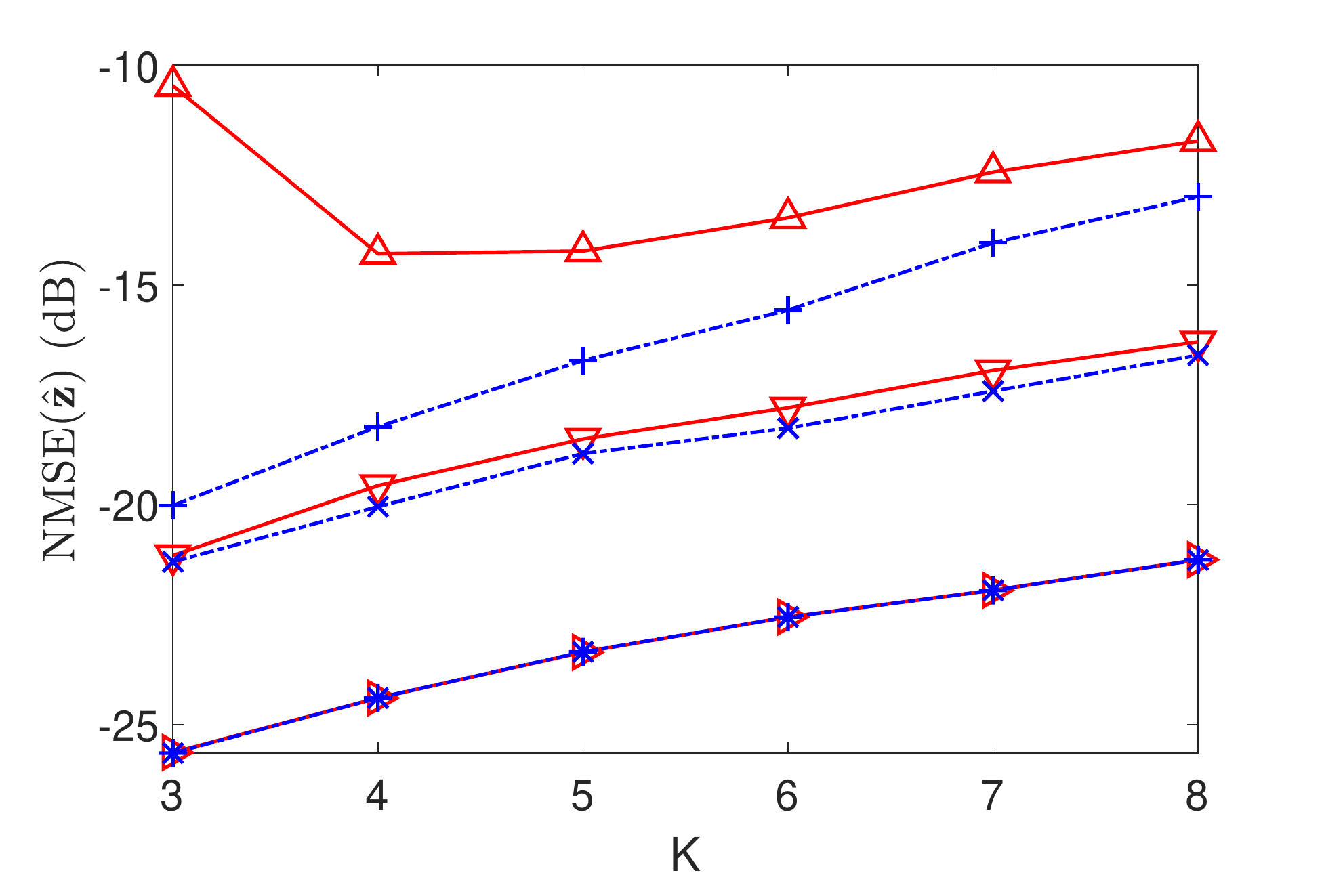}}
  \subfigure[]{
    \label{ProbvsK20dB} %% label for second subfigure
    \includegraphics[width=58mm]{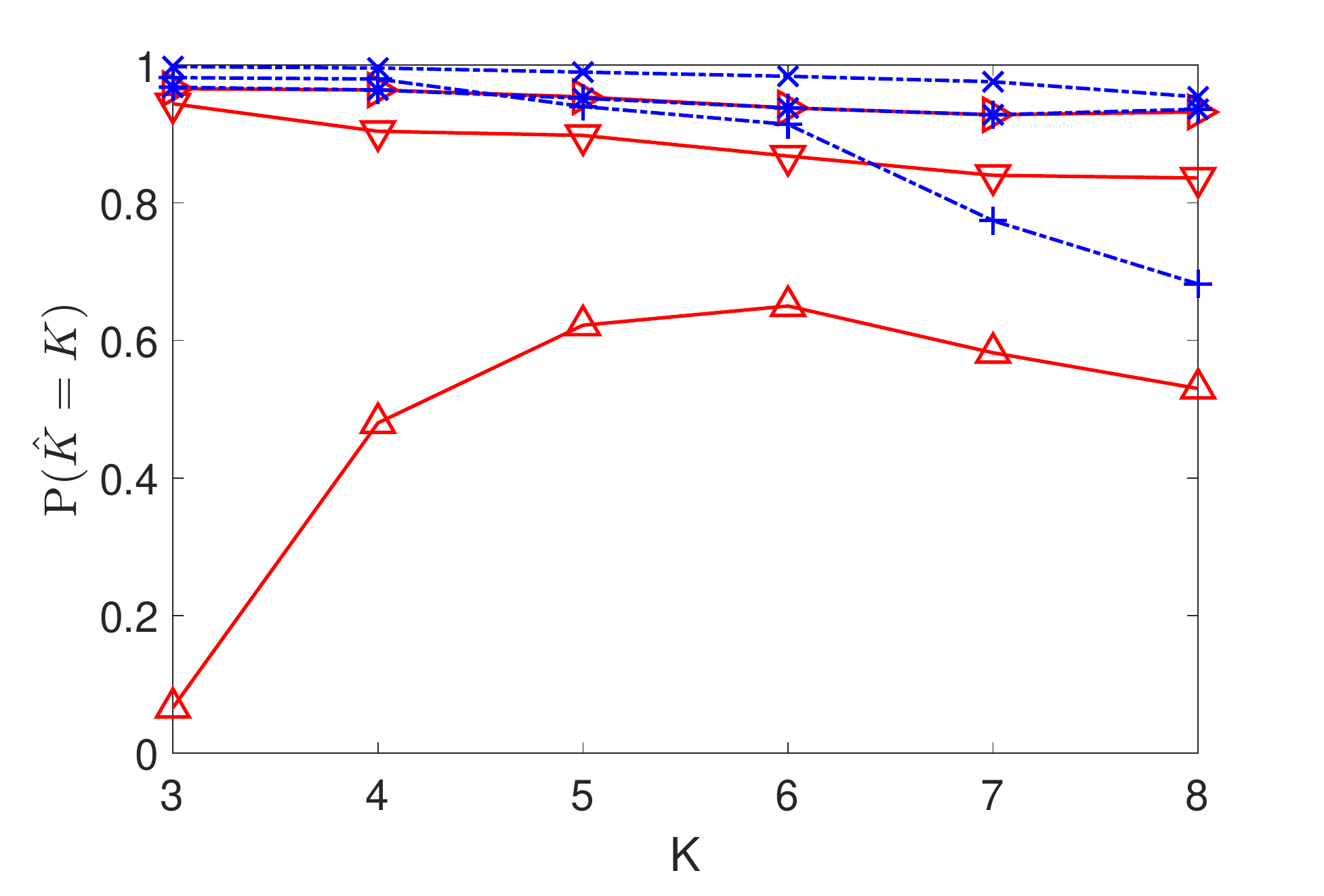}}
  \subfigure[]{
    \label{NMSEFvsK20dB} %% label for first subfigure
    \includegraphics[width=58mm]{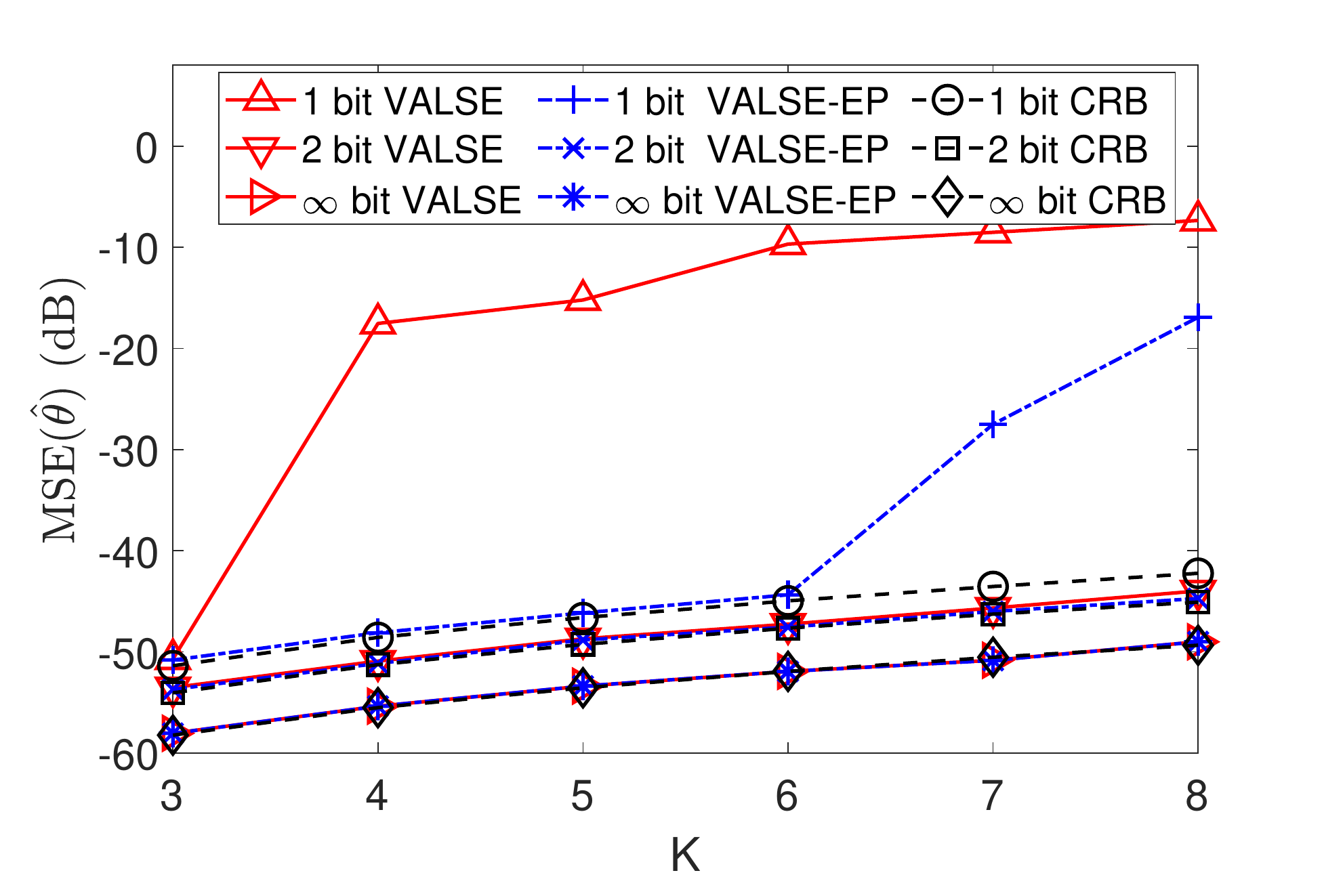}}
  \caption{The performance of VALSE-EP versus the number of spectrum $K$ for $N=160$ and ${\rm SNR}=10$ dB.}
  \label{PerbvsK} %% label for entire figure
\end{figure*}

The performance of VALSE-EP is investigated with respect to the number of spectral $K$, and results are presented in Fig. \ref{PerbvsK}. For the signal reconstruction error and success rate of model order estimation, the performances of all algorithms degrade as $K$ increases, except that VALSE under $1$ bit quantization, which first improves and then degrades. For the frequency estimation error, as $K$ increases, VALSE degrades quickly and deviates far way from CRB when $K\geq 4$ under $1$ bit quantization. When $K\leq 6$, VALSE-EP is always close to the CRB under $1$ bit quantization. As $K$ continues to increase, the frequency estimation error of VALSE-EP begin to deviate far way from CRB.
\subsection{Real Data}\label{Realdatasub}
The performance of VALSE-EP is evaluated with real data collected at sea, which was collected on a HLA (part of the Shark array) during the Shallow Water 2006 (SW06) experiment on August 5 and 6, 2006 \cite{TCYang}. The array had $32$ elements uniformly spaced at $d=15$ m (design frequency $f=50$ Hz, thus the wavelength $\lambda$ is $\lambda=c/f=1500/50=30$ m and $d=\lambda/2$). The first $1000$ snapshots ($8494$ snapshots in total) are used. The source depths are estimated to be around $12$ m. For the DOA problem, $\theta_i=-2\pi d/\lambda\sin \varphi_i=-\pi \sin \varphi_i$, where $\{\varphi_i\}_{i=1}^K$ denote the DOAs. The variance of the real data is calculated to be $0.0136$, which is used to design the $2$ bit quantizer.

\begin{figure*}
  \centering
  \subfigure[]{\label{1bitVALSE}
    \includegraphics[width=58mm]{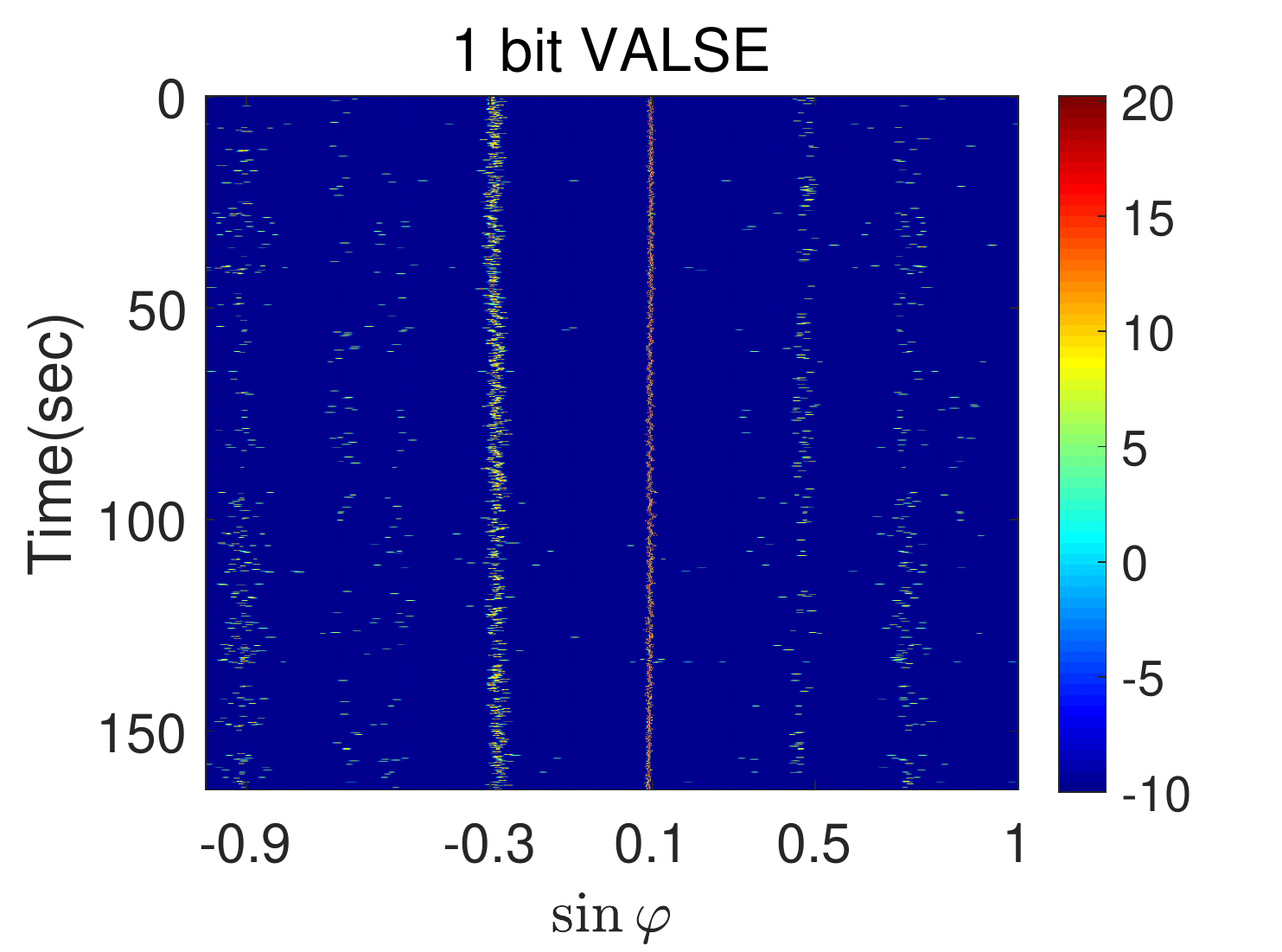}}
  \subfigure[]{\label{2bitVALSE}
    \includegraphics[width=58mm]{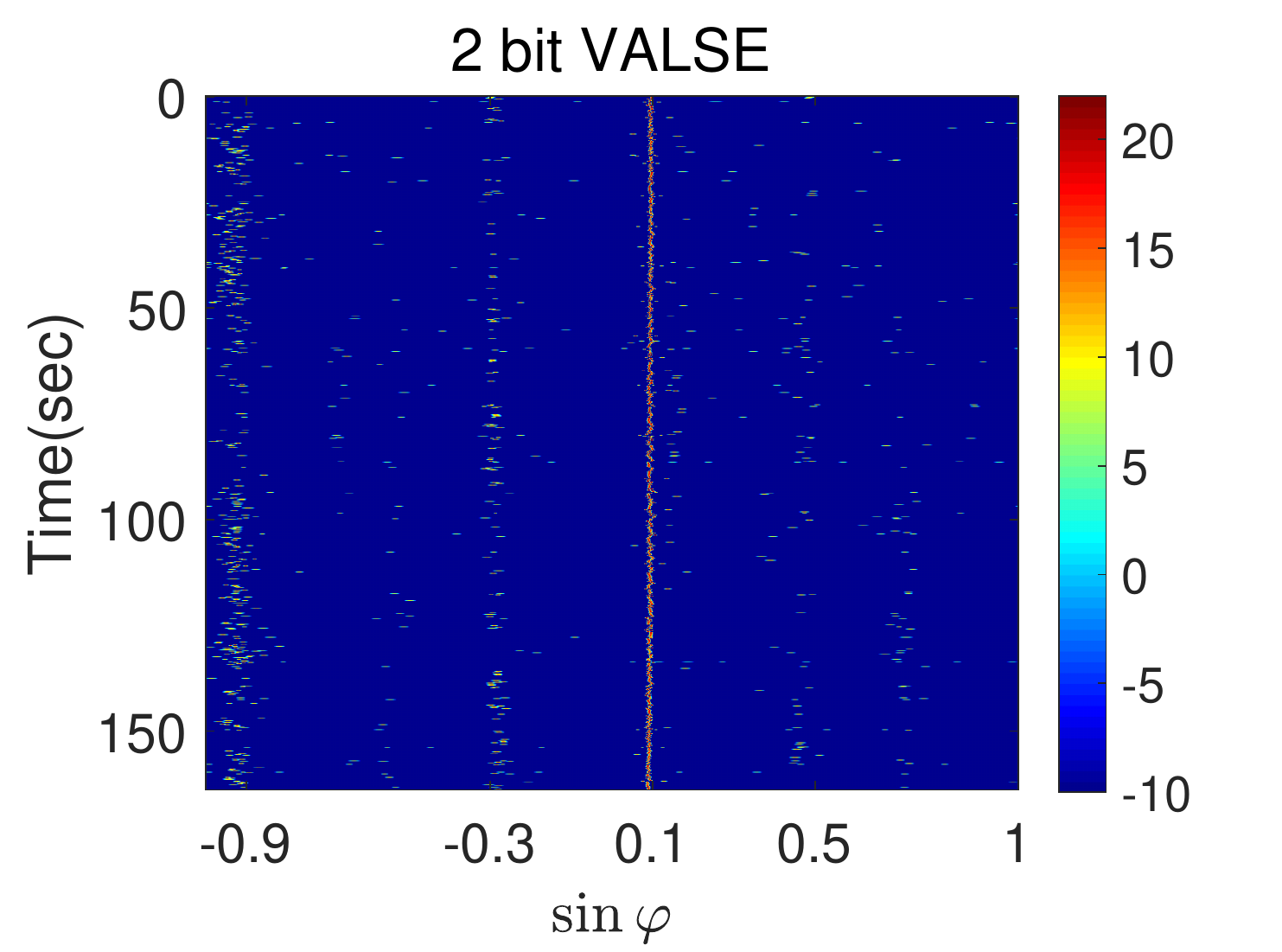}}
  \subfigure[]{\label{infbitVALSE}
    \includegraphics[width=58mm]{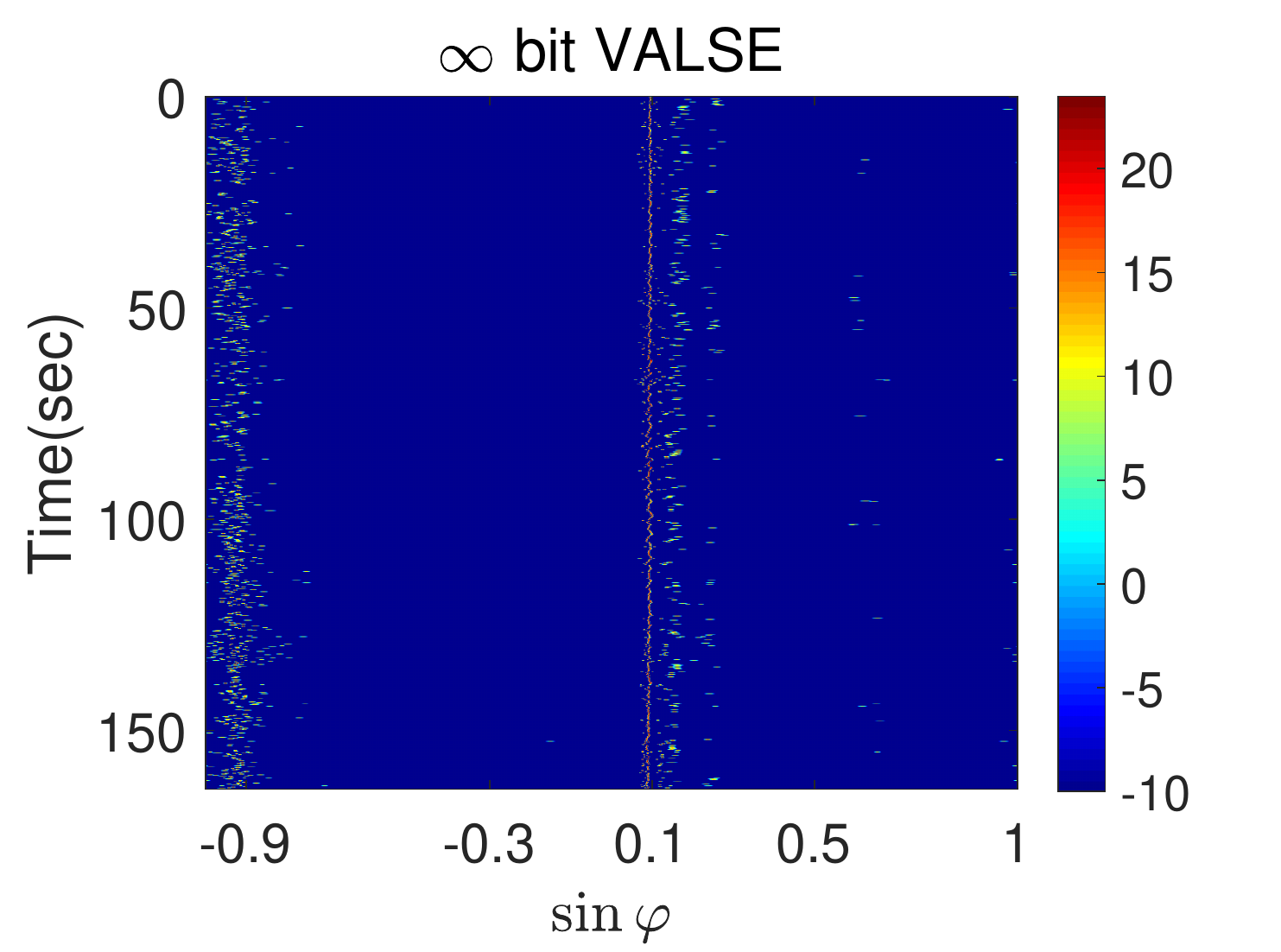}}
  \subfigure[]{\label{1bitVALSEEP}
    \includegraphics[width=58mm]{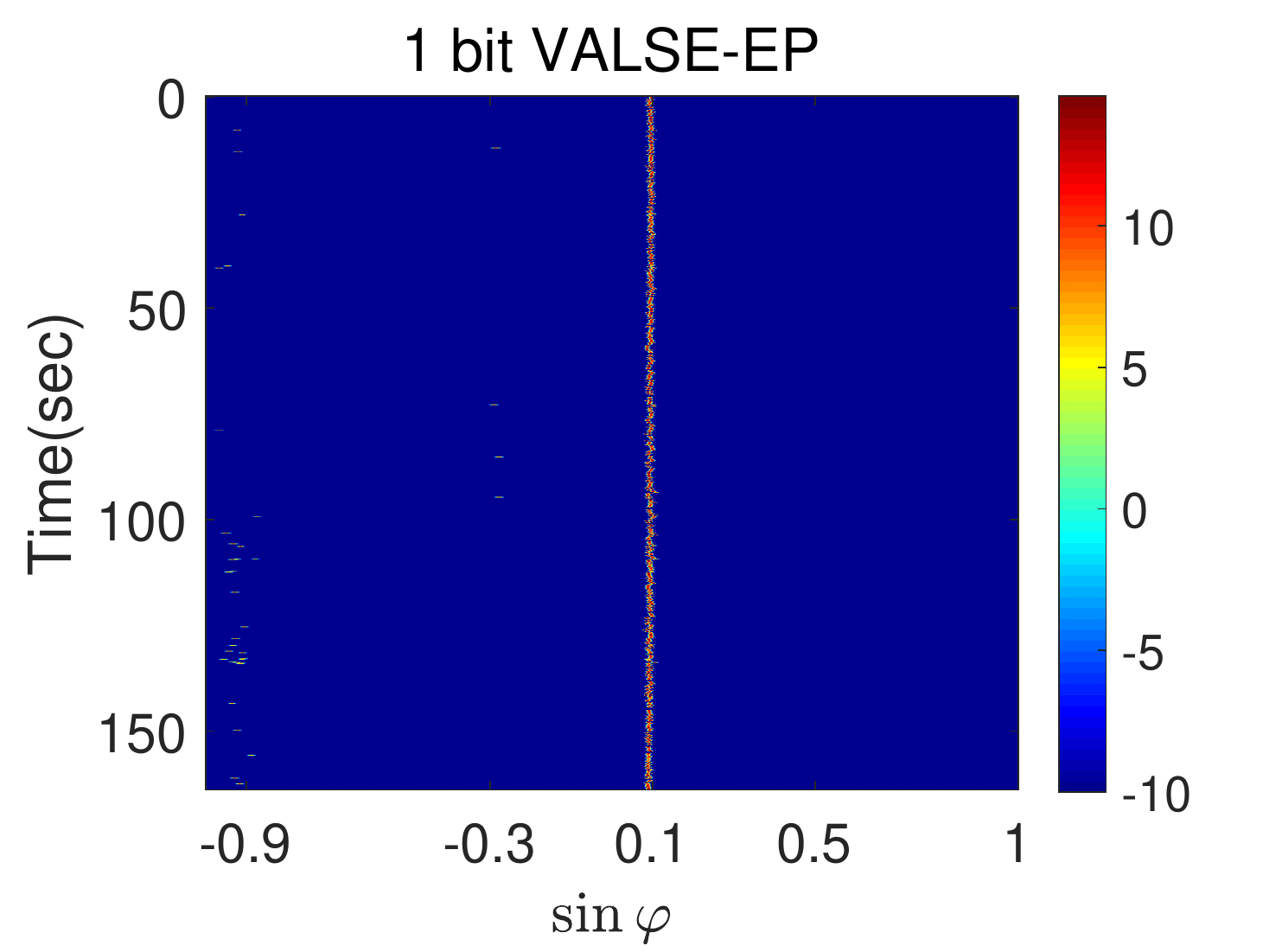}}
  \subfigure[]{\label{2bitVALSEEP}
    \includegraphics[width=58mm]{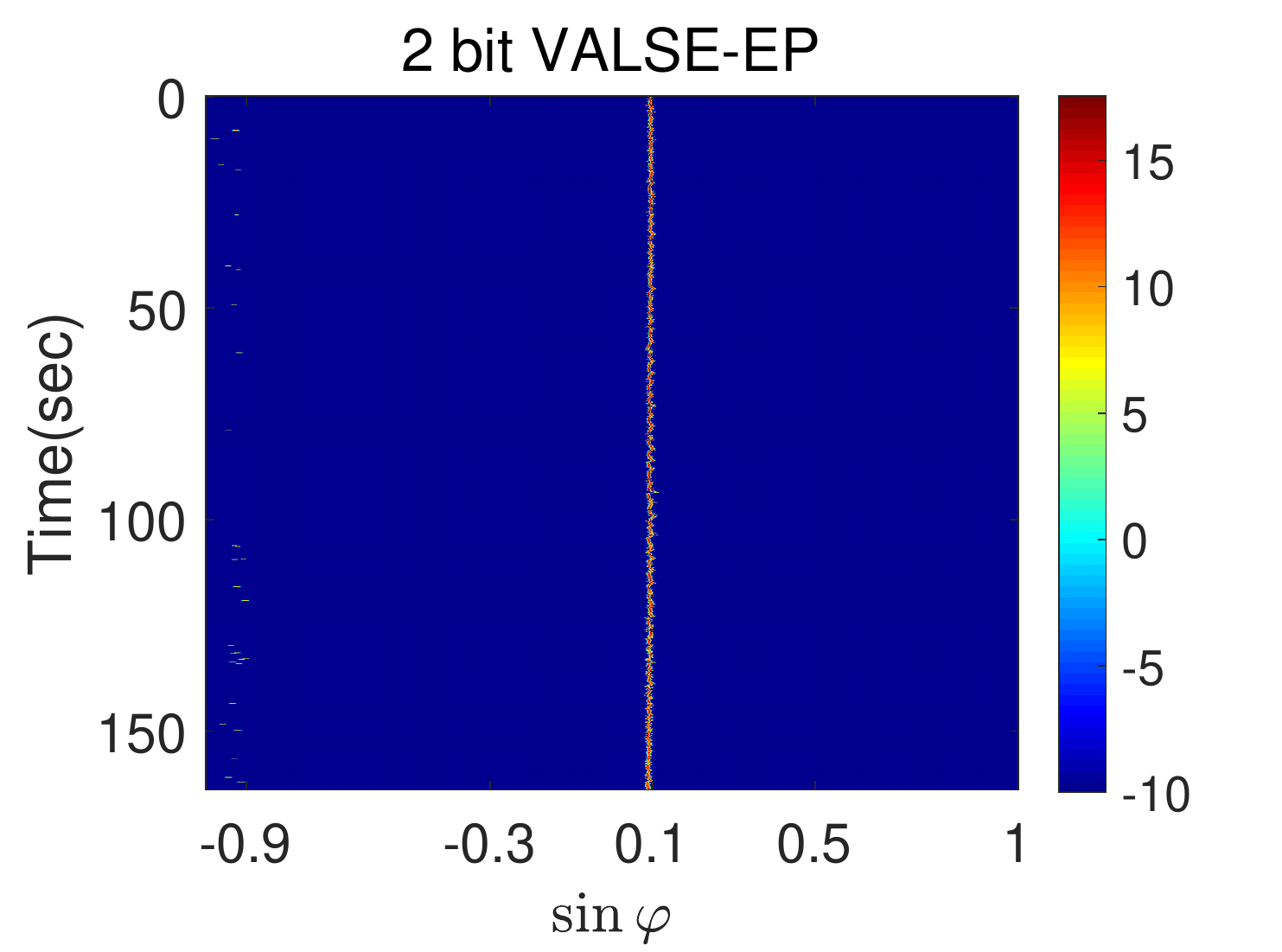}}
  \subfigure[]{\label{infbitVALSEEP}
    \includegraphics[width=58mm]{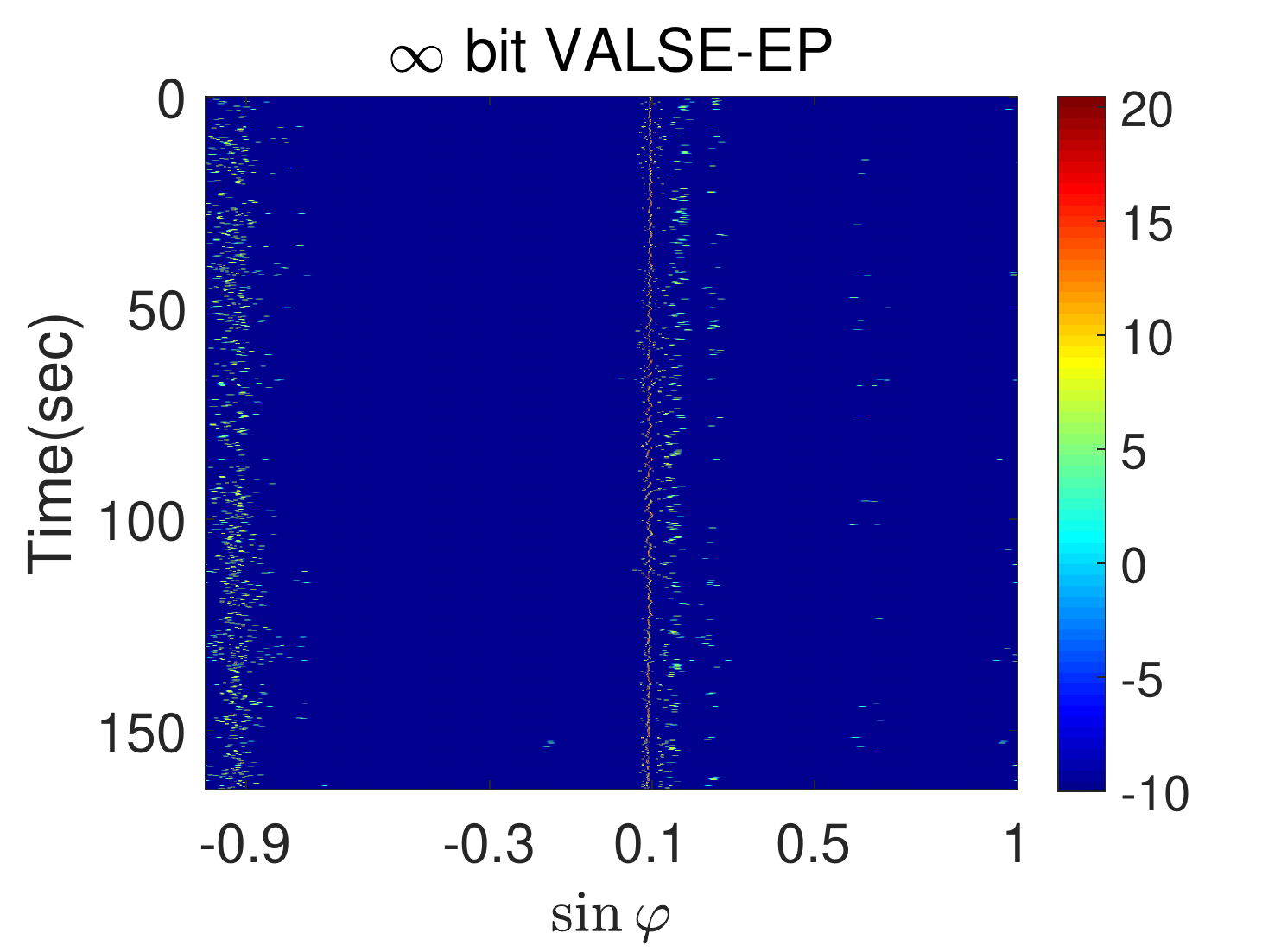}}
  \caption{The synthesized posterior PDF of $\sin\varphi$ for real data. }
  \label{RealData}
\end{figure*}

The synthesized posterior PDF of frequencies $1/|\hat{\mathcal S}|\sum\limits_{i=1}^{|\hat{\mathcal S}|}q(\theta_i|{\mathbf Y})|_{\theta_i=-\pi\sin \varphi_i}$ based on the data from one of the oblique runs is presented. In \cite{TCYang}, it is stated that the towed source signal is at $\sin \varphi_1\approx 0.1$, and a signal (the interferer source) near the endfire direction ($\sin \varphi_2\approx -0.9$) is assumed to be from a ship (The existence of this ship is unknown). For unquantized setting, Fig.\ref{infbitVALSE} and Fig. \ref{infbitVALSEEP} are consistent with the results obtained in \cite{TCYang}. While when the data is heavily quantized into 1 bit and 2 bit, the towed source signal is estimated accurately and the interferer source disappears for VALSE-EP. In contrast, the towed source signal are estimated accurately for VALSE algorithm, with an additional false source signal corresponding to the third order harmonic $-3\times (-\pi\sin \varphi_1)\approx 0.3\pi$ (located at $-0.3$ in Fig. \ref{1bitVALSE}) of the fundamental frequency $-\pi\sin \varphi_1\approx -0.1\pi$.
%\subsection{Estimating by varying the snapshots $G$}
%The performance of the VALSE-EP is investigated with varied $G$. The results are shown in Fig. \ref{fig:subfig4}. It can be seen that as the number of snapshots increases, the NMSE of the signal decreases and becomes stable. In addition, the performance of the VALSE-EP is better for ${\rm SNR}=40$ dB than that of ${\rm SNR}=20$ dB.
%The NMSE of the frequency decreases as the number of snapshots increases. As for the correct model order probability, the overall trend is upwards as the number of snapshots increases.

%%We examine the performance of the VALSE-EP by varying the number of snapshots. The results are shown in FIg. It can be seen that the performance of the VALSE-EP improves as the number of snapshots increases. In addition,
\section{Conclusion}\label{con}
In this paper, a VALSE-EP algorithm is proposed to deal with the fundamental LSE problem from quantized samples. The VALSE-EP is one kind of low-complexity grid-less algorithm which iteratively refines the frequency estimates, automatically estimates the model order, and learns the parameters of the prior distribution and noise variance. Importantly, VALSE-EP provides the uncertain degrees of the frequency estimates from quantized samples. Substantial numerical experiments are conducted to show the excellent performance of the VALSE-EP, including on a real data set.
\section{Acknowledgement}
The authors acknowledge the editor and the four anonymous reviewers for valuable comments and suggestions on this work.
\section{Appendix}\label{Appendix}
%In this Appendix, the VALSEs for the pseudo standard linear model corresponding to the SMV and MMV are derived, respectively.
\subsection{Finding the local maximum of $\ln Z(\mathbf s)$}\label{findmaxz}
A greedy iterative search strategy similar to \cite{Badiu} is adopted to find a local maximum of $\ln Z(\mathbf s)$. We proceed as follows: In the $p$th iteration, the $k$th test sequence $\mathbf t_k$ which flips the $k$th element of $\mathbf s^{(p)}$ is obtained. Then $\Delta^{(p)}_k=\ln Z(\mathbf t_k)-\ln Z({\mathbf s}^{p})$ is calculated for each $k=1,\cdots,N$. If $\Delta^{(p)}_k<0$ holds for all $k$, the algorithm is terminated and $\hat{\mathbf s}$ is set as ${\mathbf s}^{(p)}$, otherwise $t_k$ corresponding to the maximum $\Delta^{(p)}_k$ is chosen as $\mathbf s^{(p+1)}$ in the next iteration.

When $k\not\in{\mathcal S}$, that is, $s_k=0$, we activate the $k$th component of $\mathbf s$ by setting $s_k=1$. Now, ${\mathcal S}'=\mathcal S\cup\{k\}$.
\begin{align}\notag\label{delta_z1}
\Delta_k &= \ln Z(\mathbf s')-\ln Z(\mathbf s)\notag\\
&=\ln\det(\mathbf J_{\mathcal S}+\frac{1}{\tau}\mathbf I_{|\mathcal S|})-\ln\det(\mathbf J_{\mathcal S'}+\frac{1}{\tau}\mathbf I_{|\mathcal S'|})\notag\\
&+\ln\frac{\rho}{1-\rho}+\ln\frac{1}{\tau}+\mathbf h_{\mathcal S'}^{\rm H}(\mathbf J_{\mathcal S'}+\frac{1}{\tau}\mathbf I_{|\mathcal S'|})^{-1}\mathbf h_{\mathcal S'} - \mathbf h_{\mathcal S}^{\rm H}(\mathbf J_{\mathcal S}+\frac{1}{\tau}\mathbf I_{|\mathcal S|})^{-1}\mathbf h_{\mathcal S}.
\end{align}
%To compute $\Delta_k$, we introduce two formulas,
%\begin{align}\label{det-formula}
%\begin{pmatrix}
%\mathbf D & \mathbf u \\
%\mathbf u^{\rm H} & 1
%\end{pmatrix}=
%&\begin{pmatrix}
%\mathbf D^{\frac{1}{2}} & 0 \\
%\mathbf u^{\rm H}\mathbf D^{-\frac{1}{2}}  & {(1-\mathbf u^{\rm H}\mathbf D^{-1}\mathbf u)}^{\frac{1}{2}}
%\end{pmatrix}\begin{pmatrix}
%\mathbf D^{\frac{1}{2}} & \mathbf D^{-\frac{1}{2}}\mathbf u \\
%0 & {(1-\mathbf u^{\rm H}\mathbf D^{-1}\mathbf u)}^{\frac{1}{2}}
%\end{pmatrix}
%\end{align}
%\begin{align}\label{inverse-formula}
%&\begin{pmatrix}
%\mathbf D & \mathbf u \\
%\mathbf u^{\rm H} & 1
%\end{pmatrix}^{-1}=
%\begin{pmatrix}
%\mathbf D^{-1} + \frac{\mathbf D^{-1}\mathbf u\mathbf u^{\rm H}\mathbf D^{-1}}{1-\mathbf u^{\rm T}\mathbf D^{-1}\mathbf u}& -\frac{\mathbf D^{-1}\mathbf u}{1-\mathbf u^{\rm H}\mathbf D^{-1}\mathbf u} \\
%-\frac{\mathbf u^{\rm H}\mathbf D^{-1}}{1-\mathbf u^{\rm H}\mathbf D^{-1}\mathbf u}  & \frac{1}{1-\mathbf u^{\rm H}\mathbf D^{-1}\mathbf u}
%\end{pmatrix}
%\end{align}
Let ${\mathbf j}_k$ be ${\mathbf j}_k =[J_{ik}|i\in {\mathcal S}]^{\rm T}$. By using the block-matrix determinant formula, one has
\begin{align}\label{det_J}
\ln(\det(\mathbf J_{\mathcal S'}+\frac{1}{\tau}\mathbf I_{|\mathcal S'|}))=\ln{\det(\mathbf J_\mathcal S+\frac{1}{\tau}\mathbf I_{|\mathcal S|})} + {\rm ln}{\left({\rm tr}(\boldsymbol\Sigma^{-1})+\frac{1}{\tau}-\mathbf j_k^{\rm H}(\mathbf J_\mathcal S+\frac{1}{\tau}\mathbf I_{|\mathcal S|})^{-1}\mathbf j_k\right)},
\end{align}
By the block-wise matrix inversion formula, one has
\begin{align}\label{inv_J}
\mathbf h_{\mathcal S'}^{\rm H}(\mathbf J_{\mathcal S'}+\frac{1}{\tau}\mathbf I_{|\mathcal S'|})^{-1}\mathbf h_{\mathcal S'}
=\mathbf h_{\mathcal S}^{\rm H}(\mathbf J_{\mathcal S}+\frac{1}{\tau}\mathbf I_{|\mathcal S|})^{-1}\mathbf h_{\mathcal S}+\frac{\xi^*\xi}{{\rm tr}(\boldsymbol\Sigma^{-1})
+\frac{1}{\tau}-\mathbf j^{\rm H}_k(\mathbf J_{\mathcal S}+\frac{1}{\tau}\mathbf I_{|\mathcal S|})^{-1}{\mathbf j}_k},
\end{align}
where $ \xi = h_k-\mathbf j^{\rm H}_k(\mathbf J_\mathcal S+\frac{1}{\tau}\mathbf I_{|\mathcal S|})^{-1}{\mathbf h}_{\mathcal S}$.
Plugging (\ref{det_J}) and (\ref{inv_J}) in (\ref{delta_z1}), and let
\begin{align}\label{v-k-u-k}
&v_k = \left({\rm tr}(\boldsymbol\Sigma^{-1})+\frac{1}{\tau}-\mathbf j^{\rm H}_k(\mathbf J_{\mathcal S}+\frac{1}{\tau}\mathbf I_{|\mathcal S|})^{-1}{\mathbf j}_k\right)^{-1}~{\rm and}~\notag\\
&u_k = v_k\left(h_k-{\mathbf j}^{\rm H}_k({\mathbf J}_{\mathcal S}+\frac{1}{\tau}\mathbf I_{|\mathcal S|})^{-1}{\mathbf h}_{\mathcal S}\right),
\end{align}
$\Delta_k$ can be simplified as
%\begin{small}
\begin{align}\label{delta-k-active}
\Delta_k = \ln\frac{v_k}{\tau} + \frac{|u_k|^2}{v_k}+\ln\frac{\rho}{1-\rho}.
\end{align}
Given that $\mathbf s$ is changed into ${\mathbf s}'$, the mean $\hat{\mathbf w}'_{{\mathcal S'}}$ and covariance ${\hat{\mathbf C}'}_{\mathcal S'}$ of the weights can be updated from (\ref{W-C-1}), i.e.,
\begin{subequations}
\begin{align}\label{cov_update}
{\hat{\mathbf C}'}_{\mathcal S'} &= (\mathbf J_{\mathcal S'}+\frac{1}{\tau}\mathbf I_{|\mathcal S'|})^{-1},\\
\hat{\mathbf w}'_{{\mathcal S'}} &= \hat{\mathbf C}_{{\mathcal S'}}{\mathbf h}_{{\mathcal S'}}.
\end{align}
\end{subequations}
In fact, the matrix inversion can be avoided when updating $\hat{\mathbf w}'_{{\mathcal S'}}$ and ${\hat{\mathbf C}}_{\mathcal S'}$. It can be shown that
\begin{align}\label{cov_up_act}
&\begin{pmatrix}
{\hat{\mathbf C}'_{\mathcal S'\backslash k}} & {\hat{\mathbf c}}'_{k} \\
{\hat{\mathbf c}'^{\rm H}_{k}} & \hat{C}'_{kk}
\end{pmatrix}=
(\mathbf J_{\mathcal S'}+\frac{1}{\tau}\mathbf I_{|\mathcal S'|})^{-1} \notag\\
=&\begin{pmatrix}
{\hat{\mathbf C}}_{\mathcal S} & \mathbf 0 \\
\mathbf 0 & 0
\end{pmatrix}+v_k\begin{pmatrix}
{\hat{\mathbf C}}_{\mathcal S}\mathbf j_k \\
-1
\end{pmatrix}
\begin{pmatrix}
{\hat{\mathbf C}}_{\mathcal S}\mathbf j_k\\
-1
\end{pmatrix}^{\rm H}=\begin{pmatrix}
{\hat{\mathbf C}}_{\mathcal S}+v_k {\hat{\mathbf C}}_{\mathcal S}\mathbf j_k({\hat{\mathbf C}}_{\mathcal S}\mathbf j_k)^{\rm H} & -v_k {\hat{\mathbf C}}_{\mathcal S}\mathbf j_k\\
-v_k ({\hat{\mathbf C}}_{\mathcal S}\mathbf j_k)^{\rm H} & v_k
\end{pmatrix}
\end{align}
${\hat{\mathbf C}'}_{\mathcal S'}$ is obtained if ${\hat{\mathbf c}}'_{k}$, $\hat{\mathbf c}'^{\rm H}_{k}$ and $\hat{C}'_{kk}$ are inserted appropriately in ${\hat{\mathbf C}'_{\mathcal S'\backslash k}}$, and
\begin{align}\label{mean_up_act}
&\begin{pmatrix}
{\hat{\mathbf w}'_{\mathcal S'\backslash k}}\\
\hat{w}'_{k}
\end{pmatrix}=
{\hat{\mathbf C}}_{\mathcal S'}\mathbf h_{{\mathbf S}'}\notag\\
=&\begin{pmatrix}
{\hat{\mathbf C}}_{\mathcal S}\mathbf h_{\mathcal S}+v_k{\hat{\mathbf C}}_{\mathcal S}\mathbf j_k\mathbf j^{\rm H}_k{\hat{\mathbf C}}_{\mathcal S}\mathbf h_{\mathcal S}-v_k{\hat{\mathbf C}}_{\mathcal S}\mathbf j_k h_k\\
-v_k\mathbf j^{\rm H}_k{\hat{\mathbf C}}_{\mathcal S}\mathbf h_{\mathcal S} + v_kh_k
\end{pmatrix}
=\begin{pmatrix}
{\hat{\mathbf C}}_{\mathcal S}\mathbf h_{\mathcal S}-u_k{\hat{\mathbf C}}_{\mathcal S}\mathbf j_k\\
u_k
\end{pmatrix}.
\end{align}
From (\ref{mean_up_act}) and (\ref{cov_up_act}), one can see that after activating the $k$th component, the posterior mean and variance of ${w}_k$ are ${u}_k$ and $v_k$, respectively.
%
%\begin{align}\label{W_active}
%{\hat w}_i=
%\begin{cases}
%{u}_k &i=k \\
%{\hat w}_i - \hat{\mathbf c}^{\rm H}_{i}{\mathbf j}_ku_k,&i\in{\mathcal S}
%\end{cases},
%\end{align}
%where $\hat{\mathbf c}_{i}$ denotes the $i$th column of ${\hat{\mathbf C}}_{\mathcal S}$.

For the deactive case with ${s}_k=1$, ${s}'_k = 0$ and ${\mathcal S}'=\mathcal S\backslash\{k\}$, $\Delta_k = \ln Z(\mathbf s')-\ln Z(\mathbf s)$ is the negative of (\ref{delta-k-active}), i.e.,
\begin{align}\label{delta-k-deactive}
\Delta_k =- \ln\frac{v_k}{\tau} - \frac{|{u}_k|^2}{v_k}-\ln\frac{\rho}{1-\rho}.
\end{align}
Similar to (\ref{cov_up_act}), the posterior mean and covariance update equation from ${\mathcal S}'$ to ${\mathcal S}$ case can be rewritten as
\begin{align}\label{C_deactive}
\begin{pmatrix}
{\hat{\mathbf C}}'_{\mathcal S'} & \mathbf 0 \\
\mathbf 0 & 0
\end{pmatrix}+v_k\begin{pmatrix}
{\hat{\mathbf C}}'_{\mathcal S'}\mathbf j_k \\
-1
\end{pmatrix}
\begin{pmatrix}
{\hat{\mathbf C}}'_{\mathcal S'}\mathbf j_k\\
-1
\end{pmatrix}^{\rm H}=
\begin{pmatrix}
{\hat{\mathbf C}_{\mathcal S\backslash k}} & {\hat{\mathbf c}}_{k} \\
{\hat{\mathbf c}^{\rm H}_{k}} & \hat{C}_{kk}
\end{pmatrix},
\end{align}
and
\begin{align}\label{W_deactive}
\begin{pmatrix}
\hat{\mathbf w}'_{\mathcal S'}-u_k{\hat{\mathbf C}'}_{\mathcal S'}{\mathbf j}_k\\
u_k
\end{pmatrix}=\begin{pmatrix}
{\hat{\mathbf C}'}_{\mathcal S'}\mathbf h_{\mathcal S'}-u_k{\hat{\mathbf C}'}_{\mathcal S'}\mathbf j_k\\
u_k
\end{pmatrix}=
\begin{pmatrix}
{\hat{\mathbf w}_{\mathcal S\backslash k}}\\
\hat{w}_{k}
\end{pmatrix},
\end{align}
where $\hat{\mathbf c}_{k,0}$ denotes the column of ${\hat{\mathbf C}}_{\mathcal S,0}$ corresponding to the $k$th component. According to (\ref{C_deactive}) and (\ref{W_deactive}), one has
\begin{subequations}\label{C_de_appro}
\begin{align}
{\hat{\mathbf C}'}_{\mathcal S'}+{v_k}{\hat{\mathbf C}'}_{\mathcal S'}\mathbf j_k\mathbf j^{\rm H}_k{\hat{\mathbf C}'}_{\mathcal S'}&={\hat{\mathbf C}_{\mathcal S\backslash k}},\label{C_de_approa}\\
-v_k{\hat{\mathbf C}'}_{\mathcal S'}{\mathbf j}_k&= {\hat{\mathbf c}}_{k}\label{C_de_approb}\\
v_k &= \hat{C}_{kk},\label{C_de_approc}\\
{\hat{\mathbf w}'}_{\mathcal S'} - {u}_k\hat{\mathbf C}'_{\mathcal S'}{\mathbf j}_k&={\hat{\mathbf w}_{\mathcal S\backslash k}},\label{C_de_approd}\\
{u}_k&={\hat{w}_k}.\label{C_de_approe}
\end{align}
\end{subequations}
Thus, ${\hat{\mathbf C}'}_{\mathcal S'}$ can be updated by substituting (\ref{C_de_approb}) and (\ref{C_de_approc}) in (\ref{C_de_approa}), i.e.,
\begin{align}\label{cov_up_deact}
{\hat{\mathbf C}'}_{\mathcal S'} = {\hat{\mathbf C}_{\mathcal S\backslash k}} - v_k{\hat{\mathbf C}'}_{\mathcal S'}\mathbf j_k\mathbf j^{\rm H}_k{\hat{\mathbf C}'}_{\mathcal S'} = {\hat{\mathbf C}_{\mathcal S\backslash k}} - \frac{{\hat{\mathbf c}}_{k}{\hat{\mathbf c}}^{\rm H}_{k}}{\hat{C}_{kk}}.
\end{align}
Similarly, ${\hat{\mathbf w}'}_{\mathcal S'}$ can be updated by substituting (\ref{C_de_approb}) and (\ref{C_de_approe}) in (\ref{C_de_approd}), i.e.,
\begin{align}\label{mean_up_deact}
{\hat{\mathbf w}'}_{\mathcal S'} = {u}_k\hat{\mathbf C}'_{\mathcal S'}{\mathbf j}_k+ {\hat{\mathbf w}_{\mathcal S\backslash k}} = {\hat{\mathbf w}_{\mathcal S\backslash k}} -  \frac{\hat{w}_k}{{\hat{C}}_{kk}}{\hat{\mathbf c}}_{k}.
\end{align}
According to $v_k = \hat{C}_{kk}$ (\ref{C_de_approc}) and $u_k = {\hat{w}_k}$ (\ref{C_de_approe}), $\Delta_k$ (\ref{delta-k-deactive}) can be simplified as
\begin{align}\label{delta-k-deactivesim}
\Delta_k =-\ln\frac{{\hat{C}}_{kk}}{\tau} - \frac{|{w}_k|^2}{{\hat{C}}_{kk}}-\ln\frac{\rho}{1-\rho}.
\end{align}
\subsection{Calculating the post variance ${\mathbf v}_{\rm A}^{\rm post}$ (\ref{post_vars_A}) of $\mathbf z$}\label{calpostcov}
Now we calculate the variance of ${z}_n$ given $q({\mathbf w}_{\hat{S}}|\tilde{\mathbf y})$ and $q({\boldsymbol \theta}|\tilde{\mathbf y})$.
Let ${\mathbf b}_n^{\rm T}=[{\rm e}^{{\rm j}(n-1)\theta_1},{\rm e}^{{\rm j}(n-1)\theta_2},\cdots,{\rm e}^{{\rm j}(n-1)\theta_{\hat{K}}}]$ be the $n$th row of ${\mathbf A}_{\hat{S}}$. Then $z_n={\mathbf b}_n^{\rm T}{\mathbf w}_{\hat{S}}$. The variance is
\begin{align}\label{calvar}
{\rm Var}[z_n]&={\rm E}[|z_n|^2]-|{\rm E}[z_n]|^2={\rm E}[|{\mathbf b}_n^{\rm T}{\mathbf w}_{\hat{S}}|^2]-|{\rm E}[{\mathbf b}_n^{\rm T}{\mathbf w}_{\hat{S}}]|^2\notag\\
&={\rm tr}\left[({\mathbf C}_{\hat{S}}+\hat{\mathbf w}_{\hat{S}}\hat{\mathbf w}_{\hat{S}}^{\rm H}){\rm E}[{\mathbf b}_n^{*}{\mathbf b}_n^{\rm T}]\right]-|\hat{\mathbf b}_n^{\rm T}\hat{\mathbf w}_{\hat{S}}|^2.
\end{align}
where
\begin{align}\label{boutprod}
{\rm E}[{\mathbf b}_n^{*}{\mathbf b}_n^{\rm T}]=\hat{\mathbf b}_n^{*}\hat{\mathbf b}_n^{\rm T}+{\rm diag}\left({\mathbf 1}_{\hat{K}}-|\hat{\mathbf b}_n|^2\right).
\end{align}
Substituting (\ref{boutprod}) in (\ref{calvar}), one obtains (\ref{post_vars_A}).
\bibliographystyle{IEEEbib}
\bibliography{strings,refs}

\end{document}